Schweizerische Eidgenossenschaft
Confédération suisse
Confederazione Svizzera
Confederaziun svizra

Federal Department of the Environment,
Transport, Energy and Communications DETEC

**Swiss Federal Office of Energy**
Energy Research and Cleantech

**Final report from 21 April 2026**

---

# CEST-DC

## Circular Economy Synergies and Trade-offs in Data Centres

---

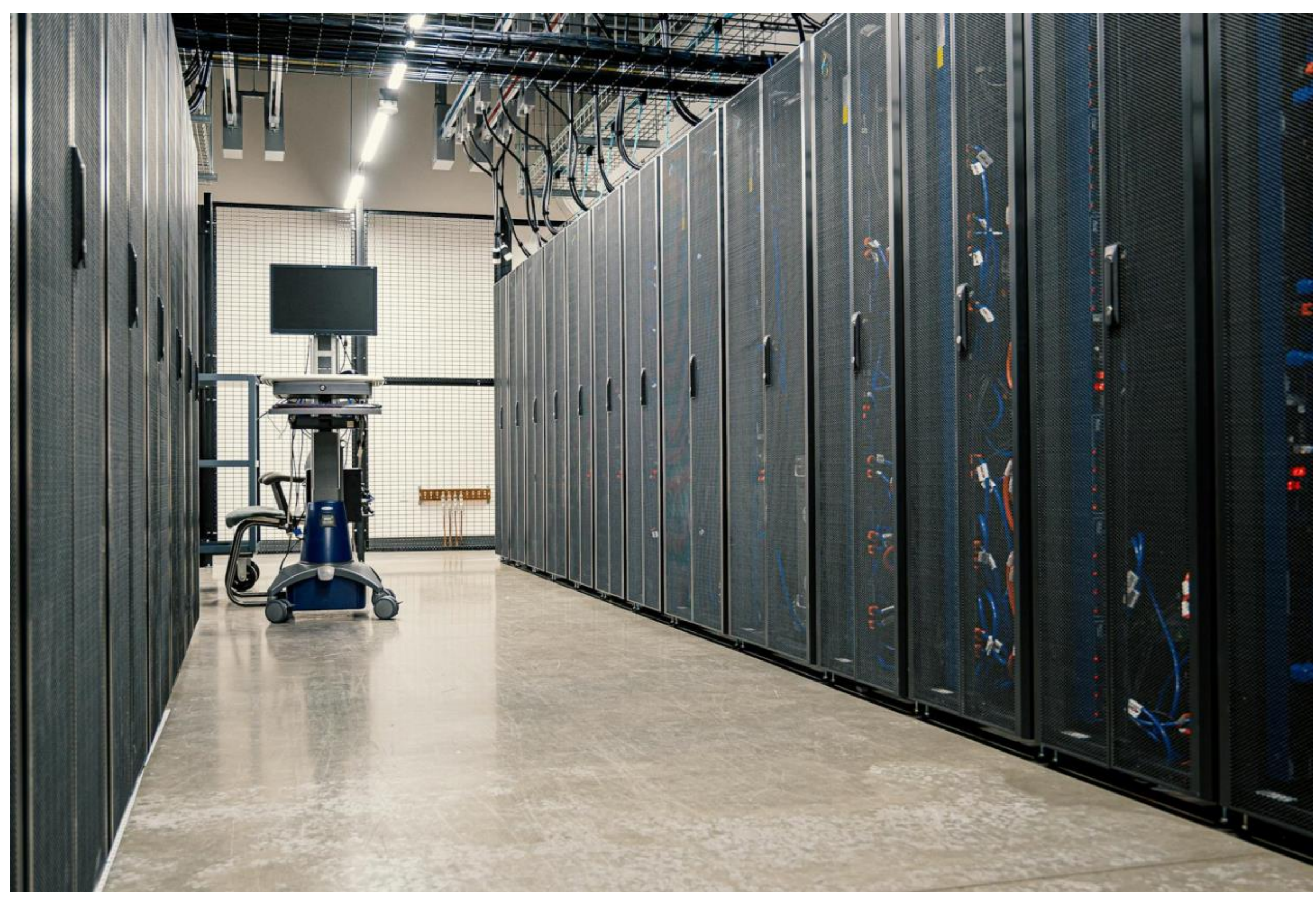

Picture source: Brett Sayles, 2020

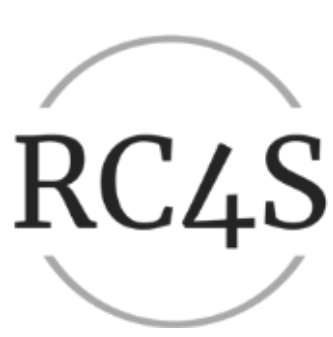

**Authors:**
Vlad C. Coroamă, Roegen Centre for Sustainability, vlad.coroama@roegen.ch
Oana Dumbravă, Roegen Centre for Sustainability, oana.dumbrava@roegen.ch

**The authors bear the entire responsibility for the content of this report and for the conclusions drawn therefrom.**

# Summary

**From a sustainability perspective, data centre (DC) growth comes with several challenges and a few opportunities**. Thise report brings these dimensions together, analysing DC sustainability from a circular economy perspective, revealing existing synergies and trade-offs.

**For energy consumption, the widely used PUE metric has several fundamental limitations**: Being a relative metric, it does not inform on the overall energy consumption of the DC. It considers the two main types of DC infrastructures (cooling and power provisioning) jointly, mixing their individual efficiencies, and is thus too coarse. It is also skewed: Due to data limitations, it attributes both consumption of server fans and transformation losses of the power supply units (PSUs) to the IT energy instead of the non-IT energy.

**The PUE does not measure compute efficiency but infrastructure efficiency**. The latter has already reached very good values and progress will only be incremental. Compute energy, however, is exploding, so a metric reflecting compute efficiency across heterogeneous loads, while challenging, is needed.

**To cope with the increased DC power densities and minimise losses, power provisioning is shifting**: from alternating current to direct current and towards higher voltages. This is the perfect opportunity to define a new metric for the power provisioning efficiency, which should be both comprehensive (including PSU losses) and disjunct from the cooling overhead. A new set of energy metrics for DCs could thus comprise i) compute efficiency, ii) transformation efficiency, and iii) cooling overhead.

**On-site water consumption depends on the cooling technologies deployed**. Waterside economisers and water-cooled chillers (via cooling towers) have a high water consumption, while dry coolers, air-cooled chillers, and airside economisers do not. However, these last technologies are often adiabatically supported, yielding them at least temporarily water consuming as well. **Upstream water consumption is mainly due to electricity production**, and its main source is the behind-the-dam evaporation for hydro power plants.

**Two trade-offs emerge between energy and water used in cooling as well as between on-site and upstream water consumption**. The two are related: Using a water-consuming technique generally lowers the energy required in cooling, which also lowers the water consumed upstream in power generation (if any). For very ‘wet’ electricity, there is little competition, and it is generally worth spending some more on-site energy to save both electricity and the related upstream water. For ‘dry’ electricity, however, there is a trade-off between the two.

**Energy circularity can be achieved through waste heat recovery**. There is a trade-off between high heat reuse (close to populated areas) and low energy consumption (far North). And the waste heat has limited uses and is not the same energy quality, a fact not reflected by current metrics. A more relevant metric would consider the avoided energy through heat recovery instead of the amount recovered.

**The production impact of microelectronics is poorly understood**. From the three production phases (mining, refining, and fabrication), the mineral refining / purification is the least understood in terms of sustainability. Prolonging lifespans is always beneficial.

**Material circularity can be achieved by interpreting general circularity principles such as the 9R framework** in the context of DCs. Possible circularity-enhancing measures exist for both the DC itself and, more importantly, for its microelectronics. The measures can be categorised into product design, process design and business models, choice of materials, and operating conditions. Together, they have effects across all circularity levels.

**The relation between DCs and the power grid is complex and multifaceted**. Modern DCs possess three crucial and novel features that present new challenges for the grid. Various measures are used to mitigate these challenges, the most important of which are workload shifting, battery storage and on-site generation. These measures have, in turn, further consequences, which are both beneficial and detrimental. They can bring offer grid flexibility as well as innovations in the field of energy. But they also bring noise, pollution, and GHGs, and compete with the energy sector for resources.

# Zusammenfassung

**Aus Nachhaltigkeitsperspektive bringt das Wachstum von Rechenzentren (RZ) verschiedene Herausforderungen, aber auch einige Chancen mit sich**. Dieser Bericht beleuchtet diese Dimensionen aus der Perspektive der Kreislaufwirtschaft und zeigt bestehende Synergien und Zielkonflikte auf.

**Beim Energieverbrauch weist die weit verbreitete PUE-Kennzahl grundlegende Schwächen auf**: Als relative Grösse sagt sie nichts über den Gesamtenergieverbrauch aus. Da sie Kühlung und Stromversorgung gemeinsam erfasst, ist sie zu grob. Zudem ist sie verzerrt: Aufgrund unzureichend feiner Messungen werden sowohl der Verbrauch der Serverlüfter als auch die Umwandlungsverluste der Netzteile (PSUs) der IT-Energie statt der Nicht-IT-Energie zugerechnet.

**Die PUE misst nicht Recheneffizienz, sondern die Effizienz der Infrastruktur**. Letztere hat bereits sehr gute Werte erreicht; weitere Fortschritte werden wohl nur noch schrittweise erfolgen. Der Energiebedarf für Rechenleistung hingegen explodiert. Deshalb braucht es – trotz aller Herausforderungen – einen Indikator der Recheneffizienz über heterogene Rechenlasten hinweg.

**Angesichts steigender Leistungsdichten verschiebt sich die Stromversorgung von Wechsel- zu Gleichstrom und zu höheren Spannungen** – eine gute Gelegenheit, neue Metriken zu definieren: i) Recheneffizienz, ii) Umwandlungseffizienz und iii) Kühlaufwand. Die letzteren zwei trennt Umwandlungsverluste von dem Kühlaufwand und sollten umfassend definiert sein (PSU-Umwandlungen und Lüfter enthaltend).

**Der lokale Wasserverbrauch im RZ hängt von den eingesetzten Kühltechnologien ab**. Wasserseitige Economiser und über Kühltürme wassergekühlte Kältemaschinen verbrauchen viel Wasser, Trockenkühler, luftgekühlte Kältemaschinen und luftseitige Economiser hingegen nicht – wobei letztere oft adiabatisch unterstützt werden und damit zumindest zeitweise ebenfalls Wasser benötigen. Der **vorgelagerte Wasserverbrauch entsteht hauptsächlich durch die Stromerzeugung**, insbesondere durch Verdunstung aus den Stauseen der Wasserkraftwerke.

**Zwischen Energie- und Wasserverbrauch für die Kühlung sowie zwischen lokalem und vorgelagertem Wasserverbrauch bestehen Zielkonflikte**. Wasserintensive Kühltechniken senken in der Regel den Energiebedarf und damit auch den vorgelagerten Wasserverbrauch. Bei sehr wasserintensivem Strom besteht wenig Konkurrenz, und es lohnt sich meist, vor Ort etwas mehr Wasser einzusetzen, um Strom und den damit verbundenen vorgelagerten Wasserverbrauch zu sparen. Bei wasserarmem Strom besteht dagegen ein echter Zielkonflikt.

**Energiekreisläufe lassen sich durch Abwärmenutzung schliessen**. Dabei besteht ein Zielkonflikt zwischen hoher Wärmenutzung (Nähe zu Siedlungen) und niedrigem Energieverbrauch (hohe Breitengrade). Eine sinnvollere Kennzahl würde die durch Wärmerückgewinnung vermiedene Energie abbilden statt der zurückgewonnenen Menge.

**Die Produktionsauswirkungen der Mikroelektronik sind wenig erforscht**. Von den drei Produktionsphasen – Bergbau, Raffination und Fertigung – ist die mineralische Raffination in Bezug auf Nachhaltigkeit am wenigsten verstanden. Längere Nutzungsdauern sind stets vorteilhaft.

**Materialkreisläufe lassen sich durch Anwendung allgemeiner Kreislaufprinzipien wie der 9R-Prinzipien** auf RZ erschliessen. Massnahmen gibt es sowohl für das RZ selbst als auch – deutlich wichtiger – für die darin eingesetzte Mikroelektronik. Mögliche Massnahmen betreffen Produktdesign, Prozessgestaltung, Geschäftsmodelle und Materialeinsatz und wirken auf allen Kreislaufebenen.

**Die Beziehung zwischen RZ und Stromnetz ist komplex und vielschichtig**. Moderne RZ verfügen über drei entscheidende und neue Eigenschaften, die das Netz vor zusätzliche Herausforderungen stellen: grosse und konzentrierte Lasten sowie steile Lastrampen. Um diese zu mindern, kommen verschiedene Massnahmen zum Einsatz, besonders Lastverschiebung, Batteriespeicher und Erzeugung vor Ort. Diese Massnahmen haben wiederum weitere Folgen – positive wie negative. Sie können dem Netz Flexibilität bringen und Innovationen im Energiebereich fördern. Zugleich verursachen sie Lärm, Verschmutzung und Treibhausgase und konkurrieren mit dem Energiesektor um wichtige Ressourcen.

# Résumé

**Du point de vue de la durabilité, la croissance des centres de données (CD) s'accompagne de plusieurs défis et de quelques opportunités**. Ce rapport examine ces dimensions sous l'angle de l'économie circulaire, en mettant en évidence les synergies et les compromis existants.

**Concernant la consommation énergétique, l'indicateur PUE présente plusieurs lacunes fondamentales** : en tant que mesure relative, il ne renseigne pas sur la consommation totale d'énergie du CD. Il traite conjointement les deux principaux types d'infrastructures (refroidissement et alimentation électrique), en mêlant leurs efficacités respectives, ce qui le rend trop grossier. Il est également biaisé : faute de données suffisantes, il attribue à la fois la consommation des ventilateurs de serveurs et les pertes de transformation des alimentations (PSU) à l'énergie IT plutôt qu'à l'énergie non-IT.

**Le PUE ne mesure pas l'efficacité de calcul mais celle de l'infrastructure**. La première a déjà atteint de très bonnes valeurs et les progrès ne seront qu'incrémentaux, tandis que l'énergie de calcul explose. Un indicateur reflétant l'efficacité du calcul sur des charges hétérogènes serait donc nécessaire.

**Face à la hausse des densités de puissance, l'alimentation évolue vers le courant continu et des tensions plus élevées**. C'est l'occasion idéale de définir un nouvel indicateur d'efficacité de l'alimentation électrique, à la fois complet – incluant les pertes des PSU – et distinct de la surcharge liée au refroidissement. Un nouvel ensemble de métriques énergétiques pour les CD pourrait ainsi comprendre : i) l'efficacité du calcul, ii) l'efficacité de conversion, et iii) le surcoût de refroidissement.

**La consommation d'eau sur site dépend des technologies de refroidissement utilisées**. Les économiseurs à eau et les refroidisseurs à eau (via tours de refroidissement) consomment beaucoup d'eau, contrairement aux aéroréfrigérants, refroidisseurs à air et économiseurs à air – bien que ces derniers soient souvent assistés adiabatiquement et consomment donc aussi de l'eau par intermittence. **La consommation d'eau en amont provient principalement de la production d'électricité**, notamment de l'évaporation derrière aux barrages hydroélectriques.

**Des compromis apparaissent entre énergie et eau utilisées pour le refroidissement, ainsi qu'entre consommation d'eau locale et en amont**. Les deux sont liés : une technique consommatrice d'eau réduit généralement l'énergie de refroidissement, ce qui diminue aussi l'eau consommée en amont pour la production électrique. Avec une électricité très « humide », la concurrence entre les deux est faible, et il vaut généralement la peine de consommer un peu plus d'eau sur site pour économiser à la fois l'électricité et l'eau associée en amont. Avec une électricité « sèche », en revanche, un véritable arbitrage subsiste.

**La circularité énergétique passe par la récupération de chaleur résiduelle**, avec un compromis entre réutilisation élevée (proximité des zones peuplées) et faible consommation (grands Nord). La chaleur résiduelle a des usages limités et n'est pas de même qualité énergétique – un indicateur plus pertinent mesurerait l'énergie évitée grâce à la récupération plutôt que la quantité récupérée.

**L'impact de production de la microélectronique reste mal compris**. Des trois phases (extraction, raffinage, fabrication), le raffinage/purification minéral est le moins documenté en termes de durabilité. Prolonger la durée de vie est toujours bénéfique.

**La circularité matérielle peut être atteinte en interprétant les principes généraux, comme le cadre 9R**, dans le contexte des CD. Des mesures existent tant pour le CD lui-même que, plus important, pour sa microélectronique. Elles se répartissent en conception produit, conception de procédés et modèles d'affaires, choix des matériaux et conditions d'exploitation. Ensemble, elles agissent sur tous les niveaux de circularité.

**La relation entre CD et réseau électrique est complexe et multidimensionnelle**. Les CD modernes introduisent de nouveaux défis pour le réseau. Les mesures d'atténuation telles que le déplacement de charge, le stockage par batteries et la production locale offrent de la flexibilité et stimulent l'innovation, mais génèrent aussi du bruit, des émissions et des gaz à effet de serre, tout en entrant en concurrence avec le secteur énergétique pour les ressources.

# Main findings («Take-Home Messages»)

- Due to their increasing power densities, DCs will transition towards direct current power provisioning at higher voltages and liquid cooling. This opportunity should be seized to define new efficiency metrics, in particular a generic compute efficiency metric and to correct the PUE by disentangling transformation losses from cooling overheads, and correctly representing IT and non-IT components.
- Water usage is most accurately represented by the total water consumption, on-site as well as upstream during electricity generation. Two relate trade-offs are between energy and water deployed in cooling as well as on-site and upstream water. These trade-offs are more pronounced for dry electricity; the "wetter" electricity production is, the more it is worth spending on-site water to save both energy and upstream water.
- While production impacts of DC devices are poorly understood, circularity principles are always beneficial. In a CapEx-dominated world, they also align with economic interests. Circularity levers exist on four levels: product design, process design and business models, materials, and operational practices. General circularity frameworks can be applied for their assessment.
- The interactions between DCs and the power gird are complex and multifaceted. The large flexible loads of DCs, their power density as well as the steep ramps of AI compute represent novel challenges for the grid. They are being met with several mitigation measures, most important of which are battery systems and on-site power generation. This mitigation brings in turn several adjacent drawbacks, but also beneficial consequences.

# Contents

# List of figures

# List of tables

# List of abbreviations

| | |
|---|---|
| a.c. | alternating current |
| AI | artificial intelligence |
| ASIC | application-specific integrated circuit |
| BESS | battery energy storage system |
| CAGR | compound annual growth rate |
| CapEx | capital experndíture |
| CE | circular economy |
| CHP | combined heat and power |
| CRAC | computer room air conditioner |
| CRAH | computer room air handler |
| d.c. | direct current |
| DC | Data Centre |
| DC-HMS | data centre modular hardware system |
| DPP | digital product passport |
| DX | direct expansion |
| EoL | end-of-life (environmental lifecycle phase) |
| EPD | environmental product declaration |
| ERCOT | electricity reliability council of Texas |
| ERE | energy reuse effectiveness |
| ERF | energy reuse factor |
| ETSI | European Telecommunications Standards Institute |
| FLAP-D | Frankfurt, London, Amsterdam, Paris, and Dublin (traditional data centre conglomerates in Europe) |
| FLOP | floating point operation |
| GHG | greenhouse gas |
| GPU | graphics processing unit (a type of processor deployed for accelerated computing, initially used in computer graphics and now increasingly in machine learning algorithms) |
| HBM | high-bandwidth memory |
| HDD | hard disk drive |
| HVAC | Heating, Ventilation, and Air Conditioning |
| IC | integrated circuit |
| IEA | International Energy Agency |
| IT | information technology |
| KPI | key performance indicator |
| LCA | lifecycle assessment |
| LFL | large flexible load |
| LLI | large load interconnection |
| ML | machine learning |
| MSB | main switchboard |
| OpEx | operational expenditure |
| PDU | power distribution unit |
| PoG | point-of-generation (for electricity) |
| PPA | power purchase agreement |
| PSU | power supply unit |
| PUE | power usage effectiveness |
| RDHx | rear-door heat exchanger |
| REE | rare earth element |
| RTR | right to repair |
| SERT | server efficiency rating tool |
| SFOE | Swiss Federal Office of Energy |
| SMR | Small Modular Reactor |
| SSD | solid-state drive |
| SST | solid state transformer |
| TDP | thermal design power |
| TPPE | total power provisioning efficiency |
| TPU | tensor processing unit (a custom-built AI accelerator, developed and deployed by Google/Alphabet for machine learning algorithms) |
| TUE | total usage efficiency |
| UAE | United Arab Emirates |
| UPS | uninterruptable power supply |
| VRM | voltage regulator module |
| WCF | water consumption factor |
| WUE | water usage effectiveness |

# 1 Introduction and context

## 1.1 Sustained computational growth and its consequences

Driven by cloud computing and artificial intelligence (AI), data centres are currently growing at a sustained pace. For about a decade until the late 2010s, efficiency gains were able to largely keep up with the growing demand (Masanet *et al.*, 2020). In recent years, however, the demand growth far outweighs efficiency progresses.

As a consequence, the energy consumption of data centres as well as various types of related environmental impacts, are steadily growing. Some of these dimensions are relatively well studied, in particular the energy consumption and related greenhouse gas (GHG) emissions during the operational phase of DCs. Others such as the water consumption of DCs are met with a growing interest. Others impacts such as the production-related impacts of microelectronics or the applicability of circular economy principles to DCs, however, are very little understood.

There are, however, not only detrimental consequences. Data centre growth can also act as a catalyst for new research and development in related domains such as the energy sector. Some of these dimensions are global, other of local importance. Greenhouse gases and resource depletion have global consequences, while the water consumption, pollution, and e-waste generation are of rather local relevance.

## 1.2 Circular economy synergies and trade-offs

Beyond the traditional focus on energy efficiency, a crucial question in this context is how circular economy principles can be interpreted and further developed for the particular characteristics of data centres. Hereby, circularity principles can include both the circularity of energy (e.g., through the recovery of DC waste heat) but also material circularity.

Given the variety of possible consequences, both beneficial and detrimental, spanning multiple dimensions, some will likely be synergistic, while others are bound to be at odds with each other. Such conflicting goals might include possible trade-offs between energy and water consumption in the cooling of data centres, between on-site water consumption for cooling and water consumed upstream during electricity generation, between energy efficiency and material consumption (e.g., by upgrading the power provisioning systems), or between minimising the energy consumption and the ability to reuse the DCs waste heat.

## 1.3 Scope and objectives

The main aim of this report is to bring all these dimensions together, highlighting the various current trends with an influence on data centre sustainability. Moreover, it also sets out to emphasise the complex interactions among sustainability-relevant dimensions and indicators, and the consequences of various DC design choices.

To the extent possible, the scope of the analysis is limited – but hence also focused – on the sustainability of a single data centre. Of course, a data centre can never be regarded entirely independent of its surrounding context. When analysing the relation between a DC and the power grid, for example, the density of other DCs in the region is highly relevant.

While such context needs to be considered, the report aims to stay as much as possible focused on single data centres. In particular, it does not aim to estimate or predict the current or future global energy consumption of all data centres or the corresponding GHG emissions. Such analyses have been the focus of numerous other analyses; they are out of scope here.

By instead staying sharply focused, the report can reveal more richness of detail across the different sustainability dimensions of individual data centres. Analysing various DC sustainability dimensions, it provides a broad understanding of existing synergies and trade-offs to its main stakeholders: data centre planners, operators, and policymakers.

## 1.4 Structure of the report

The individual dimensions addressed in the report are the following: Chapter 2 discusses data centre energy consumption and the suitability (or lack thereof) of existing metrics for its assessment. It criticises one widely used metric – the power usage effectiveness – and argues for more meaningful alternatives. Keeping the focus on DC energy, Chapter 3 addresses both legacy and novel paradigms for DC power provisioning, and argues for a novel metric for its assessment as well.

Chapter 4 focuses on all aspects of a data centre's water consumption. It first presents various cooling techniques, showing how some of them induce more water consumption than others. It also discusses the upstream water consumption due to electricity generation. After addressing various indicators for water consumption, both absolute and relative, it then discusses two related trade-offs: between on-site and upstream water consumption, and between total water consumption and energy consumption in a DC.

Chapters 5 through 7 revolve around circularity: Chapter 5 first discusses energy circularity through waste heat recovery, its limitations and induced trade-offs. Chapter 6 presents the impacts occurring during the production of microelectronics devices (including the very low data quality in this domain) and shortly the lifespans of devices. Building on these insights, Chapter 7 then sheds light on the material circularity in DCs, interpreting known circular economy frameworks in the context of data centres.

Chapter 8 embarks on perhaps the most ambitious undertaking of the entire report: Highlighting the very complex relationship between data centres and the power grid. It starts with an analysis of three crucial and novel features of modern DCs and shows how these present new challenges for the grids. It then discusses various mitigation measures being deployed against these challenges, which have beneficial consequences beyond the initial challenges that called for them. Unfortunately, however, the mitigation measures do not only bring unintended beneficial consequences, but equally unintended detrimental ones as well. Both the additional benefits and detrimental consequences are presented as well.

Finally, Chapters 9 through 11 round up the report: Chapter 9 presents DC sustainability standards and recommendations as well as voluntary certification and labelling schemes. Chapter 10 summarises the main conclusions of the work, while Chapter 9 provides a comprehensive list of open issues and presents important paths for future research.

# 2 Energy consumption and suitable metrics

## 2.1 Sources of data centre energy consumption

Data centres require power both for their core activity of delivering IT services such as computing and storage, as well as for the required supporting infrastructure. Computation takes place in servers, which are dedicated, high-performance computers specifically designed to run server software and deliver various services to other computers or devices over a network. Servers are "built with powerful processors (such as CPUs, GPUs, TPUs, chiplets, or others), extensive memory, and substantial storage, along with multiple network interfaces and redundant power supplies for maximum reliability" (Coroamă *et al.*, 2025).

The supporting DC infrastructure comprises various ancillary services required to support the reliable operation of the servers. The infrastructure thus includes DC cooling and further environmental controls, power transformation and delivery, backup power systems, and others (such as lighting and ventilation of offices, etc).

### 2.1.1. Data centre topology

Modern DCs are organised either in several server rooms, or as one large industrial-sized server "mega-hall". While the latter topology is today's standard for hyperscale DCs, the former is typically for colocation DCs (and legacy facilities often built in pre-existing buildings).

Whether in a mid-sized room or in a huge hall, servers are organised in rows of server *racks* or *cabinets*. Historically, the first ones to appear were servers racks in the 1990s (von Hollen, 2024), and their advent arguably represents one of the most notable developments towards the appearance of modern data centres (Coroamă *et al.*, 2025). Racks are open-frame structures of standardised dimensions designed to house servers, networking equipment, and storage devices (Jose, 2019), while featuring adaptable shelving to accommodate different types and sizes of equipment. Racks are essential components for optimising space utilisation, efficient cable management, and proper airflow within a DC.

Cabinets provide the same organisational benefits and supporting structure for IT equipment but are additionally fully enclosed with doors and side panels. Thereby, they offer additional security and environmental control. Locking mechanisms and access controls prevent unauthorised access, which can be important features in a colocation setting. The environmental control includes more fine-granular thermal control (inherently through the enclosure but also via e.g. built-in fans) and noise reduction. Racks, on the other hand, are less expensive. Through their open design, they can also allow the maximum of airflow, if required.

### 2.1.2. IT and non-IT energy

All the energy used in DCs (for computation, storage, networking, and supporting activities) transforms into heat. Because it cannot keep accumulating and to protect the equipment, the heat needs to be continuously removed from the building. Removing the computing heat from the data centre requires itself some additional energy, which is one of the main sources of energy consumption that is not related to computing. It is discussed in more detail in Chapter 4 jointly with the water consumption of DCs.

The other major source of non-IT energy consumption in DCs are the losses that occur during power provisioning (i.e., power supply and distribution), especially due to transformation losses. They are discussed in the subsequent Chapter 3.

In line with the literature, we differentiate between the power used in a DC by the IT equipment, $P_{IT}$, and the power used by the DC infrastructure for all such ancillary services, $P_{non-IT}$. The corresponding energy consumptions over a time period are the integrals over time of the power consumptions:

$$E_{IT} = \int P_{IT}\, dt\,; \quad E_{non-IT} = \int P_{non-IT}\, dt \tag{1}$$

Equation 1 is on purpose kept conceptual, but it obviously applies over any concrete time interval $(t_1, t_2)$. Similarly, the overall energy consumption in a DC over any time interval equals the sum of the two, as shown in Equation 2:

$$E_{DC} = E_{IT} + E_{non-IT} = \left(\int P_{IT} + \int P_{non-IT}\right) dt \tag{2}$$

where $E_{DC}$ is the total energy consumption of the DC over a certain period of time.

## 2.2 Efficiency metrics and server/cooling trade-off

While not explicitly mentioned so far, it is evident that in a data centre, $E_{IT}$ is *useful energy*, while $E_{non-IT}$ is necessary, but essentially *wasted energy*. Consequently, for the last two decades the focus of DC energy efficiency research and praxis was on minimising the wasted portion. To this end, a well-known and widely used DC efficiency metric is the “power usage effectiveness” (PUE), defined as the ration between total DC energy and its useful (i.e., IT) part:

$$PUE = \frac{E_{DC}}{E_{IT}} \tag{3}$$

The ratio can be expressed as either the momentary PUE (case in which the equation shows the ratio of power consumptions), or as average PUE over a certain period (such as a day, month, season or year), case in which it is expressed as ration of energy consumptions, as done in Equation 3.

As $E_{DC} = E_{IT} + E_{non-IT}$, and thus $E_{DC} \geq E_{non-IT} \Rightarrow E_{DC} \geq 1$. That the PUE is always supra-unitary becomes even clearer if rewriting Equation 3 as:

$$PUE = \frac{E_{DC}}{E_{IT}} = \frac{E_{IT} + E_{non-IT}}{E_{IT}} = 1 + \frac{E_{non-IT}}{E_{IT}} \tag{4}$$

The ideal PUE would equal 1.0, and it would be achieved when the wasted energy $E_{non-IT}$ would decrease and reach 0. While this ideal is likely unachievable, the PUE has been asymptotically decreasing towards 1.0 over the last two decades.

### 2.2.1. Infrastructure savings largely exploited already

When the metric was first introduced 2007 by the data centre industry association The Green Grid (DLR, 2023), the average worldwide PUE was about 2.5 (Azura, 2024). This means that a larger portion of energy was spent for supporting infrastructure than for computing. In the meanwhile, hyperscale DCs have achieved yearly PUE average values smaller than 1.1 – across all its data centres, Google, for example, has an average value around 1.09 (Google, 2025b). And large colocation DC are not far behind, despite the more challenging environment, in which the operator cannot influence the servers installed by the customers on its premises (Coroamă, 2025).

Far less efficient DCs exist as well, in particular legacy and enterprise DCs. They drive the unweighted average global PUE towards a value around 1.5 (Azura, 2024). Even hyperscale and colocation DCs in less favourable climates (i.e., warm and/or humid ones) cannot achieve the same PUEs as data centres far north in Europe or North America in cold and dry climates.

Nonetheless, for an increasing share of DCs – and in particular for the largest hyperscale and colocation ones that drive overall global DC consumption – the infrastructure overhead today already amounts to less than 10% of the IT-related energy. There is only little potential left, as also revealed by the figures published by e.g. Google, which show a largely flat curve since around 2020 (Google, 2025b).

### 2.2.2. Conceptual limitations of the PUE metric

These diminishing returns of efforts towards more efficient DC infrastructure question the adequacy of the PUE as a metric going forward. Beyond this pragmatic issue, however, the PUE also has several fundamental conceptual limitations.

**General limitations of relative indicators**

As any relative metric, an obvious limitation of the PUE is that is does not capture in any way the *absolute energy consumption* of a data centre. As an analogy, cars today are far more energy efficient than they were in the early days of the automobile. Cumulatively, however, their impact in the late 19th or early 20th century was negligible. Meanwhile, through their ubiquitousness and sheer numbers, today they represent one of the important sources of GHGs and thus culprits for the climate crisis.

Similarly, despite poorer PUE values, until recently data centres used to have a negligible global energy consumption and related environmental impact: As recently as 2018, their cumulative global energy consumption was around 200 TWh yearly (Masanet *et al.*, 2020). In the few years since, however, this has already reached around 500 TWh and will continue growing to around 700 – 1,000 TWh yearly by 2030 (IEA, 2025a; Kamiya and Coroamă, 2025). Beyond the growing global impact, the ever-increasing *power density* of data centres is a growing source of concern, as subsequent chapters will discuss. All of this happens while the PUE keeps improving and is asymptotically approaching its ideal value, as discussed above.

**Does the PUE indicate useful output?**

But even as a relative metric, it is questionable whether the PUE reflects the most relevant ratio, as it would suit the outstanding importance it is given in the DC energy and efficiency domain. Another main limitation, recognised early on (Yuventi and Mehdizadeh, 2013), is that the PUE measures *infrastructure efficiency* and not *computing efficiency*. Normally, any energy efficiency metric relates a useful output to the energy input (Coroamă *et al.*, 2025):

$$EE_O = \frac{useful\ output}{energy\ input} \tag{5}$$

With the compute energy $E_{IT}$, the PUE does indeed feature an energy input in its denominator. The numerator, though, is $E_{DC}$; or rather – and conceptually more accurate – $E_{non-IT}$ according to Equation 4. The non-IT energy consumption is, however, hardly a useful output. By relating infrastructure energy to computing energy, the PUE is purely concerned with the efficiency of the overhead. What might constitute more meaningful metrics is discussed in Section 2.3 below.

**PUE does not accurately reflect IT and non-IT energy**

Finally, even disregarding the limitations above (that it does not capture absolute values in an age when they become increasingly important, and that it does not measure the most useful of outputs), the PUE does not even accurately reflect what it is supposed to do: the ratio between IT and non-IT energy consumption.

To conserve cooling energy, numerous DC operators are increasing the server inlet temperatures, which implies less energy required for cooling, in particular when chillers are involved. A temperature raise, however, typically increases the speed of server fans that need to circulate more air to compensate for its higher temperature. And this additional fan consumption can become substantial, as it correlates with the third power of the air speed generated, albeit starting from a low value (Coroamă, 2025).

Consequently, for the overall consumption of a data centre for a given computer load, there is a trade-off between cooling energy and server fan energy. Both these consumptions, which contribute to cooling and not to compute, should count towards the non-IT-energy of a data centre. For practical reasons, however, the server fan energy is not available separately: Server fans are integrated into servers and their energy consumption cannot be separately assessed. Their energy consumption thus counts towards the denominator of the PUE ratio and not its numerator as it should.

This opens the door for conscious or unwilling PUE manipulation: Raising inlet temperatures in a data centre will decrease infrastructure energy and thus lower the numerator in Equation 4. The additionally induced server fan energy, instead of having the countereffect of increasing the same numerator, instead lower the denominator of the same fraction, as it is considered as IT energy. Taken together, these two effects thus necessarily decrease the PUE irrespective of how the overall cooling energy (i.e., infrastructure + server fans) actually evolved (Coroamă, 2025).

Another important source of non-IT energy consumption in data centres are transformation losses that occur throughout the several power transformations within a DC. Similar to cooling, a part of these transformations occurs in servers and is thus also wrongly counted as IT energy instead of non-IT one. Data centre power provisioning is discussed in the following Chapter 3, and this PUE-related issue in its Section 3.3.

These are not fundamental conceptual limitations but pragmatic ones, due to the lack of semantically appropriate data. They represent nevertheless important limitations, especially in times when the PUE approaches its theoretical ideal value and remaining differences are sensitive to fine nuances. They are also not easy to overcome limitations, due to the way servers are built with integrated power transformation and – for air-cooled servers – with integrated fans.

## 2.3 Meaningful metrics for data centre energy consumption

Despite its limitations, DC-related environmental regulations or labels often assign a central role to the PUE. The German "Energieeffizienzgesetz" (energy efficiency law), for example, requires three minimal requirements from DCs, the most important of which is a maximum PUE value (Bundestag, 2023). The Swiss Datacenter Efficiency Association (SDEA) awards the "SDEA Label" for efficiency in DCs. Until recently, its bronze, silver, and gold levels were exclusively driven by the PUE, although they now also consider the (related) GHG emissions and will soon be expanded to cover relative water consumption as well (SDEA, 2025).

The European reporting and rating scheme for data centres (European Commission, 2024a) established in 2024, on the other hand, requires a plethora of DC key performance indicators (KPIs) to be reported. They include

- general parameters such as the DCs type (enterprise, colocation, co-hosting), total floor area, and total computer room floor area,
- absolute energy, ICT capacity and data traffic metrics (total DC energy consumption, total computing performance and storage capacity, maximum traffic bandwidth and actual traffic),
- various cooling parameters (intake air temperature, type of refrigerant, and the amount of "cooling degree days"), and
- many more parameters related to the use of renewable energy, amount of heat reuse (see Chapter 5), water consumption (see Chapter 3), and power grid services (See Chapter 8).
- Finally, based on these parameters, DCs also need to report four relative "sustainability indicators", one of which is the PUE.

In this context of PUE limitations on one side and an avalanche of indicators on the other, an interesting question is what could represent a meaningful set of energy-related DC sustainability metrics. This set should be minimal yet catch crucial sustainability-related information. For now, we focus on energy and not on water and other parameters that will follow later. This question is addressed in Section 2.3.4 below, after first discussing a more meaningful efficiency metric for data centres.

### 2.3.1. The more relevant efficiency metric: Computing efficiency

As discussed in Section 2.2, one of the main limitations of the PUE is that it does not relate the useful computational output to the input energy. To do so, a metric for the *computing efficiency* is required, i.e., how much useful *computing output* can be generated per amount of energy input.

Although difficult to implement in practice, as discussed shortly, such metric would describe much more semantically accurate the efficiency of a data centre in the spirit of Equation 5. Generically, the computing efficiency can be expressed as (Coroamă *et al.*, 2025):

$$EE_{compute} = \frac{useful\ computations}{energy\ input} \tag{6}$$

The meaning of “useful computations” is hereby context-dependent. (Coroamă *et al.*, 2025) discuss it in the context of servers and not data centres, contextualising compute efficiency for three server types:

- For CPU-based general-purpose servers, the widespread “Server-Side Java operations per second” (*SSJ_ops*) is used (SPEC, 2018):

$$EE_{x86-SSJ} = \frac{SSJ_{performance}}{P}\ in\ \left[\frac{SSJ_{OP}/s}{Watt}\right]$$

- For GPU and TPU servers deployed in accelerated computing, a more technical metric was used, reflecting a focus on floating point operations (FLOPs):

$$EE_{GPU} = \frac{GPU_{performance}}{P}\ in\ \left[\frac{GFLOP/s}{Watt}\right]$$

- For the hashing operations required for proof-of-work in blockchain systems, the efficiency metric deployed was:

$$EE_{ASIC} = \frac{Hash_{performance}}{P}\ in\ \left[\frac{TH/s}{Watt}\right]$$

As can be seen in the bullets above, these definitions divide both the numerator and the denominator by time, thus indicating average *amount of computing per time per power*. This is of course perfectly equivalent to removing the division by time and indicating *total amount of computing per energy*, as required in Equation 6.

These examples further show the challenge in measuring useful computations. Some domains, such as general-purpose servers, have established metrics. When these do not exist, technical metrics such as the number of atomic mathematical operations (i.e., floating point operations) can be deployed as proxy.

For the computing efficiency of entire data centres, the European DC reporting scheme (European Commission, 2024a) requires the declaration of another widely-used server efficiency metric, called “server efficiency rating tool”, SERT (SPEC, 2022). SERT is a unitless metric that represents the weighted efficiency of a server across various types of workloads, each of which is originally measured as FLOPs/second. The European reporting scheme actually only requires the declaration of the upper part of the metric, the “ICT capacity for servers” ($C_{SERV}$), to then compute a DC-wide SERT value for the by adding all of the data centre’s compute capacity, and dividing it by the energy input of the entire DC.

### 2.3.2. Measuring compute efficiency

We argue that a metric to capture the computational efficiency of the data centre is necessary. Such relative energy metric is more important than the PUE, as it relates the computational (i.e., useful) output to the energy input supporting it. This is even more the case when the infrastructure efficiency (as measured by the PUE) has already achieved a very high level, and future gains are likely only marginal, as argued in Section 2.2.1.

It remains to be defined which metric exactly to use for the computational efficiency. Considering Equation 6, definitions are required for both the numerator and the denominator of the ratio:

- For the denominator of Equation 6, two options are conceivable: to use the IT-energy $E_{IT}$ only, or the entire data centre energy input $E_{DC}$.
- For the numerator of Equation 6, a metric is required that is applicable to various types of workloads in the DC and also to various types of servers.

The first question is easier: We argue in favour of the second option, the entire energy consumed by the DC. Otherwise, the ratio would only be a weighed average of the computing efficiencies of the servers, ignoring the non-IT energy of the DC infrastructure. Relating useful computations to the entire DC energy, however, seems a semantically more meaningful definition of the *computing efficiency of a data centre*. By including the energy wasted on non-IT overhead, it is a comprehensive DC compute efficiency metric.

A metric to quantify useful computations is more challenging. For a more homogeneous DC, for example one consisting only of general-purpose servers, a dedicated metric such as SERT (SPEC, 2022) or SSJ_ops (SPEC, 2018) is meaningful, as suggested by the (European Commission, 2024a). These metrics, however, have little meaning for accelerated computing, and today's DCs become more heterogeneous, combining volume servers with accelerated computing.

A more technical metric such as the total number of FLOPs could serve as common ground for all types of computations. However, different types of logical processing unit and corresponding server types have inherently distinct compute efficiencies. While the metric could apply to all sorts of compute, it would require the usage of some weights to reflect the shares of different computation types. This remains for now an unanswered question requiring future research.

Additionally, colocation DCs confront an additional challenge in measuring the useful computations that take place in the DC. As the colocation provider does not own the servers, which belong to its customers, it also does not have access to the amount of compute being performed by the servers on its premises. This is not a conceptual, but a pragmatic limitation; expressed in terms of uncertainties, it does not represent an ontological uncertainty, but an epistemic one.

To circumvent it, we suggest the solution put forward by the European Commission for the reporting of colocation DCs, i.e., gathering computation data from their customers: "Colocation data centre operators may gather the key performance indicators from their colocation customers, if necessary, by setting up an anonymous internal reporting mechanism" (European Commission, 2024a).

Based on these considerations, we rewrite Equation 6 for the compute efficiency of data centres as:

$$DCCE = \frac{Compute_{DC}}{E_{DC}} \tag{7}$$

where $DCCE$ is the suggested metric "data centre compute efficiency", describing the compute efficiency of the DC, $Compute_{DC}$ is a measure for the total computations performed in the DC over a certain period, and $E_{DC}$ the total energy consumption of the DC over the same period.

### 2.3.3. Previous work and future prospects for the metric

The idea of an indicator for the computing efficiency of the data centre is of course not new; it is, in fact almost as old as the PUE itself and has the same origin. Although the original document can no longer be found, (Maagøe, 2022) argues that a similar metric, called the "data centre energy productivity" (DCeP) was already put forward by The Green Grid in 2008. By 2014, the indicator had been agreed upon and published (The Green Grid, 2014) after "more than five years in the making" (Verge, 2014).

JouleX, a former software company that was specialised in monitoring and controlling the power consumption of computers and other network-attached devices, claimed in 2011 that "PUE is dead" and it should be replaced with a performance-per-Watt (PPW) metric (Davidson, 2011). To cope with the issue of heterogeneous workloads and hardware adequacy for these workloads, its suggestion was to introduce a performance indicator to normalise the individual performance types for comparability.

A 2022 study commissioned by EDNA reviewed data centre energy metrics and their relevance (Maagøe, 2022). At the end of the review, the study ends by suggesting two indicators. One of them, called "data centre functional efficiency" (DCFE), takes the same line, being designed to measure the total functions delivered over a reported period divided by the total DC energy over that period. A normalisation approach is part of the proposal, making the different types of DC services comparable within the DCFE. Via a case study, the study also shows how the indicator could work to measure simultaneously compute operations, storage, and networking for a DC (Maagøe, 2022); however, it does not compare different types of workloads.

These long ongoing efforts, however, did not translate into a broad reporting even remotely comparable to the PUE. There are likely several culprits for this:

- The numerator $Compute_{DC}$ is conceptually difficult to assess. "Useful work" strongly varies between e.g. AI inference, web serving, and HPC, and even more so if storage and networking services are to be taken into account.
- Communicating a (potentially large) set of indicators does not bring the elegance, simplicity, understandability, and comparability of a single number such as the PUE. However, collapsing heterogeneous workloads into one figure risks to "distort and oversimplify" the story, as the Uptime institute also points out (Lawrence, 2025).
- Beyond these fundamental issues, there are also organisational challenges. Especially in colocation/cloud environments, workloads need to be gathered from a variety of parties, all of which (or at least a majority, if extrapolations are permitted) need to be willing – or coerced – to reveal this data.

Despite both these challenges and the largely fruitless efforts of the past, today's situation seems more favourable than ever for a DC efficiency metric to be finally established. As discussed above, the European reporting and rating scheme for data centres (European Commission, 2024a) requires, among other parameters, the measurement and reporting of "ICT capacity for servers". A prerequisite would be the existence of a metric. Which might become even more relevant, as the European reporting scheme might also become the basis for minimum efficiency requirements (European Commission, 2025b).

In this context, the Uptime institute urges the EU to establish a compute per energy metric, believing that this might be the "data centre metric for the 2030s" (Lawrence, 2025). And in 2025, The Green Grid has adopted a metric to calculate the "IT work capacity" (ITWC) for CPUs and plans to extend it towards storage and network (The Green Grid, 2025). Additionally required, and arguably more challenging, will be to cover accelerated computing as well.

### 2.3.4. A balanced set of metrics: Crisp, yet significant and informative

This section returns to the quest for a meaningful set of metrics to capture the energy-related sustainability aspects of a data centre, while striking a balance between eloquence and comprehensiveness.

**Indicator 1: compute efficiency, $\boldsymbol{DCCE}$.** In agreement to (Maagøe, 2022), we also suggest that the compute efficiency of the DC should be the main indicator, whether called DCCE, DCeP (The Green Grid, 2014), DCFE (Maagøe, 2022), or ITWC (The Green Grid, 2025). Defining a meaningful (i.e., generic yet precise) indicators for $Compute_{DC}$ is a crucial and urgent topic for future research.

**Indicator 2: total energy consumption, $\boldsymbol{E_{DC}}$.** Complementary to $DCCE$, the overall energy consumption $E_{DC}$ is also relevant for a variety of reasons, including modelling the energy grid, understanding potential local challenges and opportunities, and aggregations towards regional or national figures. DC operators, however, are quite reluctant to disclose it, as it allows conclusions on the size of its business. If feasible, gathering this data – at least anonymously as the European reporting scheme does – would be very valuable.

Infrastructure energy consumption and efficiency remain important topics as well, the issues highlighted above notwithstanding. They are the topic of the following two chapters. Whether the PUE is the most adequate metric to assess infrastructure efficiency, will be discussed in the subsequent chapters as well.

# 3 Power provisioning and its efficiency

The largest non-IT energy uses in a data centre are typically cooling/HVAC and electrical power chain losses (power distribution and backup power supply). All other sources – e.g., DC humidifiers, lighting, security, controls, offices, etc. – usually contribute to a much smaller extent. Data centre cooling, which is an important contributor towards the water consumption of a DC, will be discussed in the next chapter while addressing the water consumption of DCs. The current chapter discusses power distribution in DCs.

## 3.1 Traditional, alternating current power distribution inside the DC

Data centres are typically supplied with power from the medium-voltage grid. While the medium-voltage level is generally defined as the range 1 – 36 kV (SwissGrid, 2025), DCs draw power from the higher range of this interval, typically between 11 – 36 kV. To this end, a local substation of the utility company steps down high-voltage electricity to medium voltage.

Very large hyperscale DCs may be connected directly to high-voltage power, having their own dedicated on-site high-voltage substation. Whether owned privately or by the utility company, however, substations are technologically very similar. This difference will thus be irgnored, and the rest of the discussion focues from the medium-voltage grid on.

Through a series of transformers, the medium voltage is successively transformed to the tension of a few volts required by the chips and other components on the server board. As discussed by several sources, such as an Nvidia whitepaper (Huntington and Tu, 2025) but also in an almost 2 decades-old Intel paper (Pratt, Kumar and Aldridge, 2007) – and reflected in Figure 7 of this latter source – an archetypal pipeline for today's power distribution is as follows:

- Transformers reduce medium-voltage alternating current (a.c.) to low-voltage a.c., typically at 480 V or 400 V, for distribution. This conversion has about 4% losses, and thus 96% efficiency (Coroamă *et al.*, 2025).
- The current then typically passes through the uninterruptable power supply (UPS), which provides backup power to the DC. UPSs usually store energy in battery banks, although they may also rely on flywheels, fuel cells or other technologies (Van Geet and Sickinger, 2024). Double conversion ("online") UPS systems, to which we refer here and as shown in (Huntington and Tu, 2025), offer the highest level of protection, as they guarantee zero transfer time, i.e., no interruptions when switching to battery power.

  Additionally, by internally converting a.c. current to d.c. and then back to a.c., they isolate sensitive DC equipment from any power anomalies, not only from grid failure (Schneider Electric, 2025). The output voltage and frequency is detached from the input supply, thus being protected from any disturbances.

  The price of this security and quality of electrical current are the transformation losses; a high-end double conversion UPS system today has an efficiency of about 96%. Most UPSs thus also have an "eco-mode", in which most of the power bypasses the double conversion inverter. The eco-mode is often deployed in areas with stable power and it has an efficiency of 98.5 – 99% (Riello, 2024; ABB, 2025).
- In the server room, the power is further stepped down to 230 / 115 V a.c. via power distribution units (PDUs), which also have losses of about 4% (Raritan, 2016).
- Finally, inside servers, the power supply units (PSUs) convert this 230/115 V a.c. power into direct current (d.c.) at 12 V, 5 V, or 3.3 V, to supply the various componnets on the server board. Depending on the electric load, this last step has an efficiency of about 90-96% (Infineon, 2020).

## 3.2 Direct current power distribution

Taken together, the losses along the a.c. power distribution presented in Section 3.1, can amount to anywhere between 13% (for an efficiency chain of 96% * 98.5% * 96% * 96%) and 20% (for a cumulated efficiency of 96% * 96% * 96% * 90%) of the total energy input; or even more, in legacy designs. A data centre design with less power transformation steps and a smaller voltage drop would help towards minimising these losses.

### 3.2.1. The relationship between voltage and transmission losses

Higher voltage inside the data centre also becomes a necessity due to the ever-increasing power density of server racks: While 1-2 decades ago, the power of a rack was a few kW, it now grew to 20-30 kW for general-purpose servers. Some GPU racks already require more than 100 kW of power, and next-generation racks are designed for up to 300 kW (Coroamă *et al.*, 2025). According to the electric power law, however, power is the product of voltage and current:

$$P = V * I \tag{8}$$

where $P$ is the power (measured in Watt), $V$ the voltage (measured in Volts), and $I$ the current (measured in Amperes). The rising power that needs to be delivered to each rack or cabinet inside the DC can thus only be achieved by raising either the voltage or the current, or both.

Using high currents in the distribution of power, however, represents an issue for two reasons:

- As reflected by Ohm's law, as current travels through conductors, the inherent resistance $R$ of the metal causes the voltage to drop by the time it reaches the server:

  $$V_{drop} = I * R \tag{9}$$

  where the voltage drop along the conductor $V_{drop}$ is directly proportional to both the current $I$ and the resistance $R$. If the current is too high, the voltage drop can be so severe that the equipment receives power below its required operating threshold.
- Arguably more importantly, however, electricity flows through a conductor are also subject to heat losses (known as "Ohmic heating"). Combining the electric power law (also known as "Watt's law") in Equation 8 with Ohm's law (Equation 9) yields Joule's law for Ohmic heating:

  $$P_{loss} = I^2 * R \tag{10}$$

  where $P_{loss}$ is the power loss along the conductor.

The power losses along the conductor are thus directly proportional to the square of the current and the resistance of the conductor. It is true that this can be partly mitigated by lowering the resistance of the conductor (which is inversely proportional to its cross-sectional area), but this requires thicker copper cables, inducing weight to the server racks, increasing costs, and adding to the building's structural requirements. Such optimisation is thus inherently limited.

And once the potential through thicker conductors has been tapped, smaller currents will lead ceteris paribus to an even more substantial optimisation potential, as the losses correlate to the square of the current (which is incidentally also the reason why long-distance power transmission uses very high voltages - hundreds of kilovolts – at relatively low currents).

#### 3.2.2. Higher voltage and direct current power distribution

To distribute the increasing power required by the racks of the DC, increasing the voltage and not the current is thus generally preferred. Given the quadratic relation from Equation 10, a twofold increase of the voltage can compensate for a fourfold increase of the power distributed at no additional losses.

As briefly mentioned in the beginning of Section 3.2, a higher voltage in the DC would have the additional benefits of smaller voltage drops during the power transformation, and consequently higher efficiencies / smaller losses. The lower part of Figure 1 in (Srivastava and Petty, 2025) shows the envisioned new DC power distribution architecture:

- A solid state transformer (SST), which transforms the medium voltage a.c. down to 800 V d.c. directly. Such SSTs perform not only step-down and a.c. to d.c. transformation, but also the stabilising function of double conversion UPS systems (Heron, 2026). They thus eliminate the need for a UPS early in the pipeline, enabling the energy storage to move closer to the servers. SSTs are still being developed, but their future efficiencies are projected to be around 98.5% (Ampersand, 2026; Heron, 2026).
- The 800 V d.c. current from the SST is sent directly to the compute racks, where it is stepped down directly to the 12 V required by the chips and further server components – or in future to 48 V, since the next envisioned transition is for 48 V at rack level (Coroamă *et al.*, 2025). In an interview with Prof. Dr. Drazen Dujic (EPFL Switzerland), who actively works on the development of both SSTs and these novel PDUs, we found out that the estimated efficiency will be 97% or higher.

Compared to four sources of losses on the traditional power distribution pipeline as discussed in Section 3.1 (which are reduced to only three in the current high-efficiency DCs, where the UPS does only convert the power to d.c. but no longer back to a.c., and the distribution is at 380 V d.c.), this future pipeline will only have three sources of losses. Compared to the 13-20% legacy losses and current losses of 11-17%, these future losses, if achieved, will only amount to 4.5% (for the two efficiencies or 98.5% and 97%, respectively).

An additional advantage of d.c. power distribution is that it is more compact than a.c. power supply, as the interview with Prof. Dujic revealed. (Heron, 2026) argues that the space savings can be up to 70%. This leaves more space for compute equipment, which is both a desirable goal and an economic incentive for any DC operator.

## 3.3 Measuring power conversion efficiencies

A question arising after the previous discussion is how to account for these power transformation losses discussed in Sections 3.1 and 3.2. Do UPSs and transformers at the different levels count as "infrastructure" and does consequently the PUE account for them? Does a different metric exist? Is perhaps even a new metric required?

#### 3.3.1. PUE levels and the included transformation losses

The PUE definition in Equations 3 and 4 in Section 2.2 seems straightforward enough: Any energy that was not used by IT equipment should count towards $E_{non-IT}$. In practice, however, the picture is less clear, for both conceptual and pragmatic issues. It turns out that the point of measurement of $E_{IT}$ is particularly relevant.

Several standards define where $E_{IT}$ can or should be measured. Part 2-2 of the European norm on DC facilities and infrastructures focuses on the power supply and distribution in (EN, 2019). In agreement to standard ES 205 200-2-1 by the European Telecommunications Standards Institute (ETSI, 2014), the norm defines three "levels" for the PUE. These levels define three points of measurement for $E_{IT}$. As also summarised in Table 1, these levels become increasingly inclusive with the losses they consider:

- Level 1 ("basic") requires a measurement after the UPS, thus including the losses of the main transformer (from medium voltage to low voltage) and the UPSs.

- Level 2 (“intermediate”) requires measurements at the output of the PDUs or remote power panels. It thus additionally includes the PDU transformation losses.
- Finally, level 3 (“advanced”) requires the measurements at the “point of connection”, i.e., the power strip / rack plug, and thus the server PUS input. This level additionally includes the losses along the conductors to the server racks.

Table 1: Overview of the three PUE levels as defined by (ETSI, 2014; EN, 2019), together with the resulting power losses that are included or not in the PUE.

| Power component losses | Level1 (basic) | Level 2 (intermediate) | Level 3 (advanced) |
|---|---|---|---|
| Main transformers (MV → LV) | Included | Included | Included |
| UPSs | Included | Included | Included |
| PDUs | **Excluded** | Included | Included |
| Cabling to rack | **Excluded** | **Excluded** | Included |
| PSUs | **Excluded** | **Excluded** | **Excluded** |

Moving from Level 1 to Level 3, the PUE is a better reflection of reality, as more and more of the non-IT energy consumption is accurately reflected in the numerator of Equation 4. Moreover, as already discussed for server fans in Section 2.2.2, not only will such consumptions not be correctly devised by the numerator, but – because by definition what is not $E_{non-IT}$ gets counted as $E_{IT}$ – it will appear on the wrong side, i.e., in the denominator of Equation 4. Any such wrongly not accounted for energy loss will thus skew the PUE for both reasons, making it appear better (i.e., lower) than it actually is.

There is thus no conceptual reason to devise any PUE level other than level 3. In the past, there might have been pragmatic reasons, as the higher levels require more thorough measurement methods. While over a decade ago, when these standards were first introduced, it was less common to measure the power at rack level, nowadays this is quite common in large DCs. Colocation DCs in particular need to be able to charge their customers for the exact amount of energy they used, so for business reasons, they need to measure the power at rack level. But for the optimisation and maintenance reasons, this is now common for hyperscalers as well. There is thus no remaining motive – neither conceptual nor pragmatic – to devise levels 1 or 2 PUEs.

As Table 1 shows, however, even this most comprehensive and widely used definition of the PUE does not correctly devise the PSU losses. As with the server fans discussed in Section 2.2.2, they are included on the wrong side of the PUE ratio.

### 3.3.2. A better metric for power supply and transformation efficiency?

Even if the PUE did devise all transformation losses correctly, it would still mix them with other infrastructure overhead, most notably that of DC cooling and further environmental controls. It would, however, be valuable to have a metric that would account for the cumulated efficiency (or, conversely, losses) within the *entire DC power supply and transformation chain*. What we would aim for could be defined as:

$$TPPE = \prod_{i \in \{T\&D\}} \eta_i \qquad (11)$$

where the newly defined metric “Total Power Provisioning Efficiency” ($TPPE$) is the product of the individual efficiencies $\eta_i$ for all power transformation and distribution processes, $i \in \{T\&D\}$.

Equation 11 would have to be differently instantiated for individual power provisioning chains. For the traditional chain described in Section 3.1, for example, the instantiation would be

$$TPPE_{trad.} = \eta_{Trans} * \eta_{UPS} * \eta_{PDU} * \eta_{cables} * \eta_{PSU} \quad (12)$$

where $\eta_{Trans}$ is the efficiency of the medium voltage to low voltage transformer, $\eta_{UPS}$ of the uninterruptable power supply, $\eta_{PDU}$ of the power distribution units, $\eta_{cables}$ the efficiency of the power distribution within the DC's conductors, and $\eta_{PSU}$ the efficiency of the servers' power supply units.

According to forum discussions, some vendors and consultants already use this or similar terms, such as "power chain efficiency" (PCE), "end-to-end efficiency", or "power path efficiency". Their usage, however, and the exact efficiencies covered, have quite some variability and are context dependent.

Sometimes, for example, the usage only accounts for transformations inside the DC building, thus not including the medium-voltage to low-voltage transformer (which admittedly has a very high efficiency). Some usages ignore the PSU efficiency, such as LBNL's "electrical power chain tool" (LBNL, 2023). Finally, some usages also include the losses within the ICs, i.e., the voltage regulator module (VRM) on the motherboard, which converts the 12V from the PSU to the roughly 1V voltage required by the chips.

To our knowledge, however, no such concept has so far been officially formalised by any norm. Some proposals cover parts of it, but also often include too much. A 2013 paper, for example, suggested a "PUE-type metric for the IT equipment rather than for the data center" (Patterson *et al.*, 2013). The idea behind this metric called "ITUE" was to distinguish within servers the energy used for computations from the one non-computationally wasted. Equivalent to Equation 3,

$$ITUE = \frac{E_{IT}}{E_{CC}} \quad (13)$$

where $E_{CC}$ is the "total energy into the compute components" (Patterson *et al.*, 2013), while $E_{IT}$ (which is the same $E_{IT}$ used since Equation 3) also comprises the energy used by server fans as well as transformation losses in the PSUs and VRMs – see Figure 1 in (Patterson *et al.*, 2013).

The ITUE metric thus represents a cleaning of the PUE from all the non-compute overhead, wrongly included in the IT energy $E_{IT}$. Combining PUE and ITUE, a second metric suggested by the authors is the "total usage efficiency" (TUE):

$$TUE = PUE * ITUE = \frac{E_{DC}}{E_{IT}} * \frac{E_{IT}}{E_{CC}} = \frac{E_{DC}}{E_{CC}} \quad (14)$$

Equation 14 shows how the TUE ultimately fulfils the original intent of the PUE, to separate the non-computational overhead from the usefully deployed energy. It does so by correcting the PUE and extracting the non-computational energy from the IT energy.

Unfortunately, the TUE never gained popularity, probably due to the challenges in assessing it, in particular of distinguishing server fan and PSU energy from the compute energy used in servers. Conceptually it might be a bit too coarse as well: While it clearly devises the compute energy $E_{CC}$, it does not distinguish nuances within the rest of energy (which could be called, similarly to $E_{non-IT}$, as "non-CC" energy, $E_{non-CC} = E_{DC} - E_{CC}$).

In particular, $E_{non-CC}$ does not distinguish between:

- cooling energy, which comprises both the infrastructure cooling and the server fans, and
- the losses due to power provisioning, both those covered by the PUE level 3, and those in the PSUs, and possibly VRMs.

As a consequence, $E_{non-CC}$ (and with it the TUE) correctly address the boundaries of $E_{non-IT}$ and the PUE. They do not introduce, however, a desired differentiation between different types of overhead, which would be valuable for detailed energy efficiency and ultimately sustainability assessments of data centres.

### 3.3.3. Suggested metric for power provisioning

**Indicator 3, total power provisioning efficiency $\boldsymbol{TPPE}$**: The suggestion of this report is thus to define the TPPE according to Equation 11, as a metric for the cumulative efficiency of power provisioning to a data centre. The unachievable ideal value for the TPPE would be 1; the higher a value that can be reached, the more desirable the power supply and distribution architecture enabling it. For each system architecture, Equation 11 needs to be individually interpreted. Additionally, a few principial questions remain open and need to be addressed:

- How can $\eta_{PSU}$ be accurately assessed?
- Should the final, on-motherboard transformation be also considered, and thus included in Equation 12 as additional $\eta_{VRM}$?

  (Patterson *et al.*, 2013) suggest so, but it is in our opinion not entirely clear whether this semantically belongs to the power transformation chain of the DC or the inherent needs of microelectronics. Additionally, this transformation is likely out of reach for DC system architects, so this point requires clarification.
- At the other end of the electricity supply chain, should the substation losses during the transformation from high-voltage to medium-voltage be accounted for as well?

  The standards define $E_{DC}$ as being measured later, at the point of handover from the utility company to the DC operator: “the consumption) from the utility (grid) supply(s) shall be measured at the input to the [medium-voltage to low-voltage] transformer” (ETSI, 2014). For its hyperscale DCs, however, Google includes these losses into its “comprehensive PUE” (Google, 2025b), and they do occur because of the DC. Whether substation – and perhaps also high-voltage power transmission from the power plant – losses should be additionally considered also requires clarification.

Based on this indicator, the absolute power and energy losses can be trivially computed as:

$$P_{PP} = (1 - TPPE) * P_{DC}; \quad E_{PP} = (1 - TPPE) * E_{DC} \tag{15}$$

where $P_{PP}$ and $E_{PP}$ are the cumulative power and energy losses due to power provisioning, respectively.

# 4 Water footprint and environmental trade-offs

Data centres use water both *directly* for operation and *indirectly* during the production of electricity. During operation, water is being deployed mainly for DC cooling and possibly also for air humidification (Lei and Masanet, 2022). For electricity generation, the source of water usage depends on the type of power plant.

On an orthogonal dimension resides the question of the semantically meaningful metrics for water use. Two indicators have been typically deployed: *water withdrawal* and *water consumption* (Coroamă and Schien, 2026). Water withdrawal denotes all water removed from its source, irrespective of what happens later to that water. By contrast, consumption refers to the subset that is withdrawn but not directly returned after usage (Peer, Grubert and Sanders, 2019). From an environmental perspective, the more relevant indicator is the definitive water usage, i.e., the water consumption (Coroamă and Schien, 2026).

The chapter first explores mainstream cooling technologies employed in DCs and qualitatively their influence on direct water (and energy) consumption (Section 4.1). It then discusses the indirect water consumption (Section 0) and a metric to assess the two (Section 4.3). Section 4.4 finishes by discussing two trade-offs: between water and energy as well as on-site and upstream water consumption.

## 4.1 Data centre cooling technologies and their impact on water consumption

Data centres are increasingly energy intensive, the energy used in DCs transforms into heat, and needs to be continuously removed from the building.

This so-called *heat rejection* is typically performed by a series of (gaseous and/or liquid) cooling circuits which pass the heat one to the next one through heat exchangers (Coroamă, 2025), until it ultimately reaches its ultimate destination. More often than not, this final heat destination is a natural sink such as the atmosphere or a body of water, although the heat can also be reused for a secondary purpose such as district heating (the subsequent Chapter 5 discusses heat reuse).

To perform the heat rejection, countless types of cooling technologies – including specific sub-flavours, additions and combinations of technologies – are deployed in data centres. Not all are suited for every climate, and some of them are more water-intensive than others. Unfortunately, there is no widely accepted taxonomy of such cooling technologies. Some literature sources distinguish three main categories (Microsoft, 2023), other seven (Lei and Masanet, 2022), and other ten (Lei *et al.*, 2025).

This report does not aim to introduce yet another taxonomy. Instead, it presents the most important technologies used in DC cooling, describing why some of them induce water consumption and to which extent.

### 4.1.1. Dry coolers

Provided outside temperatures are cold enough, the cooling liquid from the computer room air conditioner or handler (CRAC / CRAH unit), which previously absorbed the heat from a first air circuit within the server room (Coroamă, 2025), can be sent to a closed liquid loop outside the building. Through the heat-transferring pipes, the heat is transferred from the fluid in this loop to the outside air, which acts as heat sink.

There are various shapes for dry coolers: Figure 1 shows a simple schematic of a horizontal one; several such units would result in the typical table-like shape of a dry cooler. Other shapes include vertical V-shaped designs.

There is **no water consumption** involved; hence the name “dry cooler”. As the heat is transported via a fluid (typically water or a water-glycol mixture), this cooling method is sometimes also called “dry fluid cooler”. As dry coolers use the atmosphere as heat sink without additional energy consumption (or rather, with minimal energy consumption for the fans helping the air circulation and pumps circulating the fluid), dry coolers are sometimes also referred to as “free air cooling” – this term, however, is more usual in the context of airside economisers (see Section 4.1.5 below).

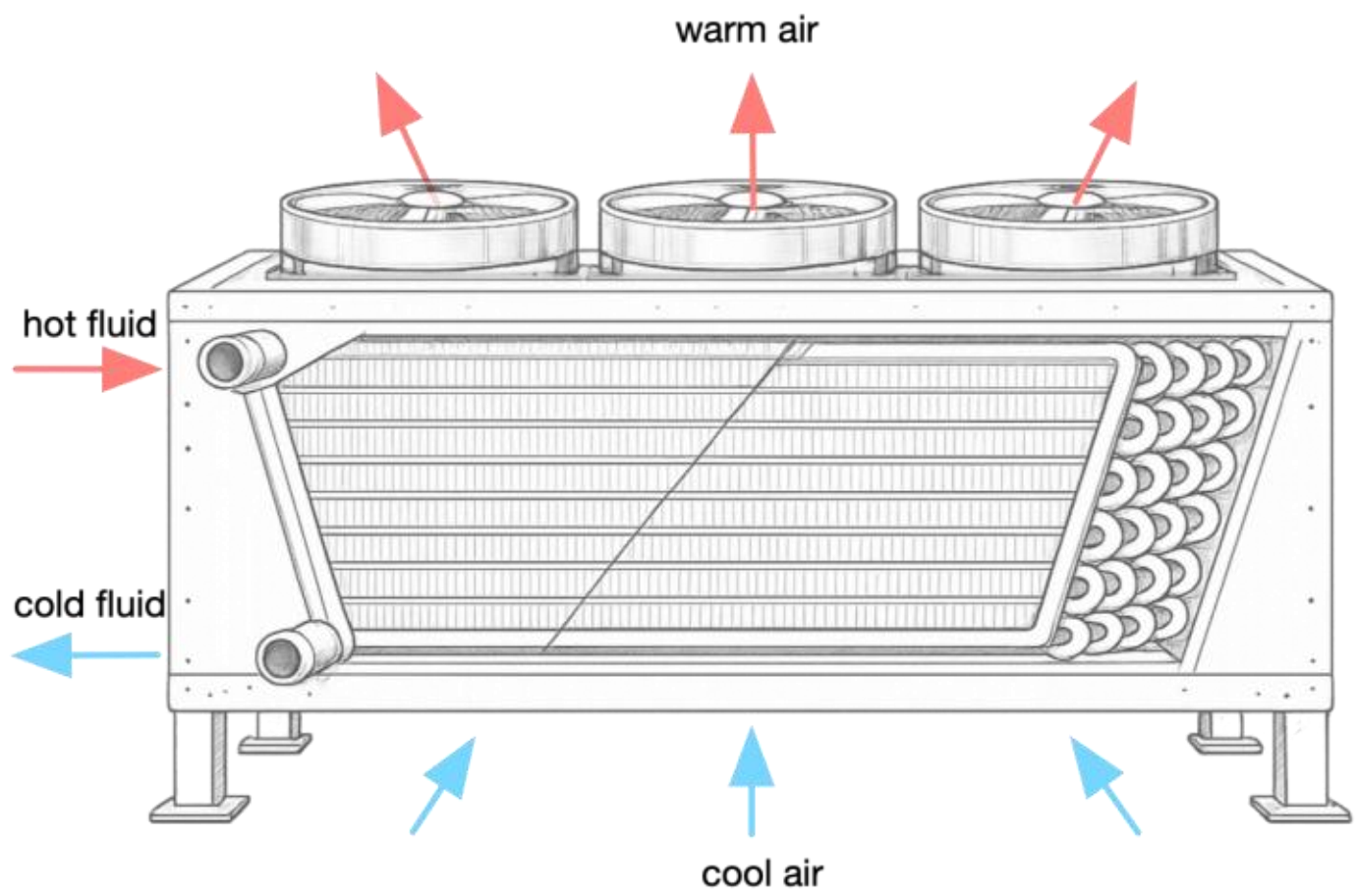

Figure 1: Simple schematic of a dry cooler (generated by Nano banana pro, corrected via MS Paint).

#### 4.1.2. Mechanical refrigeration: Chillers and direct expansion units

When outside temperatures are not cold enough, however, a way of transferring heat from a colder body (i.e., the heated cooling fluid) to a relatively warmer body (i.e., outside air) is needed. This is performed via specific types of heat pumps, such as chillers and direct expansion (DX) units. They work along the same principle, so this section describes just chillers in more detail.

Chillers are one of the most widely deployed cooling technologies – in buildings generally, and also in data centres. These mechanical refrigerators are closed thermodynamic cycles that use a compressed refrigerant to move heat from a lower to a higher temperature level. They consist of four main components (Coroamă, 2025):

1. evaporator ((low pressure, low temperature),
2. compressor,
3. condenser (high pressure, higher temperature), and
4. expansion valve (or expansion device).

The functioning principle of the chiller is sketched below; for more information see e.g. (Patel *et al.*, 2025):

- The chiller-internal refrigerant enters the *evaporator* as a low-pressure liquid–vapor mixture at low temperature; colder than the cooling liquid from the CRAH/CRAC unit that passes through a coil. Heat thus flows from the cooling liquid to the chiller refrigerant. As a consequence, and due to its low pressure and boiling point, the refrigerant evaporates and leaves the evaporator as saturated vapour.
- The *compressor* takes in this low-pressure vapor and compresses it to a higher pressure. This raises the saturation temperature of the refrigerant. The work added by the compressor also raises its temperature, bringing it to a higher value than the heat sink, enabling it thus to reject heat in the next step.
- This happens in the *condenser*, where the heat is transferred to the next cooling circuit (or, for air-cooled chiller, directly to the atmosphere, see below), the refrigerant condenses back to liquid, transferring both the heat it absorbed while evaporating in the evaporator and the work of the compressor.

- Finally, the high-pressure liquid passes through an expansion valve, where both the pressure and the temperature of the liquid further drop, part of the liquid flashes to vapour, and with this low-pressure, low-temperature liquid-vapour mix, the cycle can restart.

**Chillers by themselves do not induce any water consumption**. They are, however, not the last heat transfer step, and several of the methods used downstream on the heat rejection path often do consume water, as discussed below.

### 4.1.3. Air-cooled chillers

In the chiller's condenser, the heat is transferred either to the next cooling circuit, which is the one that will ultimately transfer the heat to the ultimate heat sink (usually the atmosphere) or directly to the atmosphere. This happens

- either via *condenser coils* exposed directly to the outside air, typically on the roof of the data centre – in this case, the *chiller is air-cooled*,
- or via *cooling towers*, case in which the *chiller is water-cooled*.

Figure 2 shows a schematic representation of an air-cooled chiller deployed in a DC. Evaporator, compressor and expansion valve are similar to those of a regular chiller. The condenser, however, is not internal to the chiller. It consists instead of the condenser coils that are in contact with the outside air, to which they reject the heat directly (very similar to how a dry cooler works). An **air-cooled chiller does not induce any water consumption** (except it is adiabatically assisted; see below).

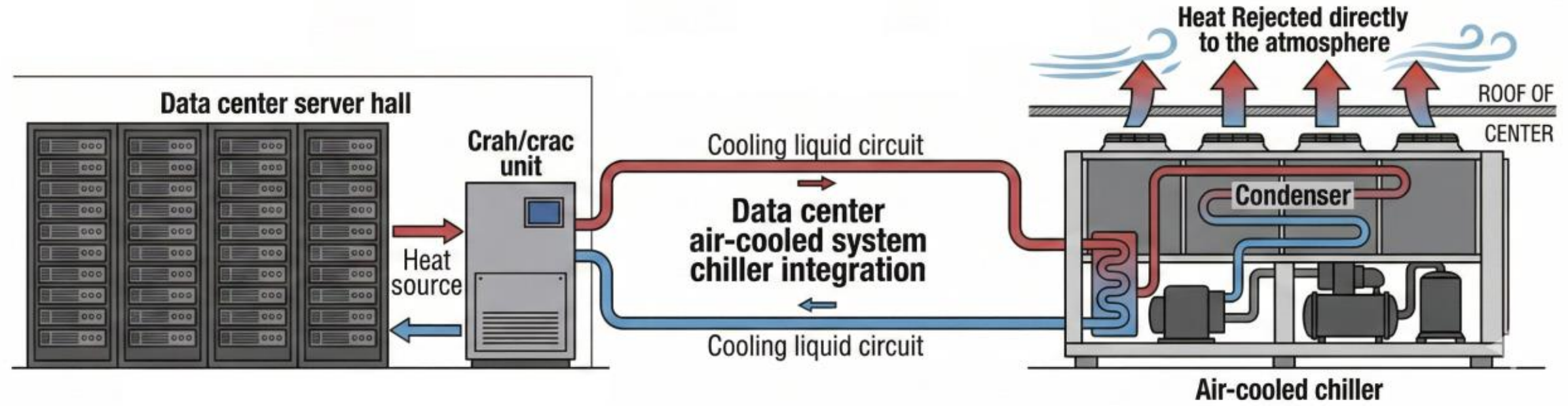

Figure 2: Schematic of an air-cooled chiller (generated by Nano banana pro, corrected via MS Paint; AI insists on using American English). The condenser is not chiller-internal but consists of the condenser coils that reject the heat directly to the atmosphere, acting like a dry cooler for the chiller.

### 4.1.4. Cooling towers and water-cooled chillers

Water-cooled chillers are quite different: The compressor and refrigerant circuit are chiller-internal and they are not rejecting heat directly to the atmosphere. Instead, the heat is transferred to the condenser water circuit, which runs between the chiller and the cooling towers, where the heat is finally transferred to the atmosphere (Coroamă, 2025).

There are two distinct flavours of cooling towers: open-loop and closed-loop towers. The main difference lies in how the condenser water circuit transfers its heat to the atmosphere (Pinnacle, 2025), as also represented in Figure 3:

- In an *open-circuit* setting, the condenser water enters at the top of the cooling tower and trickles down or is sprayed down the tower, where it is in contact with the surrounding air. Heat is lost both by direct transfer to the colder air, and by partial evaporation of the condenser water (which extracts heat from the remaining, surrounding water). The cooled water gathers in a basin at the bottom of the tower, from where it is sent back to the chiller. The path down the tower and air contact can happen in various architectures (such as counterflow and crossflow), which are beyond the scope of this report; a good overview is provided by (*How Cooling Towers Work*

*Counterflow vs Crossflow*, 2022). Obviously, the name “open circuit” stems from the condenser water circuit being open inside the tower.

- By contrast, in a *closed-circuit*, the condenser circuit circulates the water within a closed loop that never comes in contact with the surrounding air. Similarly to the condenser coils of an air-cooled chiller, the condenser water is distributed within a network of many small pipes inside the cooling tower. This is traditionally called “fill media” (Pinnacle, 2025), and it increases the surface for the heat transfer. To support the heat transfer from the condenser water inside the fill media, there is another (external) open water circuit: Water is being sprayed down inside the tower and onto the fill media. This water, which partly evaporates during the process, also gathers in a basin at the bottom of the cooling tower and is then sent back up to be sprayed again.

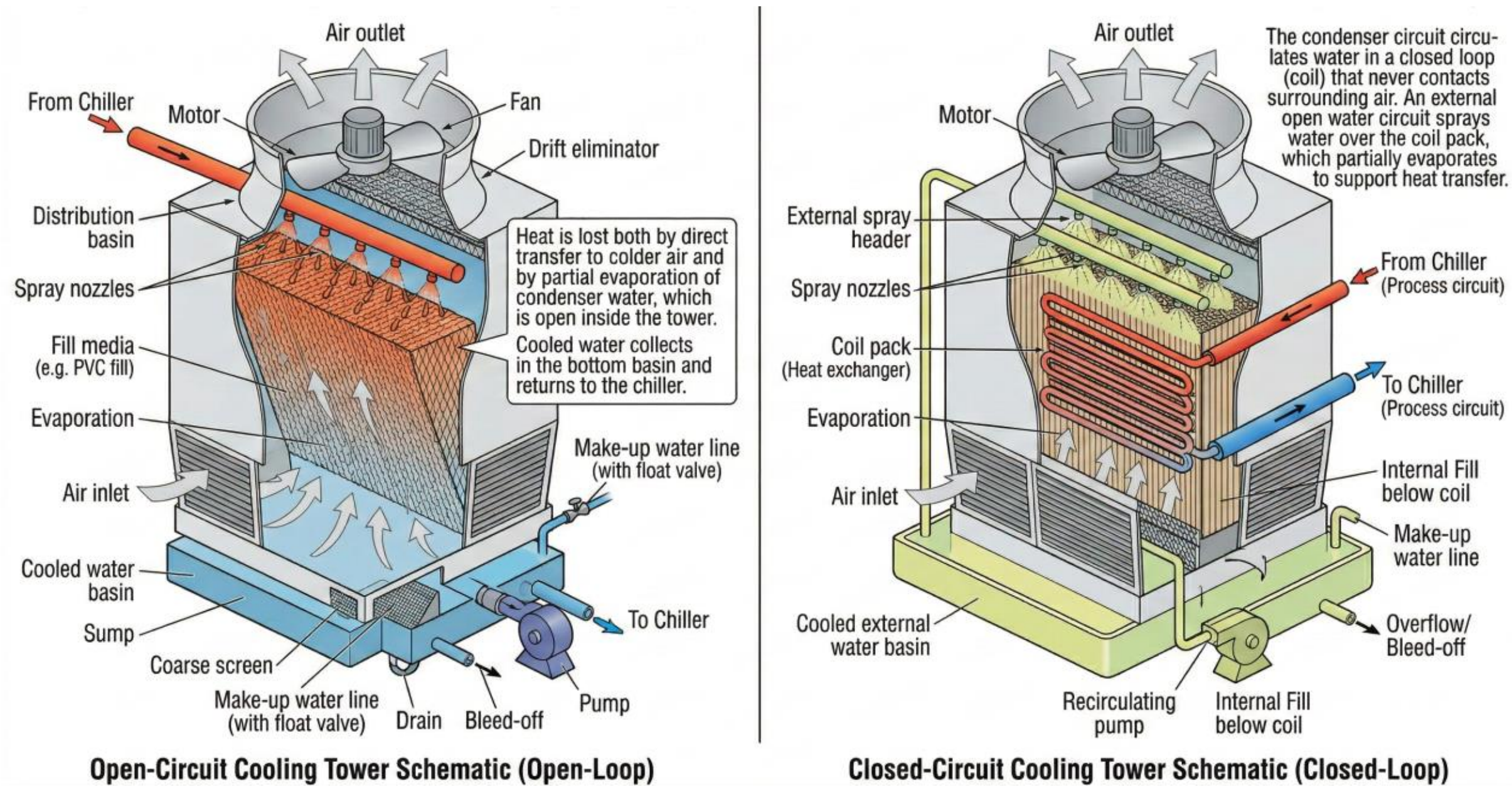

Figure 3: Schematic representations of an open-circuit cooling tower (left) and closed-circuit cooling tower (right). Both incur water consumption, mainly through evaporation. Generated by Nano banana pro based on the text in this section, corrected via MS Paint.

As is obvious from the descriptions above, both of these cooling tower paradigms consume water, albeit from different sources: from the condenser water circuit and the external cooling circuit, respectively. Hence, both systems need water replenishment, usually called “make-up water” in the technical jargon.

There are two further potential water consumption sources for cooling towers, blowdown and drift:

- Blowdown describes the water that is bled from the bottom of the cold water basin to reduce the concentration of salts and other impurities that accumulate in the water exposed to the elements. Performing it is more important for open circuits than for closed ones, as in the former case, it is the condenser circuit that is polluted, which can cause issues to the chiller if left unchecked. But even for open circuit cooling towers, the water consumed through blowdown is usually small compared to the one lost to evaporation.
- Drift is non-evaporated water that escapes from the respective open circuit to the atmosphere nevertheless, as liquid droplets hitchhiking on the airstream. To not waste this water, cooling towers are typically outfitted at the top with “drift eliminators”, which capture most of these droplets from the airstream. Drift is thus also a minor concern as compared to evaporation.

In conclusion, **water-cooled chillers consume water, mainly due to evaporation**. Open-circuit towers have a better heat transfer due to the direct air contact of the condenser water circuit. They thus tend to have less water evaporation per amount of heat rejected as compared to closed-circuit towers. Although they consume a little more through blowdown (drift being negligible for both), open-circuit cooling towers are generally less water-intensive than closed-circuit towers. Due to the risk of impurities in the condenser water circuit, they require, however, a higher level of maintenance (including chemical treatment of the open condenser water circuit).

At the same time, both types of water-cooled chillers are typically more energy efficient than air-cooled chillers. This is mainly because the compressor has to perform less work, as water-cooled chillers have lower condenser water temperatures and pressures.

#### 4.1.5. Economisers (airside and waterside)

When the outside air temperature is below the return air temperature inside server rooms, outside air can be directly used for cooling, resulting in a more efficient process than mechanical cooling. When such conditions exist, *airside economisers* circulate outside air directly through the server rooms to help cooling them. Also called "free air cooling" or "dry free air cooling", **airside economisers induce no water consumption**.

*Waterside economisers*, on the other hand, are technologically more elaborate heat exchangers, while ultimately also taking advantage of the low temperature of outside air and using the atmosphere as heat sink. Their role is to cool the warm cooling water that comes from the CRAH/CRAC units and would otherwise go directly to the chillers. To perform this, waterside economisers can be placed either in series to (and ahead of) the chillers (so they cool a part of the chilled cooling water) or bypass them altogether and carrying the full load.

The heat is transferred to cool water that is being fed to the waterside economiser from the basins at the bottom of cooling towers. The heat source of the economisers (i.e., the water coming from the CRAH/CRAC units), however, has a lower temperature than the output of chillers. The heat transfer will thus only work if the medium that absorbs the heat (i.e., the basin water from the cooling towers) is also colder than when absorbing heat from chiller water. As their airside counterparts, waterside economisers thus also only work for relatively low outside air temperatures: Only then is the water from the basins of cooling towers cold enough to absorb the heat directly from the indoor cooling units (i.e., CRAH/CRAC) without the temperature having been raised by chillers first. A very good explanation of the functioning principle of waterside economisers is provide by the video (*How Waterside Economizers Work*, 2023).

As they require cooling towers, **waterside economisers induce water consumption**. They can theoretically be used to complement both air-cooled and water-cooled chillers. As they rely on cooling towers, however, they are typically deployed complementary to water-cooled chillers. In fact, water-cooled chillers with additional waterside economisers are one of the most widely deployed cooling techniques in data centres, particularly in hyperscale DCs (Lei *et al.*, 2025).

Both types of economisers can reduce the amount of time the mechanical cooling equipment (e.g., chillers) must run. They can also reduce the intensity of chiller usage, as economisers can replace but sometimes also merely complement chillers.

#### 4.1.6. Evaporatively/adiabatically assisted cooling (direct and indirect)

A last important cooling technology that needs to be addressed is the evaporative (or "adiabatic") cooling. This is in fact more of an assisting technology than a cooling technology itself; hence the heading "adiabatically assisted".

There are two main flavours of adiabatic assistance. They both extend technologies that have already been presented, and they both rely on the same main principle: using evaporative cooling to pre-cool air that is then deployed in DC cooling. The air cooling through evaporation is also identical between these two flavours: Large, wetted pads (the "adiabatic layer") are placed in front of the air intake, and the air is blown (or sucked) via fans through these wet pads.

As succinctly but clearly outlined in (Microsoft, 2023), the two flavours are *direct evaporative cooling* (DEC) and *indirect evaporative cooling* (IDEC):

- Direct evaporative cooling is an evolution of the airside economisers presented in the beginning of Section 4.1.5 above. As in airside economisers, the ambient air is not cooling medium for a secondary loop; it enters the server rooms directly. When it is not cold enough, however, it needs to be cooled down first, so it passes through wetted pads – or a high-pressure misting chamber. The evaporation extracts heat from the air, cooling it down.
- Indirect evaporative cooling is an extension of the dry coolers discussed in Section 4.1.1 and the related air-cooling of chillers addressed in Section 4.1.3. As discussed earlier, these systems circulate the cooling fluid in cooling coils outside the building, and use ambient air in direct contact to the coils to cool it down. If the ambient air is too warm, however, it needs to be pre-cooled, which can happen via evaporative cooling: At the air intake of the dry coolers (or the chiller's condenser, respectively), wet pads are deployed to this end.

There is another commonality between the two flavours of evaporative cooling: the evaporation-based pre-cooling can be turned on or off as needed. When the temperature of the ambient air is cold enough (e.g., at night or in the respective seasons), the evaporative pre-cooling is turned off, and the cooling reverts to being airside economiser, dry cooling, or air-cooled chillers without any water consumption.

Both for this reason, but also the way water is consumed by comparison to cooling towers, evaporative cooling generally consumes less water than cooling towers for a similarly-sized DC in comparable climates (Microsoft, 2023). In temperate climates, where the evaporation can be turned off for extended periods, the savings can be spectacular. In hot and dry climates, on the other hand, evaporation is very effective (as dry air absorbs moisture faster, evaporation and the subsequent air cooling work very efficient); the price, however, is a very high water consumption, which can be on par, or surpass, that of cooling towers.

#### 4.1.7. An overview of cooling technologies and their impact on energy and water consumption

After the previous discussions, it is clear that two technologies in particular are responsible for direct (i.e., on-site) water consumption in DCs: cooling towers and adiabatic cooling. At the same time, while there are various reasons for energy consumption (e.g., pumps, fans, etc), the main source of cooling energy consumption are the chillers.

Table 2: Qualitative summary of the cooling technologies presented, together with their main sources of energy consumption (chillers) and water consumption (cooling towers and adiabatic cooling). All of the technologies indicated as consuming no water (second-to-last column) have a star, which indicates that they will often be adiabatically supported, which induces some water consumption after all.

| Technology | Flavour | Energy (chiller) | Cooling tower | Adiabatic cooling | Water consumption | Comment |
|---|---|---|---|---|---|---|
| Dry cooler | – | no | no | no | none* | seldomly single cooling tech |
| Chiller | air-cooled | **yes** | no | no | none* | w/o adiabatic support, more energy |
| | water-cooled | **yes** | **yes** | no | **moderate – high** | balance between energy and water |
| Economiser | airside | no | no | no | none* | outside air in server rooms |
| | waterside | no | **yes** | no | **moderate** | often jointly w/ water-cooled chillers |
| Adiabatic cooling | direct | no | no | **yes** | **low – moderate** | can be turned off; consumption dependent on climate |
| | indirect | no | no | **yes** | **low – high** | |

Table 2 offers a qualitative summary of the main features of the technologies discusses so far, highlighting whether they deploy chillers, cooling towers, or adiabatic cooling. In practice, many of these technologies are mixed and will be used alternatively, depending on the climate, or in parallel. A good overview of types of combinations of cooling technologies employed by DCs are provided in (Lei and Masanet, 2022; Lei *et al.*, 2025).

## 4.2 Indirect water consumption and withdrawal for electricity generation

The water used upstream in the generation of electricity (indirect consumption from the DC's perspective) can also be consumed or merely withdrawn but later returned to its source.

### 4.2.1. Thermal power plants

Thermal power plants in particular have a high amount of water withdrawal. Regardless of their fuel (coal, oil, gas, or nuclear), they work after the same principle: the fuel heats water to create steam, which drives a turbine to create electricity after which the low-pressure steam then enters a condenser where its heat is removed, converting it back into water (condensate) to be pumped back to the boiler.

The heat from this primary, closed-circuit loop of the thermal power plant, is typically transferred in the condenser to the cooling water. When power plants are placed next to large bodies of water such as oceans or large rivers), then *once-through cooling* can be used: water is withdrawn from the source, passes once through the condenser, cooling the steam of the closed-loop circuit, and is discharged warmer back to the source (e.g., downstream for a river). From the DCs perspective, once-through cooling of thermal power plants thus represents indirect water withdrawal, but not consumption.

Once-through cooling is increasingly rare (Peer, Grubert and Sanders, 2019). When thermal power plants are not placed near a large, suitable water source, their cooling often involves *recirculating cooling* with evaporative cooling towers. In this paradigm, the hot water from the condenser is sent to a cooling tower, similar to those of DCs. Air is drawn through the tower, causing a small portion of the water to evaporate. This evaporation process removes a significant amount of heat from the remaining water, cooling it down. The remaining cooled water is then recirculated back to the plant's condenser to absorb more heat. This is a semi-closed loop for the cooling water, as it involves evaporation and requires a continuous supply of "make-up water" to replace what is lost to the atmosphere as steam. As such, this cooling technique also induces indirect water consumption (from electricity generation) for the DC.

### 4.2.2. Gas turbines and water consumption for fuel extraction

For natural gas, however, further electricity generating technologies are almost waterless. One of them are simple-cycle combustion turbines (or, simply, “gas turbines”), which are essentially large jet engines adapted to generate power. Gas turbines burn the gas resulting into a hot, high-pressure combustion gas which directly drives the turbine without the detour over water and steam. The gas turbine exhaust is released directly to the atmosphere, and as there is no need for steam condensation, there is also no need for a large cooling water source (UCS, 2013). Another waterless gas-based technology are fuel cells, in which the main component of the natural gas (i.e., methane or $CH_4$) reacts electrochemically with oxygen to produce electricity (Leo, 2022). The process converts chemical energy directly into electrical energy without combustion or moving parts, thus not requiring any water (in fact, fuel cells produce water as a byproduct of the chemical reaction).

The processes mentioned so far induce varying amounts of water consumption at the point-of-generation (PoG) of electricity. However, the extraction of all these fuels (coal, oil, gas, and uranium) also typically requires water consumption (Peer, Grubert and Sanders, 2019). From the point of view of electricity generation, this water is upstream consumption (as opposed to the PoG consumption discussed above). As seen from the point of view of DCs, this consumption is (even more) upstream.

4.2.3. Renewables

Among renewable options, wind and solar are among the most water-preserving electricity sources available. They use virtually no water during operation and relatively little amounts upstream (Peer, Grubert and Sanders, 2019).

Reservoir-associated hydroelectric power plants, on the other hand, can be the worst electricity source of them all in terms of water consumption (albeit not withdrawal). This consumption stems from the evaporation from the reservoir, which would otherwise not have existed. For the US, for example, hydroelectricity generated 6% of electricity in 2015, but was responsible for 23% electricity-associated water consumption; a substantially larger amount per unit of energy compared to thermal power plants (Peer, Grubert and Sanders, 2019).

Overall, depending on the mix of generation technologies and climate, the water consumption of electricity generation can vary by orders of magnitude between individual geographies. In a 2015 estimate for the US, for example, the range among 26 regions in the US was roughly 20-fold, from 0.42 litres / kWh (or 0.42 $m^3$ / MWh) to 9.2 litres / kWh (Peer, Grubert and Sanders, 2019).

## 4.3 Measuring relative water use: The water usage effectiveness

A relative measure for the water consumption of a DC was proposed in 2011 by the non-profit organisation "The Green Grid" (Azevedo, Belady and Pouchet, 2011). The water usage effectiveness (WUE) assesses how much water a facility uses relative to the energy consumption of its IT equipment and is defined as

$$WUE = \frac{Water\ usage}{IT\ equipment\ energy} \left[\frac{litres}{kWh}\right] \tag{16}$$

where the IT equipment energy is the $E_{IT}$ defined early in Chapter 2. Given the considerations on water withdrawal versus water consumption in the beginning of the chapter, we suggest that "water usage" should be interpreted as "water consumption".

The WUE relates water usage to the useful energy consumption within the DC, not the overall DC energy consumption $E_{DC}$. It is designed to resemble the PUE (its name not being a coincidence), and relate the entire water consumption to the useful energy. This, however, can make its assessment more challenging.

After the PUE, the WUE is the second of four relative "sustainability indicators" required by the European reporting and rating scheme for DCs complementarily to the defined KPIs (European Commission, 2024a).

Water usage effectiveness can refer either to the direct water consumption only (i.e., on-site), or to both direct and indirect water consumption (i.e., on-site and upstream consumption for electricity production). To distinguish between the two options, (Azevedo, Belady and Pouchet, 2011) suggest to name them $WUE$ and $WUE_{source}$, respectively.

To avoid the confusion potentially generated by the first, unqualified name, we adopt here the terminology as suggested by later literature, e.g. (Lei *et al.*, 2025), naming the two $WUE_{site}$ and $WUE_{source}$, respectively. As they also typically refer to the more relevant water consumption and not to withdrawal, the two can be expressed as

$$WUE_{site} = \frac{Water_{DC}}{E_{IT}} \left[\frac{litres}{kWh}\right] \tag{17}$$

$$WUE_{source} = \frac{Water_{DC} + Water_{elec}}{E_{IT}} \left[\frac{litres}{kWh}\right] \tag{18}$$

where $Water_{DC}$ is the water consumption at the DC site, $Water_{elec}$ the water consumed in the production of the electricity used in the DC, and $E_{IT}$ – the same denominator as for PUE – the IT power consumption, as defined in Chapter 2. All of these values refer to the same time period; typically, an year or a month.

In both cases, the water consumption is put into relation to the direct electricity consumption in the DC. As the water consumption during electricity generation can be small but never zero, $WUE_{site}$ is a subset of $WUE_{source}$, and

$$WUE_{source} > WUE_{site}$$

Among the two, $WUE_{source}$ reflects better not only general lifecycle assessment (LCA) principles, but also the principle of causation: Whether water was consumed in the DC itself or upstream during electricity production, the computations taking place in the DC were responsible for this consumption. This study thus focuses on $WUE_{source}$, and the generic term $WUE$ is used synonymous to it.

The water consumed during electricity production can be expressed as

$$Water_{elec} = E_{DC} * WCF \tag{19}$$

where $Elec_{DC}$ is the electricity consumption of the DC over the period of assessment and $WCF$ the average "water consumption factor" of the electricity used by the DC. It is correct in this context to consider the entire DC electricity $Elec_{DC}$ and not only the IT equipment electricity $Elec_{IT}$, as the water consumed in the generation of the electricity used by the infrastructure (such as cooling) is also causally linked to the compute loads. The nowadays widely used term WCF (Lei *et al.*, 2025) was called "energy water intensity factor" (EWIF) in the original Green Grid white paper (Azevedo, Belady and Pouchet, 2011). It represents the water intensity of electricity and is measured in $[litres/kWh]$.

Substituting Equation 19 in Equation 18 yields

$$WUE_{source} = \frac{Water_{DC} + E_{DC} * WCF}{E_{IT}} = WUE_{site} + PUE * WCF \tag{20}$$

an equation showing not only how $WUE_{source}$ and $WUE_{site}$ are related, but also a more subtle dependency of the former on the latter.

## 4.4 The trade-offs between PUE and WUE as well as $WUE_{source}$ and $WUE_{site}$

Some cooling methods are more water-intensive while others are more energy-intensive, so there is an inherent trade-off between minimising the PUE and $WUE_{site}$ (Lei and Masanet, 2022; Shehabi *et al.*, 2024). It is true, nevertheless, that some classes of technologies are at the same time more energy and water efficient than others. Economisers, for example, fare better in both dimensions than indirect evaporative cooling, and so is direct liquid cooling.

But these classes of cooling technologies are not always deployable. Economisers, for example, require a relatively cold climate for most the year (Microsoft, 2023), while liquid cooling is complex, expensive, and not an easy retrofit for most DCs. More importantly, however, inside each of these classes of cooling technologies, there is typically still a trade-off between optimising for energy or water efficiency. As shown in (Shehabi *et al.*, 2024), for example, air-cooled chillers or dry coolers work almost water-free, but require more electricity than wet coolers in similar conditions.

Beyond the trade-off between on-site energy and water consumption, the picture becomes more complex when considering the indirect (i.e., upstream) water consumption for electricity production. As shown by Equation 20, the overall water usage effectiveness $WUE_{source}$ depends not only on the on-site water consumption but also on the indirect water consumption, which in turn is directly proportional to both the DC's PUE and the regional WCF.

To minimise $WUE_{source}$ thus requires minimising the overall function

$$WUE_{source}\,(WUE_{site}, PUE) = WUE_{site} + PUE * WCF \tag{21}$$

Defining the relation between PUE and $WUE_{site}$ as an (unknown and probably non-linear) function $f$:

$$PUE = f(WUE_{site})$$

yields by substitution in Equation 21 the final optimisation function with just one parameter:

$$WUE_{source}\,(WUE_{site}) = WUE_{site} + (f(WUE_{site}) * WCF) \tag{22}$$

The minimum of this trade-off curve will be in one of its – and possibly the only one – critical points, i.e., where the first derivative equals zero. The derivative of $WUE_{source}\,(WUE_{site})$ is defined as

$$WUE'_{source}(WUE_{site}) = \frac{d(WUE_{source})}{d(WUE_{site})} \tag{23}$$

and using $WUE_{source}$ as computed in Equation 22 yields

$$WUE'_{source}(WUE_{site}) = \frac{d\big(WUE_{site} + (f(WUE_{site}) * WCF)\big)}{d(WUE_{site})} = 1 + WCF\,\frac{d(PUE)}{d(WUE_{site})} \tag{24}$$

because WCF is a constant and $f(WUE_{site})$ is $PUE$.

The minimum is where these derivate equals zero

$$1 + WCF\,\frac{d(PUE)}{d(WUE_{site})} = 0 \tag{25}$$

which is equivalent to

$$\frac{d(PUE)}{d(WUE_{site})} = -\frac{1}{WCF} \tag{26}$$

$WUE_{source}$ thus reaches its minimum at the point of the PUE-WUE trade-off curve where the slope of the curve (i.e., marginal change in $PUE$ for a marginal change in $WUE_{site}$) is equal to the negative inverse of the water intensity of electricity in the local grid.

The mathematical interpretation of Equation 26 is that a high water consumption embodied in the electricity is reflected in a high $WCF$ value and thus a small negative value for its negative inverse, $-1/WCF$. The optimal point will thus be where the trade-off curve is very flat, which for a typical trade-off curve would be far to the right, corresponding to a low $PUE$ and a high $WUE_{site}$. This corresponds to expectations: when the electricity embodies high amounts of water, it may be worth spending a little more water on-site (e.g., through evaporative cooling), if this saves some electricity, including its embodied water.

When, on the other hand, WCF is low (e.g., a DC in a region with solar or wind power), the electricity is “dry”, embodying very little water. Saving energy on-site has almost no water benefit at the electricity source, and the main water impact is clearly on-site. Purely from a water footprint perspective, on-site water conservation should be prioritised.

# 5 Energy circularity through waste heat recovery

All the energy consumed by data centres for computation and other activities is transformed into heat. As discussed in Chapters 2 and 4, this heat is extracted from the DC (while consuming further energy and possibly water in the process), and often released to a natural heat sink such as the atmosphere or a body of water.

This thermal energy produced by data centres represents a particular case of "waste heat". The terminology describing this energy varies, with "the terms excess heat, surplus heat, and heat offtake being used interchangeably throughout the industry" (Groucott *et al.*, 2025). This chapter addresses the possible circularity aspect of waste heat and how it affects the sustainability of a data centre.

## 5.1 Definition and brief history

In general, waste heat is defined as a byproduct of production or consumption processes, which is unavoidable excess heat that cannot be reused within the process that generated it (Pettersson *et al.*, 2020). In the context of data centres, very little heat can be reused within the data centre itself. On the contrary, as the last chapter has shown and the literature also highlights (Yuan *et al.*, 2023), a main preoccupation is how to reject the massive amounts of heat generated within the data centre.

Crucial from a circular economy perspective, however, instead of going unused and be dissipated into the environment, waste heat could potentially be reused for other, external production or consumption processes.

This is particularly relevant for data centres: While serious efforts are being undertaken to further improve the cooling efficiency (and thus the PUE), the potential left for improvement is very limited. As discussed earlier, for large hyperscale and colocation DCs, the cooling energy is already a relatively small share of the overall energy consumed in a DC. While efforts to optimise the PUE are inherently limited to this share of energy, the potential for waste heat recovery is the entire energy consumed in the data centre, i.e., $E_{DC}$ and not only $E_{non-IT}$.

### 5.1.1.Heat recovery for district cooling

The reuse of unused thermal energy gained momentum over the past decades, driven in particular by district heating in Nordic countries. The Nordics also pioneered this integration, leveraging cold climates and pre-existing district heating infrastructure. Sweden's first district heating system, for example, was introduced in Karlstad in 1948, when the municipality started delivering heat from a combined heat and power (CHP) thermal power plant to nearby industrial buildings. This was followed by Helsingborg in 1974 when the municipal utility began purchasing waste heat from the Boliden Kemi (now Kemira) sulphuric acid plant (TOFANI, 2022).

The first reuse of waste heat from data centres dates back to the 1970s, when a data centre in Älvsjö, Stockholm, was experimentally equipped with a heat recovery system so that server heat could be captured instead of simply vented away (Fors and Lennerfors, 2018). In the 1980s, more systematic attempts were made at the Swedish Kommundata data centre, where a dedicated heat recovery installation marked the beginning of DC waste heat reuse in the Nordic countries (Saunavaara, Laine and Salo, 2022).

These early systems were relatively modest and served nearby offices or buildings. They nevertheless established the basic technological concept and feasibility. Further policy support came with the introduction of oil taxes in the 1980s and a $CO_2$ tax on fossil fuels in 1991, which incentivised the use of waste heat. Based on these learning from the Nordics, London began advancing its heat network development following the Mayor's Energy Strategy in 2004 (Groucott *et al.*, 2025).

Large-scale projects started in the late 2000s: In 2014, a 10 MW data centre in Mäntsälä, Finland, started utilising about one-third of its waste heat to provide heat to the local district heating system. And in 2018, the telecom provider Telia opened a 24 MW DC in Helsinki, designed to supply its entire waste heat – i.e., almost 200 GWh per year – to the nearby city of Espoo (Oltmanns *et al.*, 2020).

In Switzerland, DCs also increasingly recover part of their waste heat for district heating. The Swisscom DC in Zurich Herdern, for example, was 2012 retrofitted to feed waste heat into the local district heating. One of the most modern and efficient Swiss data centres, the Swisscom DC in Bern Wankdorf, also feeds part of its waste heat into Bern's district heating system (Energeia, 2023). And a new data centre in Geneva is designed to feed 100% of its waste heat into district heating (SwissInfo, 2025).

#### 5.1.2. Further uses for data centre waste heat

Beyond centralised heating, waste heat can be deployed for other, more specific, purposes as well. It can be used to heat large individual buildings or building complexes such as industry buildings (that do not produce heat themselves), greenhouses used in food production, or animal farming.

Equinix, for example, operates an urban farm on its "PA10 Paris" data centre campus, showcasing a possible model for city centre greenhouses heated by DCs (Judge, 2023). The low-temperature output of DC heat is often considered ideal for this industrial symbiosis, promoting local food production in regions such as the Subarctic / Northern Sweden (Cáceres *et al.*, 2022). Waste DC heat can also be used in various industrial processes, such as preheating feed water, industrial drying (Wahlroos *et al.*, 2018), and desalination for clean water production (Yuan *et al.*, 2023).

## 5.2 Feasibility, benefits, and challenges of heat reuse

#### 5.2.1. Technical and economic feasibility

The temperature of waste heat available for reuse varies significantly depending on the data centre's cooling technology (see previous Chapter 4). This temperature determines the maximum practical application without requiring additional energy inputs (such as heat pumps) to raise it further.

Modern district heating networks often require heat with flow temperatures of 45 – 55 °C for efficient preheating (Groucott *et al.*, 2025). Much higher temperatures of 75 °C and above (even up to 120 °C at higher pressure) are required for conventional networks (Lund *et al.*, 2010). Higher quality DC waste heat – typically stemming from liquid cooling systems – can thus be directly used in some modern district heating systems.

Lower quality heat and heat for traditional district heating systems require heat pumps and additional energy input first to raise the temperature before reusage. The transfer of large volumes of low-temperature heat also requires additional infrastructure such as new pipes. The proximity of the data centre to the users of the district heating system is another important attribute. Larger distances mean both additional piping infrastructure and larger heat losses along the way.

Technical and economic viability thus depend foremost on the existence of a district heating system. In its absence, local solutions to a few close-by large consumers are feasible, such as some farming or industrial usages (as discussed in Section 5.1.2). These, however, are generally not available, or not on the scale needed to  the waste heat of large DCs.

In this context, the major capital expenditures (CapEx) for the DC operator include heat recovery equipment, additional pipes, and heat pumps. The operational electricity consumption (i.e., the operational expenditures, OpEx, for heat reuse) also increase due to the energy consumption of the heat pumps that raise the temperatures of the water for district heating. Nevertheless, as it is obviously energetically and financially cheaper to heat up water from 35 °C than, say, 15 or 20 °C; these losses will normally be overcompensated – energetically and financially – by the savings of the utility company buying the waste heat.

It is thus often the utility providers themselves that install and manage the additional heat pumps, pay the corresponding OpEx as well as additional fees to the DC operator. Helen, a major Finnish energy provider, for example, manages a system that captures heat from multiple data centres from around Helsinki, including from Telia, Equinix, and Microsoft, and sends it to an underground heat pump station.

The Helen-operated heat pumps raise the temperature of the captured heat before distributing it through the city's district network. The process also returns used (and thus cooled) space heating water to the DCs, for renewed usage for cooling – an ideal symbiotic relation, as described in a presentation video (*Finland's Big Idea: Turning Data Center Heat Into Power*, 2025).

#### 5.2.2. Sustainability benefits and challenges

The benefits of this circularity are not only energetic and monetary, but as a consequence also climatic. In Espoo, Finland's second most populous city, for example, Microsoft is building a cluster of data centres that, when completed, is expected to supply heating to 40% of the city, or about 100,000 homes. Although not yet completed, the project has already led to the shutdown of a local coal-fired heating plant. The heat recovery facility uses water-to-water heat pumps to extract lukewarm water (about 25°C to 35°C) from the DCs, sending it to two large electric boilers to be heated to 115°C for distribution (*Finland's Big Idea: Turning Data Center Heat Into Power*, 2025).

#### 5.2.3. Challenges for heat reuse

Beyond the availability of heat consumers and the technical infrastructure Not all DC sizes are adequate for heat reuse. An analysis of investment proposals in Scandinavia showed that the scale of the DC is a major factor in profitability. Medium (from around 500 racks) and large (from 5,000 racks on) DC facilities are generally considered profitable investments for waste heat utilisation. Conversely, waste heat from smaller DCs is often unprofitable, as CapEx is harder to cover (Pärssinen *et al.*, 2019).

The largest challenge to waste heat reuse, however, is **seasonality**. There is a large discrepancy between (almost) constant DC heat production and the seasonally fluctuating heat demand. This is a complex issue, reminding of the seasonality issue of renewable energy storage (i.e., that only few storage options such as hydro are suitable for inter-seasonal balancing while batteries, for example, can very well compensate daily variations but only poorly seasonal ones). Seasonality has two main consequences:

- The colder the climate, the more suitable the reuse of DC waste heat for district heating; it is thus perhaps no coincidence that the Nordics together with Switzerland or Germany are leading in this field.
- Data centres usually require two complementary systems for heat rejection: One for heat reuse that will be used extensively in winter and only to a small part in summer (e.g., for warm water provisioning but not heating), and one able to use a natural heat sink for the share of waste heat that is not usable at any moment.

Seasonality also has consequences when modelling the energy, economic, or sustainability impact of the heat reuse: Dynamic or seasonal data need to be considered to accurately assess the environmental and economic impacts. Relying only on annual averages would be insufficient to capture the real-world operational trade-offs and potential systemic problems caused by the seasonal mismatch (Gustavsen *et al.*, 2025).

## 5.3 Measuring heat reuse

All the energy-related metrics presented so far – whether the widely-used $PUE$, the $DCCE$ and $E_{DC}$ suggested in Section 2.3.4, or the efficiency metrics discussed in Chapter 3 – stop at the boundaries of the data centre itself. None of them is suited to reflect the downstream usage of the reused heat. A new metric that expands the system boundaries and includes downstream reuse is thus required.

To assess the success of waste heat integration, the "energy reuse factor" (ERF) was the earliest suggested metric. The metric was originally defined in 2011-2012 by a "global taskforce" led by The Green Grid (that had also proposed the PUE a few years earlier) and comprising the US Department of Energy, the US Environmental Protection Agency's "ENERGY STAR" programme, the Joint Research Centre of the European Commission, and the Japanese Ministry of Economy, Trade, and Industry. The rather trivial metric simply represents the share of heat that is being reused among the total amount of waste heat:

$$ERF = \frac{Q_{Re}}{E_{DC}} \tag{27}$$

where $Q_{Re}$ is the amount of heat used downstream (although a form of energy, denote by tradition with $Q$ and not $E$), and $E_{DC}$ the total energy consumption of the data centre, as defined from the outset in Chapter 2.

The $ERF$ is an important indicator used in various contexts. It is, for example, the third of four relative "sustainability indicators" required by the European reporting and rating scheme for DCs next to the $PUE$ and the $WUE$ (European Commission, 2024a).

Another metric developed around the same time is the "energy reuse effectiveness" (ERE). It is related to the ERF, but with a different aim: Instead of relating the reused energy to the total energy consumed in the DC, **ERE aims to "correct" the PUE by factoring in circularity**. To do so, the indicator is defined as follows (Patterson *et al.*, 2010; Oltmanns *et al.*, 2020):

$$ERE = \frac{E_{DC} - Q_{Re}}{E_{IT}} = \frac{E_{IT} + E_{non-IT} - Q_{Re}}{E_{IT}} \tag{28}$$

Equation 28 can be rewritten as

$$ERE = \frac{E_{IT} + (E_{non-IT} - Q_{Re})}{E_{IT}} = 1 + \frac{E_{non-IT} - Q_{Re}}{E_{IT}} \tag{29}$$

showing immediately the "correction" mentioned above. As opposed to the PUE, the ERE can become sub-unitary, and does so when $Q_{Re} > E_{non-IT}$, in other words when the waste heat reused (which is no longer "waste") becomes larger than the energy wasted (i.e., not used for computing purposes) in the first place.

Considering the ERE formulation from Equation 28, it becomes clear that that ERE can take values in the interval:

$$ERE \in (0, PUE] \tag{30}$$

ERE equals the PUE when zero energy is reused ($Q_{Re} = 0$), and Equation 28 reduces to the PUE definition. The more energy is reused, the more ERE decreases, at some point becoming sub-unitary and in the best case tending towards zero (when almost all of $E_{DC} = E_{IT} + E_{non-IT}$ is reused). Due to the laws of thermodynamics (e.g., transformation and line losses that cannot become zero), it is impossible to recuperate (and subsequently reuse) the entire energy; however, very low values are theoretically possible.

To derive the relation between $ERF$ and $ERE$, we rewrite again Equation 28 by multiplying on the right side with an identity, which helps reformulating ERE in terms of ERF and PUE:

$$\boldsymbol{ERE} = \frac{E_{DC}}{E_{IT}} - \frac{Q_{Re}}{E_{IT}} = PUE - \frac{Q_{Re}}{E_{IT}} * \frac{E_{IT}}{E_{DC}} * \frac{E_{DC}}{E_{IT}} = PUE - \frac{Q_{Re}}{E_{DC}} * PUE = (\mathbf{1} - \boldsymbol{ERF}) * \boldsymbol{PUE} \tag{31}$$

## 5.4 Heat quality, optimisation trade-offs, and effect allocation

Although heat reuse has been discussed for almost two decades now, several aspects are still insufficiently explored or clarified. Some of the most important are: the trade-off between optimisation for low cooling overhead and maximum heat reuse, metrics that account for the quality of reused heat (and not only its quantity), and the allocation method between heat producer (i.e., the DC) and its users.

### 5.4.1. The quality of the reused heat, and the substitute in the counterfactual

Both $ERF$ and $ERE$ quantify the amount of heat reused and directly related it to $E_{DC}$ and $E_{DC}$ as well as $E_{IT}$, respectively. While all of these are energies, they are not the same form: While $E_{DC}$ and its subset $E_{IT}$ are in the form of electricity, $Q_{Re}$ is thermal energy. Not only is this not the same form of energy, it is also not the same energy *quality*: Electricity is a high-quality, versatile, and low-entropy (i.e., ordered) form of energy that can be employed across almost human activity domains.

By contrast, thermal energy is of low quality, has high entropy, and low versatility – the latter is responsible, among other aspects, for the fact that the usages for reused heat are limited and it is quite challenging to find suitable users for the waste heat produced in DCs. As the wonderful David MacKay explains: “When we ‘use up’ one kilojoule of energy, what we’re really doing is taking one kilojoule of energy in a form that has low entropy (for example, electricity), and converting it into an exactly equal amount of energy in another form, usually one that has much higher entropy (for example, hot air or hot water). When we’ve ‘used’ the energy, it’s still there; but we normally can’t ‘use’ the energy over and over again, because only low entropy energy is ‘useful’ to us” (MacKay, 2009).

These are not merely academic distinctions, but have practical implications if considering the counterfactual alternative: When reusing a quantity (say, 1kWh) of heat, for example, a nearby farm or greenhouse, what would the counterfactual have been? Perhaps the optimum would have been an air-source heat pump with a coefficient of performance of 4-5. This heat pump would thus have required only 200 – 250 Wh to produce the same heating energy, i.e., 20-25% of the reused energy.

It is insufficiently understood today whether the “credit” for reused energy should be 100% (as $ERF$ and $ERE$ suggest) or depend on the counterfactual, or some entirely different method. And if the counterfactual is relevant, then which one precisely: the energetic one, the GHG one, etc?

### 5.4.2. Trade-off between cooling minimisation and reuse maximisation

The previous discussion notwithstanding, it is most likely preferable from a circularity (and thus sustainability) perspective to have a slightly higher PUE, if a high heat reusage can be achieved. For example, if a DC can reach an ERE of, say, 0.5, a slightly worsened PUE of 1.2 is probably a price worth paying as compared to an industry-leading PUE of 1.1, if the latter would imply no reuse whatsoever.

This has concrete policy implications: Most of the best-of-class hyperscale DCs achieve their small PUEs by being placed in northern Europe or North America, where for most of the year cooling is free (e.g., via dry coolers, see Section 4.1 and Table 2). In the wilderness, however, no heat can be reused. Placing DCs a bit further south, in the vicinity of urban centres and industry, would enable heat reuse, but would also decrease the cooling efficiency.

With increasing DC construction and stagnant or perhaps even decreasing need for heat (e.g., due to climate change and better building insulation), at some point, a saturation could be reached. An important question is thus how to determine when saturation per country or region is being approached.

The question is particularly important for dictates such as the German energy efficiency regulation, which mandates that a certain percentage of heat must be reused for all new DC developments. With a tension between mandatory reuse, limited availability of infrastructure (above all, district heating) and possibly approaching saturation, the subsequent question is what this implies for new DC developments and their regulation.

These questions are worth exploring further to avoid inefficiencies and bottlenecks arising from well-meant but at some point outdated regulation.

# 6 The impact embodied at production

Historically, sustainability assessments of data centres focused mainly on energy consumption and GHG emissions during their operation. For these indicators, this focus is correct, as the use phase dominates the overall energy and GHG balance of DCs (Coroamă and Schien, 2026). A more comprehensive sustainability assessment, however, cannot ignore other types of environmental impacts such as resource depletion, toxic pollution, and e-waste.

Above all, the production process of microelectronics requires substantial quantities of finite resources, including various minerals sourced through environmentally damaging mining practices. The extraction and processing of these materials can also generate significant toxic pollution, while down the device's lifecycle, these devices generate e-waste.

The assessment of these impacts, however, faces two interconnected challenges: The supply chains for microelectronics are notoriously complex and opaque, spanning several countries and suppliers, deeming comprehensive LCAs practically impossible. Tracing a specific mineral from a mine, with its unique environmental impact, to a final chip in a DC is an intractable data problem. Additionally, there is a lack of standardised metrics and reporting. Companies rarely disclose the specific resource footprint of their hardware, and academic studies are sparse and often present estimates without access to primary data.

While the DC building and its infrastructure (cooling and power provisioning in particular) also have a material footprint, it involves fewer elements and is better understood. Section 6.1 thus focuses on the production process of microelectronics, its complex materiality, and the few particularities that sets it apart from most other sectors of human industrial production. Worsening matters are the relatively short lifespans of microelectronics; Section 6.2 briefly discusses them, and compares them to the lifespans of DC buildings and infrastructure.

## 6.1 Production impact of microelectronics

The materiality of microelectronics is driven by the high content of precious metals and critical minerals, but also by various mineral refining and manufacturing processes required for semiconductors (ARUP, 2025a). There are three main stages of the production process. These stages are shown in Figure 4 and include: the mining of raw materials, their refinement (microelectronics often require a high degree of purity, potentially resulting in substantial environmental impact), and finally the manufacturing of the microelectronic devices from the materials thus obtained. The current section is organised along these three production steps.

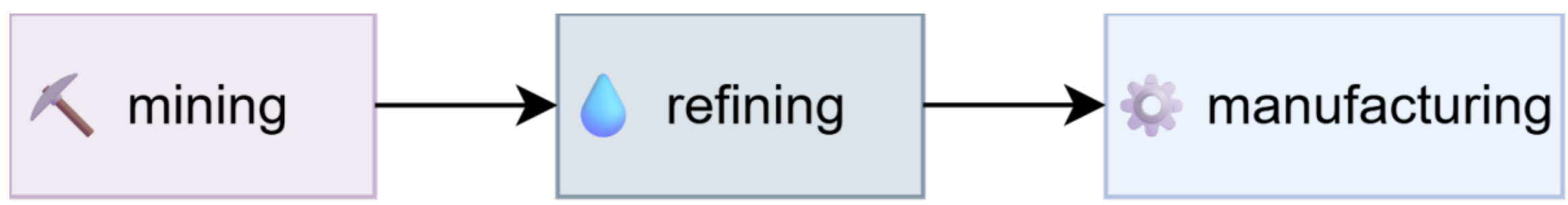

Figure 4: A simple high-level schematic of the three main phases of microelectronics production: extraction of raw materials, their refinement, and manufacturing of the devices from the refined materials.

### 6.1.1. Mining of minerals

Various sources estimate humankind's unsustainable resource use in general and that of critical minerals in particular. The impact specific to a DC (or even to digitalisation more generally), however, is usually either not singled out at all or to a relatively low degree. Quite often, the perspective is more geopolitical (e.g., in terms of supply chain concentration and related national policy interventions) and less environmental. The environmental perspective, when it does single out digitalisation, is frequently restricted to one or a few minerals only, or quite high-level and general.

According to data from the US Geological Survey, the metals mined for the technology sectors overall (including digitalisation and thus DCs) are almost negligible among the quantity of all mined metals: From the estimated 2.78 billion tonnes of metals mined in 2022, only 1.5 million (0.05%) were precious metals and metals for the technology sector (USGS, 2023), of which microelectronics in turn represent a part.

A small overall quantity, however, does not necessarily imply a small environmental footprint, as extraction and refining techniques can be very different among individual minerals. The production of microelectronics can be quite resource intensive, typically requiring between 50 – 400 times their final weight in material inputs. Up to 800 kg of materials, for example, are required to produce a 2 kg computer. There is additionally the diversity of materials: Producing a 22 kg server involves the extraction and processing of 27 key elements, including 12 critical raw materials such as gold, copper, and rare earth elements (REEs). Many of them carry a high risk to supply and significant environmental footprints despite their relatively small mass (Andrews and Kerwin, 2026), being recycled at rates as low as 1% (CEDaCI, 2020).

A 2025 report on critical minerals by the International Energy Agency (IEA) provides an in-depth analysis of critical mineral supply chains, demand trends, and geopolitical dynamics (IEA, 2025b). Data centres are explicitly recognised as major driver of critical mineral demand and playing a vital role across high-tech sectors. Copper, for example, is a critical component for both conventional and AI data centres due to its high electrical and thermal conductivity, durability, and affordability. It is primarily used in power distribution equipment, but also in cooling systems and network infrastructure (IEA, 2025b).

The report notes that the clean energy transition and digital technologies often rely on the same minerals. Materials essential for semiconductors, such as high-purity silicon, indium, germanium, and gallium, are also crucial for solar photovoltaic (PV) production (IEA, 2025b). According to IEA's projections, by 2030, DCs alone could increase mineral consumption by up to 2% for silicon, over 3% for REEs, and as much as 11% for gallium (as a share of their total demand in 2024) (IEA, 2025b).

This partial overlap between clean energy transition minerals and minerals required for the digital transition and AI, is also highlighted by UNEP's "Global Resources Outlook 2024" (UNEP, 2024) and UNCTAD's "digital economy report" (UNCTAD, 2024). The minerals and metals identified by the latter as key for digitalisation include aluminium, cobalt, copper, gold, lithium, manganese, natural graphite, nickel, REEs, and silicon metal.

These materials contribute to essential digital (as well as supporting electricity supply) functionalities. They are, however, "almost identical" to those required for the shift towards a low-carbon economy, leading to competing demands. This convergence implies that "the world is moving from dependence on fossil fuels to dependence on multiple elements in the periodic table" (UNCTAD, 2024).

#### 6.1.2. Mineral refining: Miniaturisation and the need for high purity materials

Ever since their beginnings, microelectronics have been defined by continuous miniaturisation. Gordon Moore recognised this early-on and stated his homonymous "law" (Moore, 1965). This continuously ongoing miniaturisation was possible through continued innovations in various technologies, materials, and processes, including the shape and materials used for integrated circuits (ICs), their production processes, and ways of achieving ever higher purity of materials.

Due to this miniaturisation, by the early 1990s, transistor gates had reached a size of about 1 micrometre (μm). The aluminium wires connecting individual transistors, however, were becoming a bottleneck. As they got thinner, their electrical resistance increased, and they were prone to break down under high currents; the technology would have hit its limits latest around 0.18 μm, i.e., 180 nanometres (nm) (Isaac, 2017). In 1997, however, IBM pioneered a new copper wiring process for integrated circuits (Edelstein, 2017), which enabled the miniaturisation to continue.

Subsequent innovations included immersion lithography (thus shortening the wavelength used in the printing of ICs, allowing the print of finer lines) or the introduction of fin field-effect transistors (FinFETs), which ended the era of planar transistors allowing further miniaturisation through three-dimensional channels (Zhang *et al.*, 2024). The current state-of-the-art reached now 5nm or even 3nm via gate-all-around FET (GAAFET), introduced in 2022 (Samsung, 2025).

It is predicted that around 2030, ICs will become sub-nanometre and enter what is sometimes called the Angstrom (Å) era, 10 Angstroms being 1 nm big. But the *atomic scale* has already been reached today. The nearest neighbour distance for silicon is 2.35 Å (Royce, 2002), implying that a gate of 3 – 5nm contains only about 12 – 21 atoms. Correspondingly, the electric connections also become ever thinner. This miniaturisation implies the need for both more variety among the materials used, and for more purity in some of these materials:

- As with aluminium 30 years ago, the resistivity of copper has now reached a point where it is too high, and does no longer allow enough electrons to pass. It thus either needs to be replaced by or enhanced with other elements; one of the chief reasons behind the ever-increasing need for more elements in the production of microelectronics.
- At the same time, some elements need to be as pure as possible. At the low nanometre scale, a single atom of impurity can disrupt the function of an entire transistor, leading to device failure. For the silicon substrate of the wafer (which is the basis on which transistors are built), for example, very high purity silicon is required.

The purity is often indicated by the number N of 9s required, each increment of N representing an order of magnitude. As an UNCTAD report states: "The ICT elements are often needed in extremely pure form, 99.999% (five nines or more, 5N+) purity" (UNCTAD, 2020), showing how the number N includes both the 9s before and after the dot. Among all elements deployed in microelectronics today, the highest purity is required for silicon; 11N, which means a difficult to grasp 99.999 999 999% purity (Roussilhe *et al.*, 2025).

But many other elements require purities several orders of magnitude higher than the regular industrial grade. While a few niche fields (such as space exploration) might occasionally also require elements of high purity, both the variety and the amounts of high-purity elements required by microelectronics, make this field unique (Roussilhe *et al.*, 2025).

These trends towards higher purity and material diversity induce both additional energy consumption and resource depletion. A 2008 article was already noting that "there is a general trend towards higher purity (lower tolerances) in materials and parts. The purification of input materials and the need to create low-entropy environments in manufacturing lead to significant energy and materials use" (Krishnan, Williams and Boyd, 2008). Although some of these consequences have been occasionally explored, this materiality remains largely unexplored: The industry seldomly publishes any data, few academic papers address it, LCA databases do not cover it.

#### 6.1.3. Microelectronics manufacturing

Semiconductor manufacturing is split in two broad steps: The production of wafers (i.e., the substrate) for the later ICs, and the semiconductor fabrication, which consists of printing the circuits of the IC on that substrate. The first step consists to a large extent of the mineral refining that has been discussed in Section 6.1.2 above; the current section thus focuses on the latter step, the manufacturing of the integrated circuits themselves.

This phase, often also referred to as "front-end fabrication" (Hess, 2024) or simply "fabrication" (a term which by tradition refers to building the ICs themselves when not further qualified) takes place in "semiconductor foundries", also known as "semiconductor fabrics" – or simply "fabs". Fabs are specialised manufacturers that typically only produce chips without designing them. Their customers are fabless firms (such as Nvidia or Qualcomm), which design ICs but outsource their production. Foundries provide the advanced factories, process know-how and cleanroom infrastructure needed to turn the bare wafers into working chips (Ruberti, 2024).

Foundry manufacturing is a long, multi-step sequence of lithography, etching, deposition, and cleaning operations. Highly automated, it requires "more than 50 types of equipment and around 300 types of chemicals in more than 1000 process steps" (Hess, 2024). Together, they induce a high energy use (for high-temperature tools and tightly controlled cleanrooms), high-water use (mainly as ultrapure water for washing the wafers) as well as large quantities of hazardous chemicals such as solvents, acids, and specialty gases, e.g., $SF_6$ (Ruberti, 2024).

### 6.1.4. Various environmental impacts due to mining, refining, fabrication

Given the low data quality and large uncertainties discussed above, we can only draw a qualitative discussion of microelectronics production impacts. These are often related to resource depletion, but not only; energy consumption, GHGs, water consumption, and different types of pollution can be relevant as well.

During the *mining* stage, the most worrisome impact type is *resource depletion*. As discussed in Section 6.1.1, low-ore grades imply that large quantities of rock need to be displaced for microelectronics. Consequently, another important environmental impact is pollution. Main risks include acid mine drainage heavy-metal runoff but also the mercury and cyanide deployed in open pits. Water is used for dust control and mineral processing; withdrawals can be large and contamination is a risk. Energy and GHGs are least worrisome in this phase: While diesel is deployed both for drilling and transport, it is typically less energy-intensive than later refining.

The *refining* phase is very little understood. As argued in Section 6.1.2, the decade-long miniaturisation led to a need for a high degree of purification in many materials. Smelting, electro-refinement, and ultra-high-purity processes, however, are energy-intensive, leading also to GHG emissions in the process. Pollution can also be relevant at this stage, both to the air ($SO_2$, $NO_x$, acid mist, fluorides) and to the soil from effluents containing metals or acids.

During front-end *fabrication*, water consumption can be an important concern, as semiconductor fabs consume large volumes of ultrapure water for repeated we cleaning. The water used during microelectronics fabrication is nevertheless about an order of magnitude smaller than both the DC cooling water and the water consumed for electricity production (IEA, 2025a). Energy and related GHGs are also important due to the need for cleanrooms, HVAC and dehumidification, vacuum systems, deposition and etching tools, and especially the advanced lithography (EUV) process. Pollution, on the other hand, is typically well managed in the multi-billion dollar fabs.

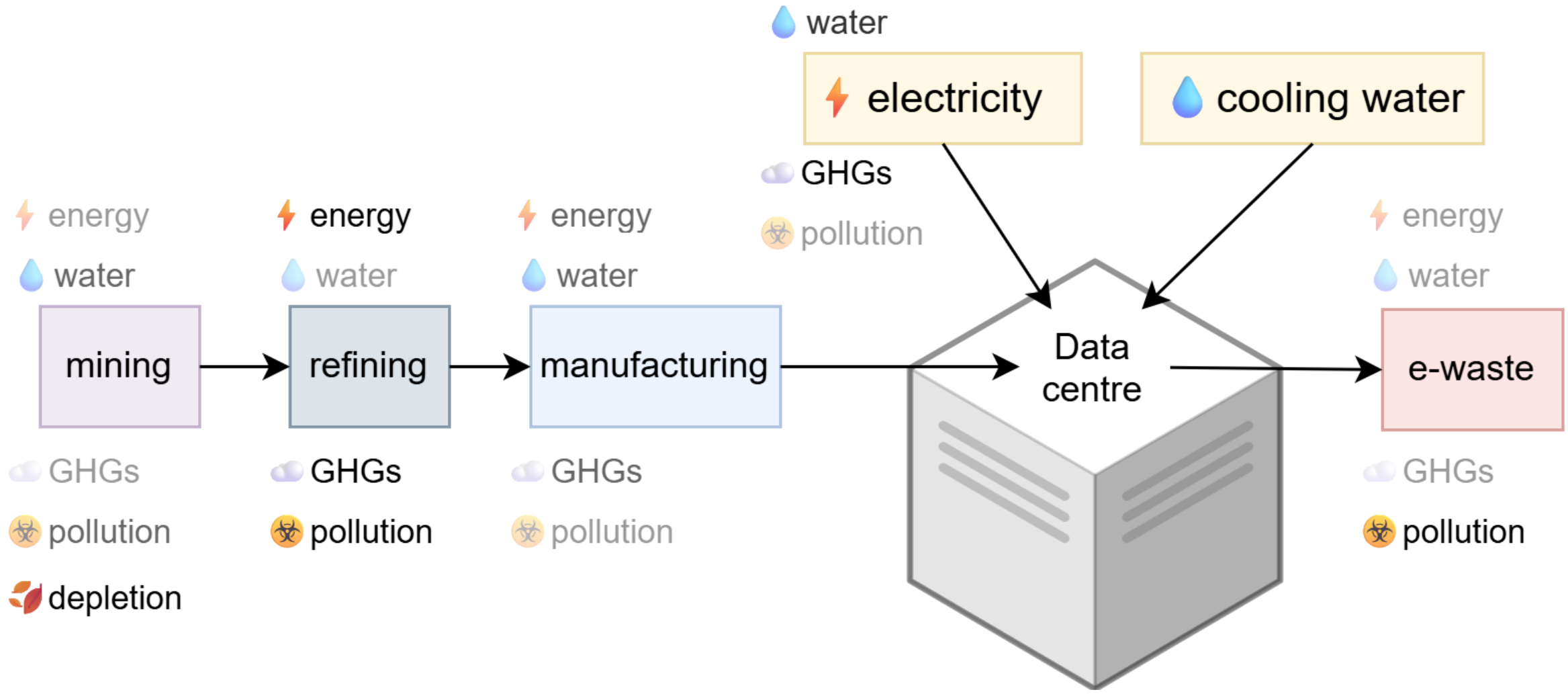

Figure 5: A qualitative depiction of a data centre's microelectronics lifecycle phases (production – with mining, refining, and fabrication), operational, and EoL, together with their main environmental impact categories. The more important a specific impact in one of the phases is, the clearer it appears. The least worrisome an impact is likely to be, the more blurred it is represented. These are only rough, qualitative assessments.

Figure 5 summarises this information in a qualitative manner. Next to the three phases of the production process, it also shows the operational phased of microelectronics in DCs (with electricity and water as main inputs) as well as their EoL, when they typically become e-waste. As the figure shows, not only resource depletion is relevant during manufacturing, other environmental impacts can be quite substantial as well.

With increasing decarbonisation of the use-phase electricity, the production-phase GHGs in particular could become relatively more important. Some studies have gone so far as to claim that the carbon footprint associated with manufacturing semiconductors has become the largest share of the total carbon footprint for computing hardware (Gupta *et al.*, 2021). Similar claims have been put forward by Microsoft, arguing that as operational energy is decarbonised through renewable procurement, the upfront embodied carbon of IT equipment and infrastructure rises to between 50% – 70% of a facility's total ten-year footprint (Alissa *et al.*, 2025).

Such assessments are likely based on misleading market-based assumptions for the carbon intensity of operational electricity, which dwarf the use-phase electricity to unrealistically tiny values (Kamiya and Coroamă, 2025). Although not sufficiently understood, the production of microelectronics is nevertheless responsible for substantial emissions. The main culprits are the important energy consumption during wafer fabrication (Ruberti, 2023) as well as the fluorinated gases employed in etching and cleaning, with global warming potentials thousands of times higher than $CO_2$. Sulphur hexafluoride, for example, is 25,000 times more potent than $CO_2$, while other gases can persist in the atmosphere for 50,000 years (Hess, 2024).

## 6.2 Hardware lifespan

For any industrial good or service – DCs, their devices and services included – the impact embodied during its production will be amortised over its lifecycle: the longer the lifespan and the more intense the usage during this period, the less per-usage impact resulting from its production. Before discussing circular economy strategies in the next chapter, it is thus meaningful to discuss the lifespans of DC devices.

Hardware lifetimes in data centres differ sharply across asset types. IT components such as servers, storage and networking are typically shorter-lived than the facility infrastructure, including the power distribution discussed in Chapter 3 and the cooling systems discussed in Chapter 4. Refresh cycles are driven by different factors across these groups, and potential circularity measures targeting them must consequently also differ.

### 6.2.1. Lifespan of IT components

The lifespan of servers varies depending on the operator’s demand (driven by workload growth), technical and performance requirements, and budget. It also depends, however, on more pragmatic considerations such as security requirements, IT support, reliability, maintenance, and efficiency targets.

Until recently, lifespans used to fluctuate from 3 years (ARUP, 2025a) to 5 years (Super Micro Computer, 2019). More recently, however, the lifespan grew towards around 6 years now (Alissa et al., 2025; Coroamă et al., 2025). This lifespan extension can be attributed to a variety of factors, such as cost consciousness, growing demand paired with supply bottlenecks, but also to diminishing efficiency gains due to a slowing of Moore’s law for CPUs and general-purpose servers (Andrews and Kerwin, 2026).

Storage (both HDDs and SSDs), on the other hand, still has slightly shorter lifespans, being usually replaced every three to five years, driven by failures and the need to increase capacity or density (CEDaCI, 2020). Networking equipment such as switches and routers is often modelled on a three to five-year refresh cycle as well, even though some devices can remain functional for 10–20 years (Andrews and Kerwin, 2026).

### 6.2.2. Lifespan of infrastructure

Network cabling is commonly rated for about 10 years, while power cables are expected to last around 20 years. The mechanical and electric systems that support IT loads are generally designed for longer service lives – often around 20 years for major equipment including UPS systems, large transformers, switchgear, backup generators, PDUs, and most cooling systems (chillers, CRAC units, pumps, cooling towers, and heat exchangers).

A key exception are batteries, which typically last only three to five years and are highly sensitive to temperature, making them both a frequent maintenance item and a reliability risk if operated outside recommended conditions (CEDaCI, 2020).

Finally, server racks and chassis sit between IT and infrastructure: while racks may have a 15-year design life, they are often replaced much earlier (sometimes every three to four years) simply because they are swapped out alongside servers (Alissa et al., 2025). Meanwhile, a reusable server chassis could last for at least 16 years (Andrews and Kerwin, 2026). Implementing modularity and standardised parts could substantially extend lifespans and thus reduce the material footprint. Modularity and standardised parts are discussed in the following Chapter 7 on material circularity in DCs.

#### 6.2.3. Consequences for circularity

These uneven lifespans create both problems and opportunities. On the one hand, short refresh cycles for servers, storage media, batteries, and some networking components can drive a steady flow of high-impact equipment through manufacturing and end-of-life pathways, increasing embodied emissions and material demand even when the hardware still functions. On the other hand, the fact that many assets around the IT stack – cabling, cooling and power equipment, and even racks – are designed to last far longer highlights a clear circularity opportunity: data centres can be treated as “mixed-lifetime systems” where the long-life frame (infrastructure and enclosures) is kept in service while only the parts that truly need performance upgrades are replaced.

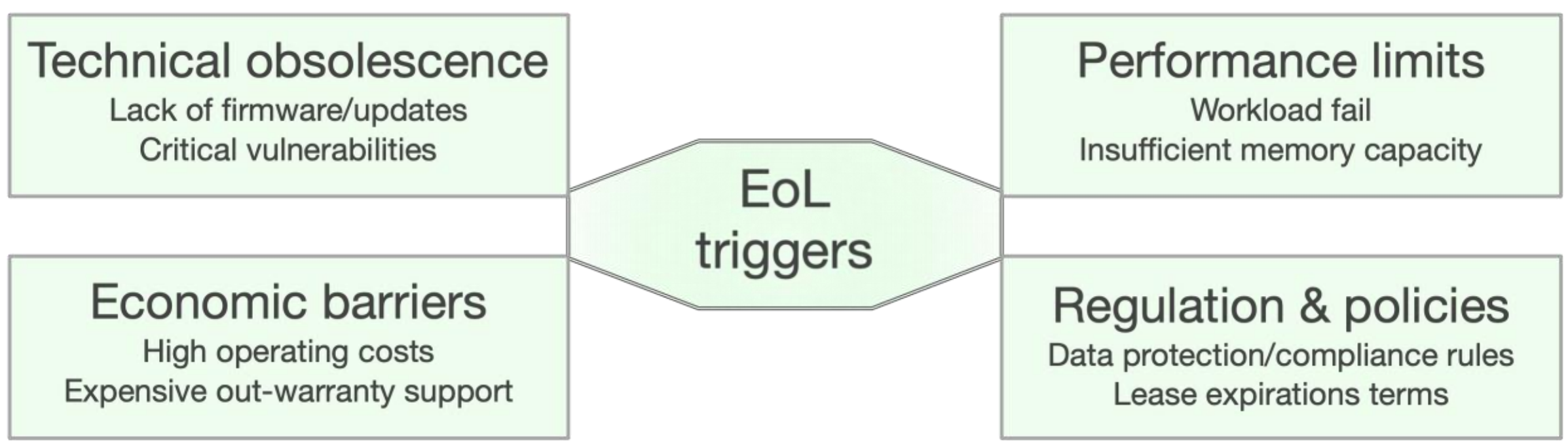

Figure 6: Various end-of-life triggers for data centre components.

Data centre hardware rarely reaches a single, clear-cut end-of-life: it reaches an end-of-role, shaped by overlapping constraints rather than age or technical obsolescence alone. As illustrated in Figure 6, equipment may remain technically operational but still be replaced because it no longer meets workload needs (performance life), falls outside vendor support or spare-part availability, becomes economically unsustainable, or is retired due to internal standards, leases, or compliance rules (policy life).

Even when specific bottlenecks such as compute capability, memory capacity, storage I/O, network bandwidth, rack power, thermal limits etc. shorten the practical service life of an otherwise functional system, there is often still remaining value that can be managed through circular strategies that will be discussed in the next chapter.

# 7 Material circularity for data centres

The processes discussed in the last chapter were implicitly presented as *linear*: Material extraction is followed by refining and production, usage, and ultimately EoL. This paradigm is not ideal from either an energy consumption or an environmental perspective (including GHG emissions, material consumption, waste generation, or pollution). Often, it is also not the most economically meaningful paradigm. The paradigm aiming to mitigate these issues of linear economic processes is typically referred to as "**circular economy**" (CE).

## 7.1 General circularity principles

The circular economy paradigm does not stem from a single source, but is a confluence of ideas that emerged from several distinct schools of ecological and economic thought starting in the 1960s (Ghisellini, Cialani and Ulgiati, 2016). Correspondingly, there are numerous existing understandings and definitions of the concept. Distilling over 100 of them, (Kirchherr, Reike and Hekkert, 2017) put forward the following definition:

> "*A circular economy describes an economic system that is based on business models which replace the 'end-of-life' concept with reducing, alternatively reusing, recycling and recovering materials in production/distribution and consumption processes, thus operating at the micro level (products, companies, consumers), meso level (eco-industrial parks) and macro level (city, region, nation and beyond), with the aim to accomplish sustainable development, which implies creating environmental quality, economic prosperity and social equity, to the benefit of current and future generations.*"

### 7.1.1. The 3R – 9R frameworks

The definition above brings together a few core concepts of the circular economy, in particular (Kirchherr, Reike and Hekkert, 2017):

- The *3R/4R framework* describes the 3-4 crucial principles of circularity, which in decreasing order of importance are:
  - *Reduce* aims at preventing or minimising inputs of raw materials and energy.
  - *Reuse* encompasses the extension of product lifecycles via e.g. initial design, repair, and refurbishment.
  - *Recycle* closes the loop by dismantling products at their EoL, salvaging as many resources as possible, and thus turning waste back into raw materials.
  - *Recover energy*, which is sometimes included as fourth principle, describes the incineration of materials with energy recovery.
- The *systems-level approach*, which identifies opportunities for circularity at three different levels:
  - Micro level, where the focus lies on individual products and consumers (eco-design).
  - Meso level, which centres around regional cooperation such as eco-industrial parks.
  - Macro level, concerned with national and international policies.
- The triple bottom line, which brings together economic prosperity, environmental quality, and societal equity.

The original reduce-reuse-recycle 3R framework in the enumeration above originates in the 1970s and describes various circularity strategies in decreasing order of value. It has progressively been expanded to 4R first, then to subsequent 5R, 6R, 7R models. Today, the arguably best-known conceptualisations are the 9R framework (Potting *et al.*, 2017) and the two-layered taxonomy (with main and sub-categories) put forward by the EN 45560 standard of the European Committee for Electrotechnical

Standardization (CENELEC, 2024). The EN 45560 standard references several further European standards, which address individual stages of circularity.

Figure 7 compares the terminologies of these frameworks, showing overlaps and differences. The R0 and R1 ("refuse" and "rethink") levels from (Potting *et al.*, 2017) cover rather non-technical aspects. They include sufficiency paradigms of avoiding consumption in the first place and developing new business models (such as sharing) that require less products overall sharing aspects, respectively.

| 3R-4R framework | 9R framework | | EN 45560 standard | |
|---|---|---|---|---|
| | R0 | Refuse | | |
| | R1 | Rethink | | |
| Reduce | R2 | Reduce | Use less / narrowing flow | Consume less input resources to make a product |
| | | | | Reliable products that last longer |
| Reuse | R3 | Reuse | Use longer / slowing flow | Product reuse |
| | R4 | Repair | | Product repair, upgrade |
| | R5 | Refurbish | | Product and parts refurbishment & recondition |
| | R6 | Remanufacture | Use again / closing loops | Products & parts remanufacture |
| | R7 | Repurpose | | Products & parts repurpose |
| Recycle | R8 | Recycle | | Materials recycling |
| Recover | R9 | Recover | Material is lost | Energy recovery |
| | | | | Waste treatment |

Figure 7: Comparison of well-known circularity taxonomies: the traditional 3R/4R framework from the 1970s, and two more detailed modern interpretations, the 9R framework (Potting *et al.*, 2017) and the EN 45560 taxonomy (CENELEC, 2024).

### 7.1.2. Circularity levers

The literature highlights several levers for circularity, which are sometimes also called "enablers" or "drivers". Although there is no widely established taxonomy, the following levers stand out:

- **Product design**: Not only does product design determine up to 80% of a product's environmental impact (European Commission, 2020). This earliest stage of a product is also crucial to support circular actions later down the product's lifecycle (Ellen MacArthur Foundation, 2013; Kirchherr, Reike and Hekkert, 2017; Potting *et al.*, 2017). Design includes, for example, the use of modular or standardised components that allow only parts of the system to be upgraded, which enhances the reusability of the swapped components and longer lifespans for those remaining. Design thus often addresses the higher (and thus more valuable) levels of circularity from Figure 7 above, in particular R2 (*reduce*) and R3 (*reuse*).
- **Process design and business models** can also enhance circularity. Shifting, for example, from traditional ownership to performance-based, leasing, or service-based business models may encourage the creation of durable, easily maintainable products while generating stable, long-term revenues (Ellen MacArthur Foundation, 2013). This may in turn contribute to a *reduction* (R2) of the products being built. Furthermore, internal-company processes can be setup from the outset towards internal *reuse*, *repair* and *refurbishment*, and all the way to *repurposing* and *recycling* (R3 – R8).

- The **choice of materials** is fundamentally important for a circular economy both upfront (using recycled materials yields a direct *reduction* of material inputs, R2), and because it dictates the potential for products to be *refurbished* or *recycled* at their end-of-life (R5 – R8). Using non-toxic inputs additionally prevents the contamination of recycling streams, extending product longevity and increasing material productivity (Ellen MacArthur Foundation, 2013).
- **Operating conditions** are also crucial for circularity because they directly affect the lifespan of the product (as well as its efficiency, if it is energy-consuming during operation). By adhering to manufacturer-recommended environmental conditions and enhanced maintenance regimes, operators can prevent premature degradation and significantly extend the operational life of products (ARUP, 2025a), which leads to a *reduced* (R2) need for new materials and energy to build new products.

These levers are intentionally not exhaustive, and the boundaries between them not rigid. Design in particular could be understood in a wide sense as encompassing everything else: the design of products, of processes, the choice of materials used for the products, and the “design” of operating conditions. In the context of this report, we discuss design in a narrower sense, while explicitly distinguishing between product and process design. For the other levers, we focus not on their design aspect but on its results (materials and operating conditions, respectively). These levers are nevertheless inter-connected and – although discussed individually – should be understood as a system rather than linear and independent. Upstream choices, for example, largely determine the feasible downstream measures.

### 7.1.3. Relevant dimensions for the analysis of circularity in data centres

The circularity of data centres can be discussed based on these two dimensions, the general circularity principles presented in Section 7.1.1 and the levers from Section 7.1.2. They are discussed in Sections 7.2 and 7.3 below for two categories: a data centre’s IT components (i.e., all electronics contained in servers) and the rest of the DC (i.e., the building itself as well as its cooling infrastructure and the server racks or cabinets), respectively.

## 7.2 Circularity of data centre microelectronics

To address the circularity of a data centre’s servers and their microelectronic components, this section addresses each of the four circularity levers from Section 7.1.2 in turn.

### 7.2.1. Product design: Modularity and standardisation of servers

In early 2022, six companies, Intel, Dell, HPE, Google, Meta, and Microsoft (and as more recent addition AMD) started a collaboration within the open compute project (OCP) “data centre modular hardware system” (DC-MHS). Through six base specifications, DC-MHS documents several types of host processor modules (HPMs) and how they interface with the further building blocks needed to configure a full server (CEDaCI, 2020). With HPMs, the same module can be deployed in multiple configurations at higher scale (Aspnes, 2022) The benefits include reduced costs, reduced custom designs, faster time to market, but especially reusable building blocks across multiple designs and markets, module replacement, easier reuse or recycling (Leddy *et al.*, 2024).

In the current context of rapid AI expansion, however, the new servers use specific logical processing units (i.e., GPUs or ASICs), with different mechanical and electrical properties (Coroamă *et al.*, 2025). Traditional multiple-purpose servers can thus hardly be reused for AI. Moreover, the DCs themselves are undergoing fundamental changes, for example with regard to their power densities (see Section 8.1) or cooling systems (see Section 4.1). While modularity and standardisation are thus bound to have little effect in the transition from traditional DCs to AI data centres, the development of circularity standards within AI is all the more important. Such standards allowing modularity and the interplay of components from different manufacturers, would be particularly relevant for a server design inducing allowing both the ***reuse*** **(R3)** of swapped components as well as longer lifespans (and thus material ***reduction*****, R2**) for those server parts that do not require upgrade.

Additionally, **ease of dismantling** enables “chip harvesting” (i.e., the surgical extraction, testing, and reuse of functional chips), which can save an average of 6.4 kg $CO_2$ per chip as well as precious materials (Schröder, Charter and Barries, 2025). If reuse is not possible, the dismantling of printed circuit boards (PCBs) enables the recovery of critical raw materials such as gold, palladium, and copper (Schumacher and Green, 2023). Product design allowing easy dismantling for component ***reuse*** **(R3)** as well as high-value material recovery during ***recycling*** **(R8)** is thus very relevant in the context of DC microelectronics.

### 7.2.2. Process design and business models

Given their substantial volumes and often homogeneous hardware, several large DC operators have created hardware refreshment cycles focused on circularity processes that lead to a maximum of value recovery. By organising reverse logistics, triage, secure data handling, and refurbishment at scale, they can prioritise higher-value outcomes (redeploy, refurbish, parts harvesting) before resorting to material recycling. Examples include:

- Microsoft’s circular centres are on-site hubs within DC campuses where retired hardware is triaged and routed to the highest-value path (internal *reuse* → resale and external *reuse* / *repurpose* → donation → *recycling*). High-value parts (e.g., CPUs, memory) are harvested, tested/refurbished, recertified, and then reintegrated into operations. Data-bearing devices are handled under strict security requirements before leaving controlled sites. Microsoft links the program to its zero-waste by 2030 commitment and reports strong outcomes such as a 90% reuse and recycling rate in 2024, 3.2 million components reused, and harvested inventory covering 85% of demand for obsolete spares (Welsch, 2025).
- In Google’s refurbished inventory loop, decommissioned servers are dismantled into components, which are inspected and returned to an internal *refurbished* inventory. Refurbished parts are used for repairs and to build *remanufactured* servers that can be redeployed. Google reports a growing share of deployments and maintenance supported by reused components, growing from 21 in 2022 and 29% in 2023 to 44% in 2024 (Google, 2025a).

Such **internal value-recovery loops** cover all circularity levels between ***reuse*** **(R3)** and ***recycle*** **(R8)**.

For smaller enterprise data centres, some of the circularity-enhancing measures for end-user devices (such as the right to repair and service, no forced destruction, firmware access, and especially the transferability of licenses for a second life use, takeback and reverse logistics obligations etc.) could be relevant as well. They could help towards longer lifespans (and thus material ***reduction*****, R2**), ***reuse*** **(R3)**, ***repair*** **(R4)**, and ***refurbish*** **(R5)**.

**Sustainability trade-offs**

However, there are also trade-offs between different sustainability-relevant dimensions of DC microelectronics. The hardware refreshment cycles, for example, represent such a trade-off between improving the energy efficiency through shorter cycles (as newer devices are typically more efficient) and longer lifespans (Coroamă *et al.*, 2025), which are inducive of less embodied material and thus more circularity.

For AI inference, this choice might be even starker. A startup has recently announced the possibility to physically print AI models (once trained) onto a dedicated integrated circuit (IC), a so-called “application-specific integrated circuit” (ASIC). These types of integrated circuits, also deployed in cryptocurrency mining (Coroamă, 2021), have narrow application usages (the ASICs deployed in cryptocurrency mining can only perform hash functions, for example), which they perform highly efficiently (Coroamă *et al.*, 2025).

Physically etching AI model weights into an integrated circuit eliminates the memory bottleneck and removes the need to move data between the logical processing unit and the HBM. Consequently, the inference via this dedicated ASIC is 1-2 orders of magnitude more energy efficient than with traditional software models (Bajic, 2026). Once the model is no longer needed, however, the IC needs to be discarded, so there is a strong trade-off here between operational energy efficiency and long hardware lifespan together with the corresponding material consumption. This example thus shows a strong trade-off between energy efficiency and circularity.

Finally, certified data-wiping software can avoid the useless shredding of storage devices, HDDs and SSDs alike (CEP, 2022), thus allowing its ***reuse*** **(R3)**.

#### 7.2.3. Choice of materials

Changes in the design of the equipment and its material composition towards more “stripped down equipment with fewer components and embodied materials” (CEDaCI, 2020) addresses the ***reduction*** **(R2)** goal directly. A good example is the shift in storage technologies from hard disk drives (HDDs) to solid-state drives (SSDs). The former not only have a higher operational energy consumption than the latter, but at the same time also a higher embodied material consumption (CEDaCI, 2020). The shift from HDDs to SSDs is thus beneficially from both a circularity and an energy efficiency perspective, not inducing a trade-off such as the one discussed in Section 7.2.2 above.

Moving away from glues and permanent adhesives towards screws, clips, or click-fit solutions can facilitate rapid disassembly for repair, refurbishment, and eventual component recovery (CEP, 2022), thus fostering ***recycling*** **(R8)**. Minimising non-recyclable plastics and avoiding hazardous chemicals like flame retardants during server design, ensure that components can be safely handled and recovered at the end of their life cycle, further enhancing ***recyclability*** **(R8)** but also the safe incineration and thus ***recovery*** **(R9)** of energy (Andrews and Kerwin, 2026).

#### 7.2.4. Operational conditions

Operational prevention and maintenance refer to the adherence to recommended operating conditions (such as thermal, humidity, power quality), which can prevent premature failure and replacement (ARUP, 2025a). Continuous monitoring (e.g., via sensors) serves the same purpose, allowing the identification of potential part failures before they occur and of subsequent predictive maintenance. This not only reduces downtime but also protects other components from being affected, thus reducing the total cost of ownership (Baumann, 2022) and maximising the lifespan of those other components.

Operational prevention and maintenance thus address the most valuable circularity principle of the circularity framework, the ***reduction*** **(R2)** of material and energy inputs by minimising premature failures, they extend the lifespan of devices, working against their premature replacement. Additionally, preserving the value of the assets also improves the feasibility of later *reuse* (R3) and *repair* and *refurbishment* (R4–R5). The use of open-source software might also be beneficial for keeping hardware in use beyond the original manufacturer's support period (Andrews and Kerwin, 2026).

#### 7.2.5. Overview of data centre microelectronics circularity

Figure 8 presents an overview of the circularity options for microelectronics discussed in Sections 7.2.1 – 7.2.4 above. It uses therefore the two dimensions introduced in the beginning of this chapter: the principles of circularity introduced in Section 7.1.1 and the four levels of circularity levers from Section 7.1.2.

Figure 8 shows that various measures are possible, reaching across all levers or circularity. As discussed in Section 7.1.1, the principles with the lowest numbers are the most valuable from a circularity perspective: A reduction (R2) of deployed materials is always preferable to a later reuse (R3), repair (R4), or recycling (R8), for example.

Levels R0 and R1 (refuse and rethink), however, are oriented towards sufficiency goals. While a meaningful paradigm from an individual’s perspective, there are no obvious ways how this paradigm could be extended for data centre operators. Correspondingly, the reviewed literature did not identify any, and these levels have not been represented in Figure 8 in the first place.

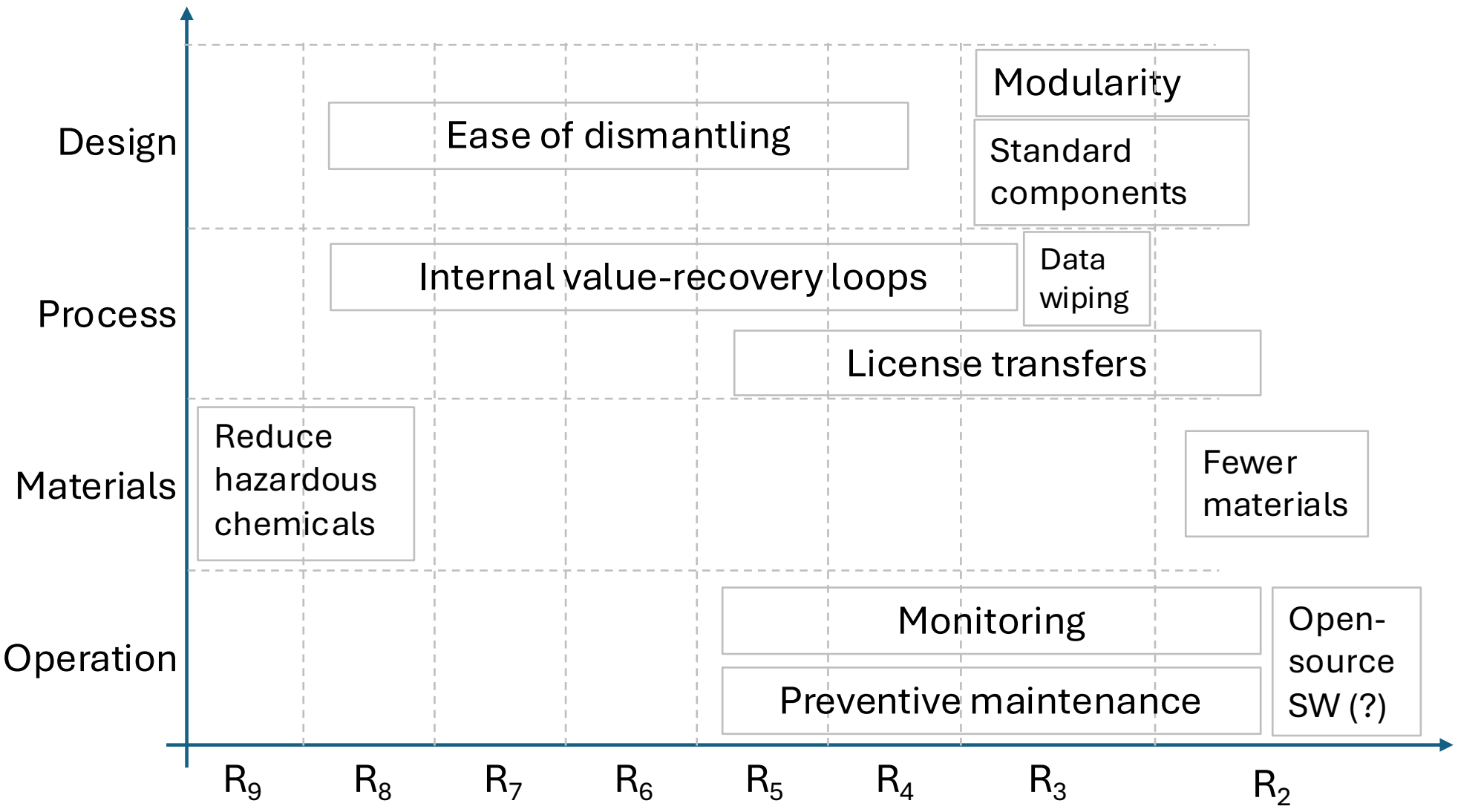

Figure 8: Overview of the circularity-enhancing measures for microelectronics discussed in this report.

## 7.3 Circularity of the data centre building and infrastructure

Similar to the circularity of microelectronics above, this section also follows the four circularity levers introduced in Section 7.1.2, showing some of the measures could be undertaken and where in the circularity taxonomy from Figure 7 their results would fit. As compared to microelectronics, however, the one on building and infrastructure level is both less relevant, and offers fewer actionable options.

### 7.3.1. Product design

Section 7.2.1 showed how modularity and standardisation, which also imply more interplay between components from different vendors, can lead to both component ***reuse*** **(R3)** – for the components that are upgraded – as well as the ***reduction*** **(R2)** of material use through the longer lifespans of those components that do not require updates.

It has further been argued that this is particularly relevant in the context of the new AI-oriented data centres deploying accelerated computing and the new standards that are being established. This principle is not only relevant for servers and their microelectronic components but also for the infrastructure: both for the new cooling systems (especially liquid cooling) as well as new power supply paradigms, as discussed in Section 3.2.

### 7.3.2. Process design and business models

In some contexts, **leasing** can serve as an alternative procurement model to buying that encourages circularity. Manufacturers could thus be incentivised to maintain longer-lasting systems components, extending their utility, while leasing and regularly upgrading the ones with a shorter lifespan. As all lifespan-extending measures, this also leads to material ***reduction*** **(R2)**, if less new products need to be built.

The used ones, in turn, can be properly ***repurposed*** **(R7)** after being returned (ARUP, 2025a). While this principle is valid for servers as well, it can also be applied to the cooling infrastructure, in a paradigm known as “cooling as a service” (ARUP, 2025a).

Furthermore, circularity-oriented DCs of large operators, with in-house triage centres as discussed in Section 7.2.2 above, extends beyond microelectronics. In its dedicated recovery hubs, Amazon, for example, routes retired racks similarly to the other components through a reverse-logistics network (Amazon Staff, 2023). This leads to ***reuse*** **(R3)** of rack infrastructure.

### 7.3.3. Choice of materials

On a building level, substituting carbon-intensive data centre construction materials with alternatives such as low-carbon concrete, mass timber (to the extent applicable), or recycled steel can significantly lower both upfront emissions and ***reduce*** **(R2)** the need for new materials (ARUP, 2025a). Materials such as timber can additionally lead to energy circularity, as they are better insulators and help reuse a larger share of the waste heat (Hoosain *et al.*, 2023).

Similarly, on infrastructure level, Meta emphasizes upstream circularity levers. Using recycled metals and post-consumer recycled plastics in server racks leads to a similar ***reduction*** **(R2)** in new materials required (Meta Sustainability, 2025).

### 7.3.4. Operational conditions

Operational conditions can also yield material ***reductions*** **(R2)**. Increasing the temperature difference in the cooling system – which may (or may not) also save overall cooling energy (Coroamă, 2025) – means less water needs to be used for cooling, which enables the usage of thinner pipes with less materials (ARUP, 2025a). For batteries as well, the usage of health management systems can also guarantee a long lifespan and prevent early disposal (ARUP, 2025a).

### 7.3.5. Overview of data centre building and infrastructure circularity

As above for microelectronics, Figure 9 summarises the circularity options for DC infrastructure and (to a lesser extent) buildings discussed above. Overall, they are not only fewer, but also in general of lesser impact than circularity goals for microelectronics. Some are even rather questionable. As discussed in Section 7.3.4, for example, higher inlet temperatures might lead to less cooling water needed and thus thinner pipes. This one-off benefit, however, seems absolutely marginal compared to the operational considerations for raising (or not!) inlet temperatures (Coroamă, 2025).

Overall, the most promising circularity measure for the DC infrastructure seems the modularity and standardisation of the new AI cooling and power provisioning infrastructures. As discussed in Section 7.3.1, these could bring about substantial sustainability benefits.

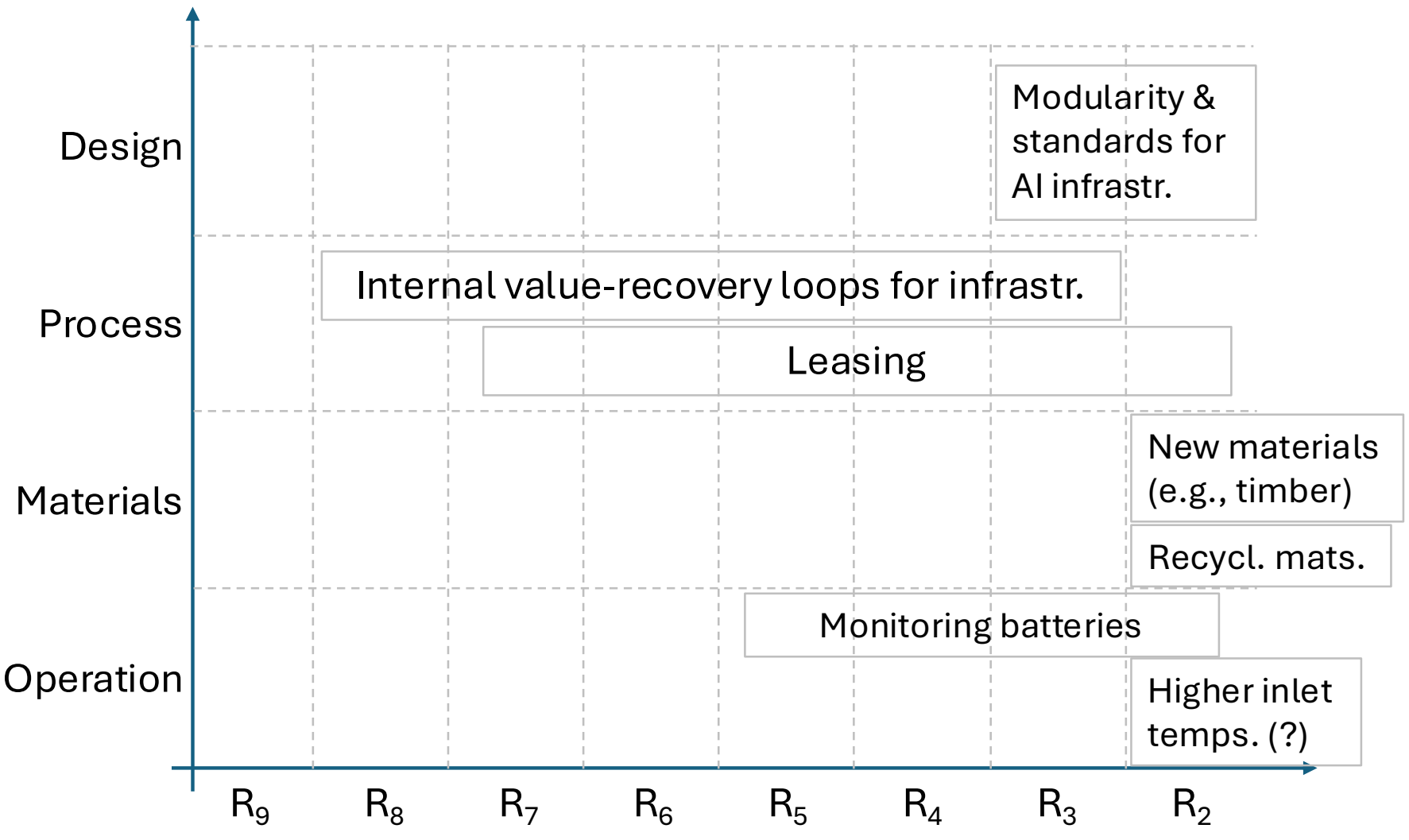

Figure 9: Overview of the circularity-enhancing measures for DC infrastructure and buildings discussed in this report.

## 7.4 Towards more material circularity in data centres

For a sustainable digital infrastructure, it is crucial to move beyond the linear *take-make-waste* model and shift from whole-asset replacement toward modular upgrades, repair, reuse and recovery. It has thus been argued that a circular electronic product has three main attributes: it is made from verified circular resources, designed for use-phase optimisation and recovery, and its materials are actually recovered at the end of its life cycle (CEP, 2022). However, as argued throughout this chapter, prolonging the lifespan of hardware might actually be the primary lever for amortising its environmental investment.

### 7.4.1. Circularity frameworks and circularity levers

Based on the principles presented so far, the circularity of DCs has been framed via the four levers that correspond to different stages in the development of a data centre. This structure is intentionally not exhaustive, and the boundaries are not rigid. The stages are interconnected and should be read as a system rather than linear and independent.

Upstream choices – especially the requirements set before equipment selection (e.g., standardisation, modularity, and serviceability) and the materials and EoL routes planned – largely determine the feasible downstream measures. These range from repair and safe reuse to refurbishment, and (as last resort) recycling.

In parallel, day-to-day operational practices both depend on those upstream choices but also influence circular outcomes by extending asset life, enabling controlled redeployment, and ensuring secure handling at transition points. Circularity is achieved through the interaction of upfront requirements, material decisions, recovery pathways, and operations across the full asset lifecycle.

### 7.4.2. Importance of data transparency

DC circularity is shaped by a complex, multi-tier supply chain that is often hard to trace. Companies may play multiple roles across the lifecycle, and design, material sourcing, and assembly are frequently subcontracted across several actors. Detailed knowledge across this complex supply chain is a prerequisite for circularity: “The first step to achieve this circular economy is to know the precise composition of the equipment and the availability (lifespan of equipment, location and amount) of end-of-life equipment generated by the data centre industry. This information is not easily accessible to designers or manufacturers, but it is fundamental to know before establishing the best end-of-life strategy (CEDaCI, 2020).

Data gaps, on the other hand, affect decisions across the entire product lifecycle. As (Wagner, 2025) notes, such gaps include:

- limited supplier identification and supply-chain traceability during resource production,
- missing long-term failure and in-service data,
- weak internal communication,
- unknown material mixtures during design and production,
- lack of repair guidance,
- uncertainty about secure data deletion,
- unclear product history or remaining lifetime during use and reuse, and
- missing dismantling instructions plus inadequate declarations (e.g., flame retardants or critical raw materials) at EoL.

One way to achieve data transparency is the prioritisation of products with digital product passports (DPPs). According to the Ecodesign for Sustainable Products Regulation (ESPR), DPPs include material components information that can be electronically accessed and shared among supply chain businesses, authorities and consumer (European Commission, 2024b).

## 7.5 CapEx versus OpEx implications

Capital expenditure (CapEx) refers to the costs due to investments and maintenance of fixed assets of an organisation, while operating expense (OpEx) encompass the running expenses necessary to sustain business operations, such as utilities (electricity and water), salaries, and recurring maintenance costs (Gupta *et al.*, 2021).

Traditionally, OpEx used to dominate the lifecycle costs of a data centre. The rapid expansion of DCs over the last years (both AI and traditional general-purpose ones), however, made CapEx also grow substantially over the past years. When OpEx are the most significant cost factor, it makes sense to have short lifespans and replace after a relatively short time the old hardware with newer, more efficient one. The increased CapEx will be outweighed by OpEx savings.

In a CapEx-dominated world, however, the balance shifts towards longer lifespans to reduce investments, even if operational expenses are comparatively higher due to the aging hardware. More than in the past, there is thus an **increasing synergy between circularity principles in DCs and economic interests**. Increasing the lifespan of devices not only serves environmental goals but can be a strategic lever to reduce CapEx – and thus lifecycle costs – as well.

Correspondingly, a decade ago, the lifespan of servers used to be around three years, but it has been progressively prolonged to 5-6 years (Coroamă *et al.*, 2025). Google, for example, extended average server lifespans from three to four years back in 2021 and then again to five years in 2023. Consequently, depreciation dropped by $3.9 billion annually, while net income rose by $3 billion. Integrating refurbished and remanufactured equipment into the supply chain can further reduce CapEx. Through such methods, including lifespan extension of servers and networking devices from four to five years, Meta achieved $1.5 billion in savings in 2022, including a $860 million reduction in depreciation expenses (Celestica, 2025).

On building and infrastructure level, applying circular principles (e.g., the reuse of structural materials and deploying modular elements) also reduces both material costs, and simultaneously the exposure to raw material price volatility and geopolitical risks. Building and infrastructure CapEx can be consequently reduced by 15–20 % (ARUP, 2025b).

Large DC operators have the scale to think through and apply circularity processes and business models, avoiding hardware replacement and lowering costs. Smaller operators with less levers, control, and experience, however, typically reach the breakeven point faster, from which on the OpEx of old machines (due to power, cooling, labour, downtime) make their operation economically unsustainable.

As (CEDaCI, 2020) thus notes: “It is understandable, however, that reuse, refurbishment and collection rates are higher in big DCs (which have internal reuse mechanisms), than in small and private data centres”. To improve circular outcomes, smaller operators who cannot building in-house capabilities, therefore need to rely on secure, well-connected supply chains – through trusted providers, clear take-back pathways, and audited reverse-logistics partners that can deliver equivalent refurbishment and recovery services.

Finally, with the rapid expansion of AI, CapEx are rising even faster, primarily driven by the accelerated computing hardware, but also by facilities, power delivery, cooling systems and other equipment. Therefore, the largest hyperscale DC operators currently incur CapEx of dozens, sometimes hundreds of billions annually. McKinsey estimates that by 2030, roughly 78% of projected global DC CapEx will be due to AI, while only about 22% for traditional IT workloads (McKinsey, 2025). The synergy between circularity principles and economic interests might thus be getting stronger.

# 8 Data centres and the power grid

A variety of challenges and opportunities arise for the power grid due to the rapid growth of data centres, and the specifics of this growth. It turns out that, not dissimilar to electric vehicles, data centres too can be both a blessing and a menace for the power grid. Challenges and opportunities are varied and quite intertwined.

In an attempt to systematise the landscape, this chapter is organised in 5 sections as follows: It starts by presenting the three features of the current DC growth with the strongest impact on the power grid. It then shows which immediate challenges for the grid follow from these characteristics. Section 8.3 presents current mitigation measures for these challenges. The mitigation measures themselves, however, bring along some detrimental side-effects, discussed in Section 8.4. The same mitigation measures, however, also present further opportunities for the grid. And even one of the DC growth characteristics presented early in the chapter brings not only challenges but also opportunities. These chances are the topic of the last Section 8.5. These topics are also highlighted in Figure 10 that summarises the content of the entire chapter but is placed at its beginning to serve as visual aid throughout.

## 8.1 Data centre growth and its features

Due to cloud computing, streaming, and in particular AI, we are witnessing an accelerated DC growth since the late 2010s (Shehabi *et al.*, 2024). Consequently, in its “base case” scenario, the IEA expects DC electricity demand to more than double from 416 TWh/year in 2024 to 946 TWh/year by 2030. Despite this accelerated growth, DCs will likely not be one of the main drivers for the global growth of electricity consumption. This growth of 530 TWh/year over 6 years would only be the 5$^{th}$ most important growth driver after industry, electrification of transport, appliances, and space cooling (IEA, 2025a).

Locally, however, the highly localised power demand of data centres is a new phenomenon for the grids, and one that can lead to various challenges. Three characteristics are of particular relevance: the *power density* of data centres (and in particular of AI-heavy DCs), the *large and quite flexible loads* they represent, and the *steep power demand ramps* of AI compute.

### 8.1.1. Power density of modern data centres, in particular AI

AI companies are currently engaged in what has been described as one of the largest infrastructure build projects in human history (‘Inside OpenAI’s Stargate Megafactory with Sam Altman’, 2025). The rapid scaling and the associated power capacity required can be detailed through specific company actions and industry projections. A few examples follow below.

**OpenAI’s Stargate**: In January 2025, OpenAI and its partners committed to investing 500 billion USD over four years into US-based AI infrastructure requiring 10 GW of power; by July 2025, they expected to exceed this initial commitment (Open AI, 2025). An new agreement with Oracle alone should require an additional 4.5 GW.

**xAI's Colossus:** xAI built its main AI data centre, Colossus – which was the world's biggest supercomputer at the time – in just 122 days, far outpacing initial estimates of a project lasting for 24 months. The data centre houses 200,000 GPUs and consumes around 300 MW of power (xAI, 2025). The next data centre campus is expected to house 1 million more powerful GPUs, requiring around 2 GW of power (Yong, 2025).

**Meta’s Prometheus and Hyperion**: Meta announced several multi-GW clusters. The first one called “Prometheus” should be online in 2026; the largest project “Hyperion” should scale up to 5GW of power and occupy a size roughly compared to all of Manhattan (Zuckerberg, 2025).

**Google**: Google’s vice-president for cloud infrastructure is reported to have said in an internal meeting that Google should double its AI compute capacity every 6 months for the next 4-5 years, which would lead to a capacity 2-3 orders of magnitude higher than today (Butler, 2025).

Legend
DC development
Challenges
Mitigation
Negative side-effects
Grid opportunities

AI-driven DC growth
Power density ↗
Large flexible loads ↗
Steep load ramps ↗
Power provisioning ↘
Peak load ↗
Grid stability ↘
Grid congestion ↗
Marginal price ↗
Transmission losses ↗
Price ↗
Local communities
Noise & pollution
On-site generation
Competition to grid generation sources
GHGs
Overprovision
Resource consumption
Workload shifting
New algorithms
Battery energy storage system (BESS)
Competition to
EV batteries
Grid-level batteries
Renewable curtailement ↘
High voltage transmission ↗
Grid development
Demand-side management
Peak demand production
Grid flexibility potential
New storage development
New el. gen. development
Grid innovation

Figure 10: Overview of the complex relationship between data centre (DC) growth and the power grid: Gray are the 3 main characteristics of the AI-driven DC growth. Red boxes show the resulting challenges. Mitigation measures for these challenges are blue, and blue arrows represent the resulting mitigation (while all the other arrows in the graph represent consequences). Yellow boxes contain the detrimental side-effects that result not from DC growth directly, but indirectly from the mitigation measures. Green are the beneficial consequences of both mitigation and the DC growth itself. Mitigation thus has three types of effects: Alleviating the issues it was originally designed for (red), detrimental side-effects (yellow), and beneficial side-effects (green).

**Anthropic**: Anthropic estimates that 2 – 5 GW data centres would be needed for the training of its most advanced AI models in 2027-2028 (Anthropic, 2025) and has already signed to use up to one million of Google's TPUs in a deal worth tens of billions. For all AI in the US, Anthropic estimates 20-25 GW required for training by 2028 and "at least as much" for inference, yielding a minimum requirement of additional 50 GW for the US in 2028 (Anthropic, 2025).

**Worldwide developments**: The drive to scale is part of a larger, global AI race (Anthropic, 2025). Data centres are also being built in e.g. China, Malaysia, Japan, and France ('Inside OpenAI's Stargate Megafactory with Sam Altman', 2025). Google will invest 2 billion USD in DCs in Turkey and over 6 billion in Germany (Butler, 2025). In the Middle East, the United Arab Emirates (UAE) and Saudi Arabia are planning large AI investments: G42 is reportedly building a 5 GW data centre campus in the UAE in partnership with Microsoft, while Data Vault, a Saudi company, has broken ground on a 2 GW data centre (*The Geopolitics of AI Infrastructure*, 2025).

The power for each of the projects mentioned above is concentrated in one or a few campuses. This implies a new level of power density, comparable to only sporadic examples in the past (for example, aluminium smelters).

#### 8.1.2. Grid-connected large flexible loads (LFLs)

Traditional DCs, even large ones, are quite stable electricity consumers. In enterprise DCs, and more so for cloud computing and colocation DCs, the diversity of customers, users, and their tasks tends to work against large compute demand variations, and consequently also against stark power demand fluctuations. Even considering skewing phenomena such as reoccurring end-of-month peaks or one-time changes of the inlet temperature (which increases the server fan consumption to compensate), a 2025 case study showed that the power drain in the server rooms of a large colocation DC stayed in a narrow band of just 6%, varying between 97 – 103% of the average consumption (Coroamă, 2025).

This is quite different for hyperscale and AI data centres. These novel DC electrical loads are not only highly concentrated, they can also be quite volatile, which yields novel challenges for the grid: "Unlike residential or legacy commercial loads, digital loads can materialize or vanish with little notice. Traditional load forecasting tools fall short in tracking this volatility" (Kamwa, 2025).

Such loads can be witnessed very well in Texas, US, where most of the grid (about 90% of it) is governed by the state's non-profit "Electricity Reliability Council of Texas" (ERCOT). Given its deregulated market, business-friendly state policies, exemption from regulatory federal oversight, and resulting low electricity prices, Texas has been an early magnet for large digital power demand – first for cryptocurrency miners, nowadays for AI (Neel, 2025), and serves as a good case study.

In line with its liberal, laissez-faire principles, the interconnection process with ERCOT had traditionally been quite lean with some minimal notices and feasibility studies required for the sporadic connection of large loads. From mid-2021, however, with the advent of cryptocurrency miners first and AI data centres afterwards, there has been "an unprecedented growth in large load interconnection (LLI) requests" (Sahni *et al.*, 2025).

The development was so abrupt that ERCOT quickly had to adapt and change its process. These loads were recognised as not merely a quantitative load growth, but a qualitative shift in behaviour. They were given a distinctive name, "large flexible loads" (LFLs) (Kamwa, 2025), and an LFL task force was established in 2022 (Sahni *et al.*, 2025).

LFLs have been defined by ERCOT as loads larger than 75 MW (Deb and Halbe, 2025). In a stark contrast to the earlier, lean registration process, LFLs now need to be submitted to the "LLI queue" of ERCOT, where they are subject to a mandatory – and increasingly process-delaying – review (Sahni *et al.*, 2025). If approved, any LFL operator must agree to curtail its consumption if instructed so by ERCOT to ensure grid stability (Deb and Halbe, 2025).

#### 8.1.3. Steep power load ramps due to synchronous AI compute

AI model training frequently save model checkpoints, which can mean a sudden and synchronous power drop across thousands – or tens of thousands – of processors. For current state-of-the-art power consumptions of around 1kW per GPU (Coroamă *et al.*, 2025), this can induce steep ramps of megawatts or dozens of megawatts. When training resumes, the same power load ramp occurs, in the opposite direction.

These ramps are not a mere theoretical construct – they have been observed in practice both by AI researchers (Morrison *et al.*, 2025) and consultants for power grid operators (Quint *et al.*, 2025). The latter study, looking specifically at LLIs, observes that "data center load have shown how unconventional they are compared with historical end-use loads, with activity like AI training runs leading to extremely intermittent power consumption" (Quint *et al.*, 2025).

In one example of one 50 MW data centre building (from a very large 200 MW DC), the authors observed a "power consumption jump from 6 MW to 30 MW in a matter of 290 ms (about one-quarter of a second). Power consumption is then highly variable for about 5 mins, with numerous 5+ MW spikes. Then it drops briefly to low power consumption levels before returning to high power consumption" (Quint *et al.*, 2025).

## 8.2 Resulting challenges for the grid

The three particularities discussed in Section 8.1 result in several challenges for the grid. They are discussed in the current section, grouped into three main resulting challenges, and a few additional ones that follow indirectly from one of the three main challenges.

#### 8.2.1. Grid stability issues

The steep power ramps addressed in Section 8.1.3 can lead to grid stability issues if left unmitigated, especially given the size and concentration of loads, as discussed throughout the entire Section 8.1 above. In the sub-second, 24 MW ramp presented in Section 8.1.3, the facility was not interfaced through an UPS with the grid, so the grid was directly exposed to this steep power ramp without any mitigation (Quint *et al.*, 2025).

This sudden 24 MW drop due to the synchronicity of AI training loads equated to about 50% of the corresponding DC building's total 50 MW. For a DC of hundreds of megawatts or even gigawatts, this could imply correspondingly larger ramps. AI-scale DC campuses (in the order of hundreds of megawatts), with their steep, sub-second ramps are an entire new challenge for power grids. In the ERCOT context, (Kamwa, 2025) cites an engineer with the words: "The grid was never designed for subsecond, 300 MW load ramps. That's now a daily reality" (Kamwa, 2025).

And these ramps are not confined to AI model training. Although not to the same extent, they can also occur during AI inference. For efficiency reasons, AI inference processes increasing numbers of batched queries in parallel (Coroamă and Schien, 2026), which exhibit a similar synchronicity as model training.

#### 8.2.2. Increasing peak loads

Returning to the ERCOT case study, submissions to its LLI queue add to an impressive total amount. In November 2024, the queue of projects submitted and in different stages of approval (most being still under review) by 2028 was of around 54 GW, with another 9 GW announced but not submitted yet (Neel, 2025). Together, these submitted and announced projects yield 63 GW for Texas alone.

Not all these projected large loads are due to DCs, some being industrial or hydrogen (via electrolysis) projects. Nonetheless, more than half of the total projected additional power would come from DCs. Taken together, they would almost double Texas' peak power load from 85 GW in 2023 (Sahni *et al.*, 2025) to about 150 GW by 2030 (Neel, 2025).

Overall, over the coming decade, peak load in Texas is projected to rise by 15% in summer and 18% in winter, largely due to LFL growth (Kamwa, 2025).

#### 8.2.3. Resulting provisioning challenges

The rapid growth of data centres, and especially their large loads and power density as discussed in Sections 8.1.1 and 8.1.2, has exposed several constraints and bottlenecks in the energy production and transmission infrastructures. In DC-intensive regions, there are increasing power deficits: The US, for example, is too slow in devising new electricity generation capacity (Anthropic, 2025) and might lack up to 63 GW of power based on DCs currently under construction and planning ('Inside OpenAI's Stargate Megafactory with Sam Altman', 2025).

Due to power bottlenecks, some DC-intensive regions already imposed in the past temporary moratoria on the development of new data centres: Singapore was the first in early 2019 (The Business Times, 2020); as a small island nation with limited land and a tightly balanced grid relying heavily on imported natural gas, the government imposed it when DC consumption reached 7% of the national demand (Savage, 2026). Amsterdam followed later in 2019, when due to grid congestions, the local utility could no longer guarantee new connections to households or schools (Goh, 2025). And the Irish national grid operator EirGrid imposed a de-facto moratorium in 2021 (Judge, 2022), when DC consumption had already reached 18% of the Irish national electricity consumption (CSO, 2023).

After a few years, all these moratoria have been lifted, but the corresponding authorities have become much more selective in giving out new construction permits (Laforga, 2022; Goh, 2025). Arguably the most restrictive conditions are in Ireland: Any new DC must encompass its own on-site generators or batteries able to cover its entire power needs. Additionally, operators must use these part of this generation capacity not for their own needs but to provide back power to the national grid when required. Finally, at least 80% of each DC's yearly electricity demand must come from newly developed renewable energy projects (Bloomberg, 2025).

#### 8.2.4. Further issues following from the provisioning challenges: Congestion, losses, prices

From the provisioning challenges above, three further detrimental effects may follow directly (Mamkhezri, Sun and Yang, 2025):

- Increasing system electricity price,
- grid congestion, and
- increasing marginal transmission losses.

Power generation sources are dispatched in merit order, i.e., in ascending order of production costs. A higher overall load can thus lead to the dispatch of more expensive production and thus to an increased average system-wide electricity price.

Congestion appears when some transmission lines are at capacity due to high demand; as a consequence, either longer, less efficient routes need to be taken or different generators need to be dispatched that can transmit over non-congested lines (Deb and Halbe, 2025); both situations increase the local price of electricity (as opposed to the system-wide base priced due to merit order dispatch discussed above).

Finally, the marginal transmission losses along power lines increase when they approach their capacity. When marginal losses increase, so do (albeit to a lesser extent) the average losses, and thus the average cost as well.

As can be seen in the paragraphs above, all these effects can ultimately lead to increased power prices – either system-wide or in the vicinity of the congested grid. Such effects can thus be particularly relevant in regions of high DC concentration, such as the "data centre alley" in Northern Virginia (US), the new concentrations in Texas, or the traditional European DC locations Frankfurt, London, Amsterdam, Paris, and Dublin (FLAP-D).

For the data centre alley, econometric difference-in-differences approach estimated a 7.30% power price increase due to DCs. For Texas, an empiric analysis found a 19.70% increase (Deb and Halbe, 2025). In both cases, grid congestion was the main culprit.

Acknowledging this effect, and their responsibility, some hyperscale operators agreed to bear the responsibility and cover the price increases for the entire local community. Anthropic, for example, recently wrote in a press release: “Data centers can raise consumer electricity prices in two main ways. First, connecting data centers to the grid often requires costly new or upgraded infrastructure like transmission lines or substations. Second, new demand tightens the market, pushing up prices. We’re committing to address both” (Anthropic, 2026), to then pledge to entirely come up for grid upgrades and where this has not happened yet to estimate and cover the price difference.

## 8.3 Mitigation measures

### 8.3.1. New algorithms and battery storage for grid stability

One option to mitigate the abrupt power ramps that can lead to grid stability issues, is to target their source: the synchronous compute load variability which triggers the synchronous power ramps for a large number of processors. First efforts to this end exist already (one software package aptly called “Pytorch_no_powerplant_blowup”), but more comprehensive solutions that are more widely adopted would be required (Morrison *et al.*, 2025).

For the moment at least, the more promising path seems to be not to act against the cause, but against the propagation of its effects to the grid. In the context of power provisioning, Chapter 3 presented a novel, more efficient architecture based on SSTs and higher voltage (e.g., 880 V), direct current distribution. This architecture, supported by various sources (Huntington and Tu, 2025; Srivastava and Petty, 2025), also favours the (partial or total) replacement of UPSs with battery energy storage systems (BESSs) and supercapacitors places much closer to the (AI) servers.

This architectural feature does not stem from a preoccupation with the power grid. It is, however, in the operators’ best interest to absorb these ramps as quickly and efficiently, protecting the own on-site generation (see below) from the fluctuations. Ideally, too many min-discharge cycle of batteries (whether UPS or BESS) should also be avoided; hence the occasional inclusion of supercapacitors for the sub-second fluctuations. As a side effect, this architecture will also be able to protect the grid.

### 8.3.2. Workload shifting to shave peak loads

The computing loads that need to be executed on a server or a cluster of servers all undergo a scheduling process. In the computing domain known as “distributed systems”, workload scheduling is a decade-old concept. Traditional schedulers prioritise goals such as maximising resource utilisation to reduce idle capacities, thereby enhancing efficiency and cost-effectiveness (Gao, Liu and Kaushik, 2021) but also load balancing among servers, which is an important feature for system stability (Wu, Lin and Peng, 2017).

With growing energy consumption of data centres and concern for their growing GHG emissions, a new scheduling paradigm with a different aim emerged: to minimise the GHG emissions of DC workload execution. This paradigm was coined “carbon-aware computing” in a seminal paper by Google (Radovanović *et al.*, 2023). Carbon-aware computing uses two options for workload shifting: temporal and geographic.

In temporal shifting, non-critical workloads can be put on hold until a later moment. For carbon-aware computing, this is typically performed when less carbon-intensive electricity is expected at a later moment, e.g., due to the expected change in renewable generation. Spatially, workloads can be shifted to a data centre in a different geography, where less carbon-intensive electricity is available at the moment. This can happen either among data centres of the same hyperscale DC operator or even across different platforms and operators via tools dedicated to the scheduling of containerised workloads across distributed cloud and edge environments. A well-known and widely deployed such platform is Kubernetes (*Kubernetes Documentation*, 2024).

While highly relevant from a sustainability perspective, carbon-aware scheduling is however not immediately relevant for grid stability, including the mitigation of peak demands addressed here. In some contexts, carbon-aware computing might even induce the opposite effects and worsen peak demands;

for example, when low-carbon electricity is available somewhere and many workloads would be shifted there for carbon minimisation.

The principles, analyses, and tools developed by carbon-aware computing, however, can be deployed with different aims as well. Both temporal and spatial shifting, for example, can be useful for peak shaving, either by postponing workloads until after the peak or by moving them to another geography, ideally another continent in an entirely different time zone and thus complementary peaks of the respective power grids.

#### 8.3.3. On-site generation to mitigate both power provisioning challenges and peak loads

The power provisioning challenges presented in Section 8.2.3 represent substantial obstacles for DC operators – and in particular AI developers – to further develop their operations. They are thus increasingly inclined to fundamentally shift their energy procurement strategies, circumvent the power grid entirely and build their own on-site power plants.

As established generation and transmission infrastructure increasingly fails to satisfy the electricity needs of DCs, operators of data centres – hyperscalers in particular – are exploring on-site generation as an alternative to the lengthy LLI queues, becoming themselves important backers of energy technologies.

While relatively shy until recently, this trend has grown very strong over the year 2025, with increasing power demand for AI data centres and increasing challenges in securing it from the grid. (Thomas, 2026) found for the US 46 new DC project totalling 56 GW that plan to build their own "behind the meter" power plants to avoid the grid at least initially – and perhaps connect later, conditions permitting. For now, however, as the Washington Post noted, they are building their own “shadow power grid” (Halper, 2026).

Most of these projects are a very recent development: approximately 50 GW of the 56 GW (about 90%) have been announced in 2025. As the author puts it: “A year ago, behind-the-meter data center power was a curiosity, embodied by xAI's controversial decision to truck mobile generators into Memphis. Now it's an increasingly common development strategy” (Thomas, 2026).

Several energy sources are being envisioned. From the project whose source could be identified, and that cover 30.87 GW of the announced 56 GW, however, the majority rely on gas turbines. The individual identified energy sources are:

- 22.8 GW (74%) gas,
- 6.4 GW (20%) nuclear,
- 1.3 GW (4%) fuel cells, and
- 355 MW (1%) battery storage.

Examples in use include xAI’s Colossus and some of OpenAI’s stargate DCs, for example, which are both powered by large gas turbines of dozens MW of power each: 29 General Electric turbines of 34 MW each, for example, provide roughly 1 GW of power (SemiAnalysis, 2025).

From the perspective of DC operators, gas turbines have two large advantages:

- they are quickly deployable, and
- able to run almost entirely autonomously with very little infrastructure required: they do not require power grid or water connectivity nor complex means of cooling. Instead, they run locally, are air-cool and reject their heat directly into the atmosphere.

Their lacking water consumption represents an advantage also from a sustainability perspective. Running a DC on gas turbine-generated power induces zero upstream water consumption. Gas turbines, however, also induce various detrimental consequences from a sustainability perspective, as discussed in Section 8.4.1 below.

## 8.4 Detrimental side-effects of mitigation

Unfortunately, the mitigation measures addressed in Section 8.3 also bring their own detrimental side-effects and consequences, which need to be weighted against the advantages they provide.

### 8.4.1. Drawbacks of on-site generation

The most substantial drawbacks come undoubtedly from the on-site generation, especially as it is overwhelmingly supported by gas, as shown in Section 8.3.3 above. As the study cited before notes: "This buildout is overwhelmingly powered by natural gas. Nearly every project we reviewed mentions renewables, hydrogen, or nuclear in its public announcements. But the equipment actually being installed in 2025 and 2026 is almost entirely gas-fired. Renewable capacity, where committed, is scheduled for 2028 or later. Nuclear is a decade away" (Thomas, 2026). This brings several drawbacks:

- Electricity generated via fossil fuels has the obvious disadvantage of GHG emissions (EDF, 2024). Even more so, as the employed gas turbines are not the most efficient ones either. Those would be combined-cycle gas turbines, but they have become themselves a bottleneck, as the entire production of large producers such as General Electric or Siemens Energy is already sold for 5-7 years in advance. They can thus hardly represent a solution for the lengthy grid interconnection queues.

  To mitigate this bottleneck, various solutions are employed. For its upcoming 2 GW data centre which will house 1 million GPUs, xAI is reported to have bought an entire power plant – presumably also based on gas turbines – overseas, and will be shipping it to the US (Yong, 2025). But xAI has also used mobile gas generators strapped to semitrucks. Others deploy refurbished turbines acquired from industrial operations. Even old jet engines from scrapped airplanes such as 747s are being repurposed for electricity production, as they are able to generate similar amounts of power – the largest of them generating 48 MW each (Robb, 2025).
- Another negative externality is the competition DCs pose to grid turbines. For a 2 GW data centre, for example, xAI has secured five huge 380 MW turbines. These turbines, however, are typically deployed by utility companies to provide baseload power to the power grid. If they increasingly disappear behind the fences of DCs, subject to a financial competition that the utilities cannot win, they might indirectly – but very concretely – affect grid security after all.
- Overprovisioning is another issue. Since gas turbines can only guarantee around 90% availability, DC operators need to overprovision. This, in turn, not only increases the production footprint, but leads to even more competition with other application domains.
- Finally, on-site generation produces vibrations, noise and possibly pollution (EDF, 2024). These, in turn, affect the local communities around the DC.

While gas turbines, to the extent they are available, represent a quick fix that does not rely on grid availability, they thus also have several important drawbacks.

### 8.4.2. Detrimental consequences of BESS in DCs

The battery storage discussed in Section 8.3.1 also has one important drawback: As DCs compete utility companies in gas turbine procurement, their massive foreseeable demand for BESS will also likely directly compete with two important economic sectors, both of which take part in the circular economy: electric vehicles and grid-level batteries. Grid-level batteries, however, are essential in the energy transition, in order to smoothen the volatility of renewable electricity sources.

## 8.5 Opportunities for the grid: Flexibility, innovation and development

Beyond their mitigating intent and their negative consequences, the mitigation measures from Section 8.3, however, also present further opportunities for the grid. These go beyond the implicit benefits of addressing the original challenges that had triggered them. Additionally, the LFLs themselves might also present an opportunity and not only challenges. They are presented in this section.

#### 8.5.1. Data centre flexibility services for the grid

Data centres can provide flexibility services to the grid via two mechanisms: The temporal and spatial workload shifting (Section 8.3.2) considered for peak load shifting of the DC load can also be deployed to shave loads not caused by the DC itself and thus offer *demand-side management* to the grid. And both the available on-site generation (Section 8.3.3) introduced as mitigation for the provisioning challenges and the BESSs (Section 8.3.1) necessary to absorb fluctuations can offer *peak-demand production*.

**Demand-side management through spatio-temporal shifting of DC workloads**

Data centres may be able to adapt their consumption to the availability of energy, in particular to help flatten peak demands, in a paradigm known as "demand-side management". Concepts and principles such as the joint spatio-temporal task migration introduced by (Yang *et al.*, 2023) can likely also be deployed for grid flexibility by substituting the optimisation criteria. Additionally, and although it is only a few years old, the domain of carbon-aware computing (Section 8.3.2) has already produced various analyses that would be relevant for offering grid flexibility as well. To better estimate the potential for grid stability, an analysis of the *shiftability* of workloads is required: which tasks can be moved across locations or rescheduled for a later moment and under which conditions?

For carbon-aware computing, (Wiesner *et al.*, 2021) have provided precisely such an analysis, which distinguishes between short-running (majority of jobs executed in DCs), long-running (e.g. ML training, scientific simulations, or big data analysis) and continuously running workloads (e.g. continuous services such as user-facing API or computationally intensive workloads such as blockchain mining, protein folding, etc.). Among these three types, long-running, non-time critical workloads have the highest potential for workload shifting.

Overall, various principles, analyses, and insights from carbon-aware computing could be directly leveraged for DC-based grid flexibility services. Other insights and concrete tools (such as existing carbon-aware schedulers) might be reused with fairly minimal adaptation efforts such as redefining the optimisation criteria. Novel analyses are required for the synergies and trade-offs between carbon awareness and grid flexibility via workload shifting.

**Peak-demand production via on-site generation and battery systems**

While curtailing their own demand can be helpful, arguably the larger potential for grid flexibility of DCs lies in their ability to quickly deliver large amounts of power – an ability developed for their own needs, but that could be leveraged for grid services as well. What might make them even more valuable for the grid is the fact that many DCs have on-site both the potential to deliver quick energy for primary grid control and slower but longer-lasting generation for either secondary and tertiary control or peak demands.

As discussed, all DCs have UPS systems – immediately available energy storage in case of grip interruptions, which can take over the entire DC energy load for up to 1-2 minutes until the backup diesel generators are working at capacity. In case of major grid issues, the DCs would of course require the power from their UPS systems firstly for themselves. However, UPSs are often over-provisioned, surveys showing that industry representatives estimate surplus capacities of their UPSs of between 10-50% of the total capacity; a reserve that would not be required for themselves in case of an emergency (Takci *et al.*, 2025).

This is especially the case during the development phase of a DC, whether hyperscale or colocation: Battery infrastructure is often designed from the outset for the foreseeable final demand, but for the period it takes to reach full capacity (often many years), there will be spare capacity available. A question that needs to be addressed, however, is whether the new architecture, in which UPSs are being increasingly replaced by BESSs deeper into the DC – and which do provide benefits for the large and steep power loads, as discussed in Section 8.3.1) – is as suited for this kind of primary control as UPSs are.

Section 8.3.3 presented the massive development of on-site generation that takes place today. Most of these projects are not in the desert, entirely detached from the grid, as e.g. the “GW Ranch” in Texas (envisioning up to 7.5 GW detached from the grid) seems to be (Pacifico Energy, 2025). By contrast, most developers are developing own solutions to bypass the long interconnection queues, envisioning on-site generation as a temporary crutch in the quest for power. Once grid connectivity does become available, these assets could be repurposed as backup generators or to offer flexibility services.

Both by shifting its own loads and by injecting power into the grid, a DC can help alleviate peak power demands. These two potentially beneficial mechanisms also explain better the parallel to electric vehicles from the beginning of the chapter, as the grid-related benefits of electric vehicles also lie in both demand-side management and vehicle-to-grid services.

### 8.5.2. New electricity generation and storage development

Although most of the on-site generation starts with gas turbines (see Section 8.3.3) with all their drawbacks (Section 8.4.1), there is simultaneously a push for carbon-free technologies, i.e., renewables and nuclear (both traditional and newer generation). Given the massive spending in search of energy sources that DCs undertake, they might thus spur a new era in electricity generation and help ramp up carbon-neutral technologies.

Before the AI demand explosion, many IT giants such as Google and Microsoft had pledged to become carbon-free by 2030. Although this has now become impossible, tech companies claim to only postpone these goals by a few years but maintaining them in principle. They have an interest in firm, round-the-clock, carbon-free sources. These can be battery-supported renewable sources. A large US-based colocation provider, for example, develops DC campuses based on various complementary renewables (simultaneously wind, solar, and geothermal for the baseload) with battery backup (Switch, 2025).

For the last two years, however, it has been most notably massive investments into nuclear power. Nuclear power has the advantage of being an operationally carbon-free (and from a lifecycle perspective very low carbon) technology with constant, predictable, high-power output capable of matching the high-density round-the-clock baseload demand generated by AI in particular.

Since the beginning of 2024, virtually all large US-based tech companies have announced investments into nuclear technology. Meta's request for proposals targets the acquisition of up to 4 GW of new nuclear capacity (Meta, 2024), while Microsoft's power purchase agreement (PPA) is supporting the economic feasibility of restarting an existing large reactor, Three Mile Island Unit 1 (Martucci, 2024).

This tendency is likely to also bring a boost – and perhaps technological maturity and economic feasibility – to Small Modular Reactors (SMRs), which are championed for their faster deployment and smaller footprint. Over the past two years, Google (Terrell, 2024), Amazon (Amazon Staff, 2024), and Oracle (Butler, 2024) have all announced significant initiatives to invest in or utilise power derived from SMRs. By contrast to large, traditional nuclear reactors, SMRs have the added benefit that they can be installed locally, independently of the grid – thus offering the advantages of gas turbines without the GHG, noise and pollution drawbacks (but, of course, with the open issue of nuclear waste).

Similarly, the sustained demand from data centres might also help deliver new impulses towards the development of new battery storage systems. If this were to happen, it could help compensate for the competition they represent to electric vehicles and grid-level batteries discussed in Section 8.4.2.

### 8.5.3. Further grid developments: Support in renewable integration and in power transmission

The drawbacks of LFLs have been discussed in detail in Sections 8.2.2 (increasing peak loads) and 8.2.3 (provisioning challenges). Their existence, however, could also yield advantages. A very concrete one is the mitigation of the (undesirable) renewable curtailment.

Renewable curtailment is the intentional but involuntary reduction of output from wind, solar, or other renewable generation sources below their maximum potential, typically ordered by grid operators. It occurs when electricity supply exceeds demand or transmission capacity, which could lead to grid congestion or stability issues, thus making curtailment necessary (Laimon, 2025).

The US Energy Information Administration (EIA), for example, estimates that by 2035, wind curtailments could rise to 13% (from 5% in 2022), while solar curtailments reach 19% (EIA, 2023). In Europe and China, with their substantial renewable investments, the numbers could be even higher. Data centre LFLs could be an option to absorb much of this surplus energy: "Strategically placed and well-managed [[LFLs]] could play a crucial role in reducing these curtailments by absorbing surplus energy, thereby enhancing grid stability and optimizing the use of renewable resources" (Deb and Halbe, 2025).

Finally, LFLs could also be a catalyst for the further development of the high-voltage transmission network. Although this potential is much more uncertain than renewable curtailment, if it were to happen, it would of course benefit economy and society as a whole through the added capacity and stability.

# 9 Standards, certifications, and operational governance

Various standards, certification schemes, and voluntary agreements relevant to DC energy efficiency and sustainability exist. They reflect the integrated nature of data centres (covering physical infrastructure, IT equipment, and software) and varied sizes and ownership structures, such as enterprise versus colocation models.

While they can be grouped according to several criteria, this chapter distinguishes among two main categories: standards and recommendations (international and European), and voluntary certification and labelling schemes.

## 9.1 Global and European DC sustainability standards and recommendations

A core international DC sustainability standard is the ISO/IEC 30134 suite (ISO/IEC, 2016), which defines essential key performance indicators (KPIs) such as the PUE (Part 2 of the standard, see Chapter 2), REF and ERF (Parts 3 and 6 of the standard, respectively; see Chapter 5 for both), and WUE (Part 9, see Chapter 3). Thereby, the standard defines the structure, terminology, and common requirements for data centre KPIs (Harryvan, 2023), and establishes a common language for global reporting and comparisons (Brocklehurst, 2024).

In Europe, the EN 50600 series is equivalent to ISO/IEC 30134. Its technical specification (TS) CLC/TS 50600-5-1 establishes a model for energy management and environmental compatibility of DCs (Brocklehurst, 2024) that aligns with best practices such as the EU Code of Conduct (Acton, Booth and Paci, 2025).

ETSI's ES 205 200-2-1 (ETSI, 2014) standard focuses on KPIs for energy efficiency in data centres. Unlike most other standards, it includes one KPI for "task efficiency" ($KPI_{TE}$), which is a measure for compute efficiency along the lines similar to the one this report has also been arguing for in Section 2.3.2.

A special mention belongs to the European DC reporting scheme (European Commission, 2024a) and the German energy efficiency law (Bundestag, 2023). Both were already introduced in Section 2.3: The European reporting scheme requires that starting in 2025, all EU-based data centres with an installed power demand of at least 500 kW report yearly a plethora of indicators and KPIs, which include both absolute parameters such as total energy consumption, water consumption, and total computing performance as well as four relative KPIs. The German energy efficiency law sets three minimal requirements for DCs, which are differentiated between newly build and already existing DCs.

These as well as further important standards are listed in Table 3, which also lists their most important features. Such standards govern the entire measurement ecosystem, specifying boundary conditions, metrics and equations to be used, measurement procedures, and reporting requirements to ensure clarity and validity across different DC scales and operational states.

Table 3 Important standards and recommendations related to DC consumption, efficiency, and related metrics, together with their respective jurisdictions or geographic scopes and key topics covered.

| Standard / framework | Geographic scope / jurisdiction | Key topics |
| --- | --- | --- |
| ISO/IEC 30134 standard | Global | Specifies KPIs for DCs. Defines metrics such as PUE, REF and RUE, WUE, "IT Equipment Energy Efficiency for servers" (ITEEsv), "IT Equipment Utilization for servers" (ITEUsv) (Part 5), Cooling Efficiency Ratio (CER), and Carbon Usage Effectiveness (CUE). Establishes basis for objective energy efficiency policy and reporting globally. |
| CLC/TS 50600-5-1 technical specification | Europe | Establishes a maturity model for energy management and environmental sustainability, aligning with recommended best practices such as the EU Code of Conduct (CoC). |
| ETSI ES 205 200-2-1 technical specification | Europe | Standard that focuses specifically on KPIs for energy management in DCs. Specifies four key KPIs: energy consumption (EC), energy reuse (REUSE), renewable energy (REN), and task effectiveness (TE). Predated ISO/IEC 30134 and informed many of the latter's energy-related KPIs, particularly REN (which matches REF in ISO/IEC) and REUSE (matching ERF).<br>The energy consumption focus of ETSI was changed for a relative metric (i.e., PUE) in ISO/IEC. Task Effectiveness, which tried to measure the amount of useful work per unit of energy, was overly complex and specific and thus not reflected in ISO/IEC (albeit ITEEsv has similar goals). |
| ITU-T L.1303 recommendation | Global | Outlines the logical architecture and functional requirements for an energy-saving management system within a DC. Establishes how such a system should measure, collect, and store energy consumption and operating information on ICT equipment, such as CPU, memory, storage, or network utilisation.<br>Specifies how the system should execute control strategies to optimise energy consumption across the entire facility. |
| ANSI/TIA-942-B standard | US | Defines minimum requirements for the telecommunications infrastructure inside DCs. Covers not only IT-specific telecommunications, but also those for the facility infrastructure (e.g., power and cooling). |
| ASHRAE 90.4 standard | US | Does not discuss at all ICT equipment, but focuses exclusively on the facility infrastructure, i.e., cooling and power distribution systems. Sharpens the general building standard ASHRAE 90.1 for DCs specifically. Applies to DCs with a power density > 20 W / ft$^2$ (about 200 W / m$^2$) and IT equipment loads > 10 kW.<br>Contains specific requirements for mechanical and electrical systems installed in new DCs or in DC additions/alterations that require new mechanical or electrical systems.<br>Does not prescribe specific equipment, but establishes maximum allowable values for the cooling overhead (chillers, cooling towers, pumps, CRAH/CRAC units, and fans), and maximum losses for electrical components (UPS, transformers, cables). |
| EC 2024 / 1364 | Europe | European reporting scheme, requiring that starting in 2025, all EU-based DCs report various indicators and KPIs.<br>The indicators include i) general parameters such as the DCs type (enterprise, colocation, co-hosting), total floor area, and total computer room floor area, ii) absolute parameters such as absolute energy, ICT capacity and data traffic metrics (total DC energy consumption, total computing performance and storage capacity, maximum traffic bandwidth and actual traffic), various cooling parameters (intake air temperature, type of refrigerant, and the amount of "cooling degree days"), and several other parameters related to the use of renewable energy, amount of heat reuse, water consumption, and grid services.<br>The four "sustainability indicators" are the PUE, WUE, and ERF (all discussed in this report) as well as the "renewable energy factor" (ERF), which is computed as the amount of renewable energy among the DCs entire energy consumption. |
| Energieeffizienzgesetz (energy efficiency law) | Germany | Law not specific to DCs, but highlighting them as a particularly relevant technology, one of the very few worthy of a dedicated section of the law comprising 5 articles of the law's total number of 21.<br>Establishes maximum PUE value as well as minimum values for heat reuse (ERF) and share of renewable energy. Required values differ |
| | | between newly developed and already existing DCs, but generally, all are becoming stricter over the years. |
| EC 2019 / 424 | Europe | One of the few binding EU instruments that brings circularity into data centre hardware via ecodesign. Requires manufacturers to provide information regarding the material efficiency of servers and data storage products, ensuring that the joining, fastening, or sealing techniques used on their product do not prevent the repair or reuse of various components like, motherboard, CPUs, chassis, data storage devices etc (European Commission, 2019).<br>A formal review process of 2019/424 within the context of the newer ESPR (Ecodesign for Sustainable Products Regulation) framework (EC 2024/1781) is intended to enable stronger circular requirements. |
| EC 2024 / 1799 (“right to repair”) | Europe | This is EU’s “right to repair” (RTR) law, which entered into force on 30 July 2024, but is applicable from 31 July 2026, as member states must first transpose it into national law (European Council, 2025; Andrews and Kerwin, 2026). It is designed to make repair the normal option when a product breaks, pushing manufacturers to make repair practical – by providing clear repair information and ensuring spare parts are available at reasonable prices.<br>While primarily targeted at consumer devices, it has relevance for DCs as well, due to these mandates for repairability and spare parts availability. This is likely to force manufacturers to support longer lifecycles, which impacts procurement, maintenance, and asset disposal strategies for DCs. |

## 9.2 Voluntary certification and labelling schemes

Legislation is often a lengthy and difficult process, given the agendas and lobbies of various stakeholders. While not as challenging, standardisation is also relatively slow, as it also needs to take into consideration the needs of a variety of stakeholders. When it sets minimum requirements, it additionally often results in a lowest common denominator, which is far from the ambitions of industry leaders, who desire their extra mile to be acknowledged otherwise.

Voluntary certification and labelling schemes are instruments that can be employed to encourage and validate DC energy efficiency and sustainability, beyond the mere mandatory regulatory compliance. These schemes are broadly categorised by their function: Endorsement labels provide a binary pass/fail indication, meaning a DC either meets all necessary requirements or it does not, without declaring specific performance values (e.g., US Energy Star or the EU Ecolabel).

By contrast, comparative labels display the DC's energy performance relative to a specified scale or range, often using categorical classes (such as star ratings in Australia/Japan or alphabetical ratings such as G to A in the EU). In the context of data centres, most of these labels rely on the de-facto standard infrastructure metric, the PUE, as a core criterion for assessment. As discussed in Chapter 2, though, the PUE has various limitations, which relativises the usefulness of these labels as well.

Because voluntary labels operate outside of mandatory regulatory constraints, they typically have a broader scope and can address more complex issues than mandatory labels, such as assessing IT system efficiency through a combination of design features, equipment efficiency requirements, and adherence to specific good practices. However, they are limited in market impact, as information is only available for participating DCs, usually those with better performance, meaning that they do not provide transparency across the entire market (Brocklehurst, 2024).

Table 4 presents a selection of important worldwide DC-related labels and certifications, highlighting their key characteristics. As discussed in the table, some of the certifications can influence public procurement decisions, which partly questions their theoretically voluntary character.

Table 4: Key characteristics of selected data centre-related labels and certification schemes.

| Scheme Name | Geography<br>Year established<br>(updated) | Key features |
|---|---|---|
| NABERS | Australia<br>2014 | Categorical label (1 to 6 stars) used to measure and verify the environmental impact of buildings, including DCs. Requires actual historical data, not models. Sometimes mandated in public sector procurement policies, requiring a rating of at least 4.5 stars. |
| Austrian Ecolabel<br>(“Österreichisches Umweltzeichen”) | Austria<br>2023 | Endorsement label that requires compliance with minimum PUE and cooling efficiency ratio (CER) requirements, differentiated by the DC commissioning date. CER relates the work done by the cooling system to the heat rejected.<br>Requires disclosure of metrics such as PUE, CER, ERF, and WUE, and is intended for use in government procurement. |
| Blue Angel DCs<br>(“Blauer Engel Rechenzentren” DE-UZ 228) | Germany<br>2012<br>(updated 2023) | Endorsement label, mandating specific minimum PUE and CER requirements based on the DC's age and requires 100% of electricity consumption to be renewable. Blue Angel-certification is often mandatory for contracting by the German government. |
| BEAM Plus New/Existing Data Centres | Hong Kong<br>2021 | Green building rating system for data centres. Awards points across categories such as energy use, water use, CER, and health and wellbeing. Integrates infrastructure metrics (e.g., PUE or total air flow efficiency) with best practices for IT equipment deployment. |
| Certified Energy Efficient Data Center Award (CEEDA) | Global<br>2010 | Independently evaluated global certification program. Assesses DCs based on adherence to best practices and design features (e.g., hot/cold aisles, liquid cooled IT equipment) rather than relying solely on PUE measurements. |
| Green Data Center Certification | Korea<br>2021 | Categorical label specific to DCs, launched in 2012. It is based on PUE and green operational practices.<br>Certification can help qualify for Korean government green incentives. |
| Green Building Index (GBI) for Data Centers | Malaysia<br>2012 | Building sustainable rating system that includes specific conditions for DCs. Integrates PUE as the only DC-specific metric into a broader green building evaluation. |
| BCA-IDA Green Mark for Data Centres | Singapore<br>2013<br>(revised 2020) | Building label extensively customised for DCs. Uses a points system across various criteria, with stringent minimum PUE requirements that vary by certification level.<br>As Singapore has a strict cap on the total energy available for DCs, the highest platinum rating (i.e., PUE < 1.25) is a mandatory (but not sufficient) requirement for building permits for new DCs. |
| Swiss Datacenter Efficiency Association (SDEA) Label | Switzerland<br>2020 | Categorical label focused mainly on the PUE, awarding “bronze”, “silver”, “gold”, and (as of recent) “platinum” labels. The carbon intensity of the electricity used in DCs can earn a “plus” to the label. |
| Energy Star Score for DCs | US<br>2010 | Endorsement label that grants certification to buildings with an Energy Star score of 75 or higher, determined by correlating their PUE against a reference PUE derived from statistical data. Offers a PUE-based score relative to the set of national buildings. |

# 10 Conclusions: Overall sustainability assessment and sustainability trade-offs

## 10.1 Current metrics for energy and water

### 10.1.1. The PUE and its limitations

Chapters 2 and 3 revealed several limitations of the widely used PUE metric:

- As relative measure, it does not tell on the absolute consumption.
- It does not describe the efficiency of computing, which is the main source of energy consumption in a DC, but the efficiency of the supporting infrastructure.
- And it does not do so correctly, but missing parts both of the power transformation overhead (the PSU losses) and of the cooling overhead (the consumption of server fans).
- The computation overhead is not only covered incompletely, but also too coarsely. The PUE mixes the two main sources of infrastructure overhead: DC cooling and power supply chain.

As discussed in Section 2.3, due to the heterogeneity of both compute workloads and different types of hardware (each with its own application domains), it is fundamentally challenging to design one generic metric for compute efficiency as put forward by Equation 7 of this report:

$$DCCE = \frac{Compute_{DC}}{E_{DC}}$$

Based on these insights, Sections 11.1.1 and 11.1.2 propose new metrics for computational efficiency and the two major infrastructure overheads, respectively.

### 10.1.2. Water consumption and two related trade-offs

Section 4.3 presented the definition of the "water usage effectiveness" and introduced the two system boundaries: on-site (i.e., in the DC alone) for $WUE_{site}$ and comprehensive (including both on-site and upstream consumption during electricity production) for $WUE_{source}$. It also established the relation between the two, while taking into account the "water consumption factor", which represents the relatively amount of water consumed during electricity production (i.e., litres per kWh). The resulting relation is:

$$WUE_{source} = WUE_{site} + PUE * WCF$$

which, as derived in Section 4.4, reaches its minimum at the point of the PUE-WUE trade-off curve where the slope of the curve (i.e., marginal change in $PUE$ for a marginal change in $WUE_{site}$) is equal to the negative inverse of the water intensity of electricity in the local grid:

$$\frac{d(PUE)}{d(WUE_{site})} = -\frac{1}{WCF}$$

***This equation elegantly encompasses two trade-offs: Between local and upstream water consumption, and between total water and energy consumption.***

The equation relates the sensitivity of $PUE$ to changes in $WUE_{site}$ to the inverse of the water intensity of electricity production (which the WCF represents). In other words, if the electricity source embodies high amounts of water, it is worth spending more water on-site (e.g., by switching on adiabatic cooling support) if this saves some electricity (and relatively much water) upstream.

The minimum overall water consumption is thus only reached when marginal increases to the local water consumption no longer trigger energy savings. As a consequence, there is almost no trade-off between water and energy consumption: The water optimum is reached when for a high on-site and low upstream water consumption. But this also reduces the PUE and thus the energy consumption.

By contrast, dry electricity induces a stronger trade-off between overall water and energy consumption: As the electricity embodies little water, $WCF$ is small and its inverse large. The optimal overall water consumption is thus quickly reached, for a relatively large $PUE$ that would still have plenty of room for improvement. From a water perspective, however, it is worth keeping on-site consumption low, as spending more of it on e.g. adiabatic support would decrease upstream electricity, but barely the related upstream water. Optimal water consumption means a relatively high PUE; hence the trade-off between the two.

## 10.2 Circular economy aspects

### 10.2.1. Circularity of energy through waste heat recovery

There are two metrics deployed to assess the recovery of DC waste heat: The “energy reuse factor” (ERF) simply denotes the percentage of heat reused from all the energy that went into the DC, whether for computation or the infrastructure overhead:

$$ERF = \frac{Q_{Re}}{E_{DC}}$$

The “energy reuse effectiveness” (ERE) is designed to “correct” the PUE by subtracting the amount of heat reused from the total energy, before dividing the result by the IT energy:

$$ERE = \frac{E_{DC} - Q_{Re}}{E_{IT}} = \frac{E_{IT} + E_{non-IT} - Q_{Re}}{E_{IT}}$$

ERE can take values between close to 0 (if almost all thermodynamically possible energy is recovered and reused) and PUE (if no energy is reused, $Q_{Re}$ is zero, and $ERE = PUE$). As soon as the reused heat is larger than the energy used by the DC’s overhead, the ERE becomes smaller than 1; hence the “correction”.

As shown in Section 5.3, the relation between the two indicators is easy to derive:

$$ERE = (1 - ERF) * PUE$$

Section 5.4.2 discussed the trade-off between minimising cooling energy and maximising its reusage potential. This is mainly related to the siting of the DC: Generalising, a placement far North guarantees good PUE values with zero reusage potential. Building DCs near urban centres with district heating, allows for heat reusage. As cities tend to be in friendlier climates, it will raise the PUE, however.

### 10.2.2. Material circularity

While data centre buildings themselves also have material impact that would benefit from circularity, this topic involves relatively few material types and is similar to general building sustainability and circularity. The production footprint of microelectronics, however, is insufficiently understood. Section 6.1 split the production of microelectronics in three main phases: mining of minerals, their refining, and the manufacturing of the integrated circuits. It has also highlighted a particularity of the field: Due to the extreme miniaturisation, very high levels of purity are required for some materials, most notably the silicon used later for substrates. All these three production phases yield various sorts of environmental impacts, most of which are poorly understood.

As the materiality of production is uncertain, so are the exact potentials for resource circularity. Based on general principles and frameworks – in particular the 3R-framework (i.e., reduce, reuse, recycle), circularity principles for DCs can be drawn. They include the design for circularity (standardised parts, modularity), material choices (recycled or low impact materials, ideally from certified origins), and operational practices to ensure long lifespans.

Some of these principles are already well-established in the DC supplier industry. As Section 7.2.1 for example discussed, major server producers have standardised several basic server architectures. This enabled a modular approach to server design, allowing for greater flexibility, scalability, and reusability of components across different suppliers.

10.2.3. Possible trade-offs between resource consumption and energy efficiency

Modern AI models – and in particular large language models (LLMs) are too large to stay entirely withing the cache of the logical processors. During the decode phase of inference, model parameters are thus constantly being pulled from memory (Coroamă and Schien, 2026). This not only leads to compute lags (the “memory bottleneck”) but also represents the largest source of energy consumption during inference.

As discussed in Section 7.2.2, a current proposal to cope with these drawbacks is to create a dedicated integrated circuit that physically represents the AI model, having all the parameters hard-wired in the chip’s structure. Deploying this custom-made IC for inference would remove the need to pull parameters from memory, and indeed entirely circumvent the need for memory during inference. It would both speed up computation and reduce its energy consumption by over an order of magnitude.

This comes to little surprise: For their specific task, such application-specific integrated circuits (ASICs) are much more energy-efficient than general-purpose computation ICs (Coroamă *et al.*, 2025), as also illustrated in Figure 11. ASICs designed for one task only – performing mathematical hashes – have, for example, been deployed for blockchain and cryptocurrency mining proof-of-work for over a decade (Coroamă, 2021).

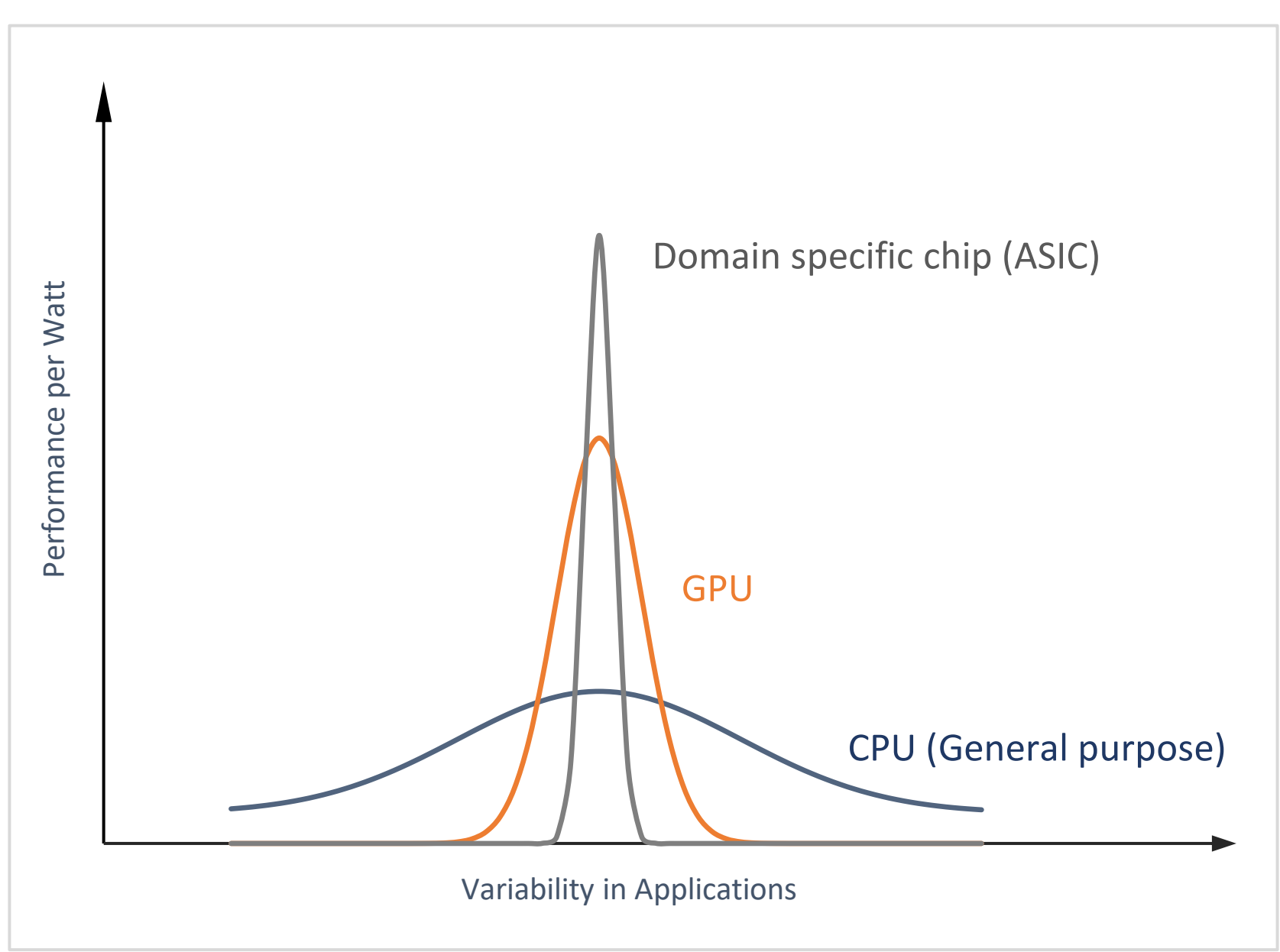

Figure 11: Performance of workload-specific ICs vs. general-purpose CPUs. Qualitative illustration from (Coroamă *et al.*, 2025); reprinted with permission.

The price for this efficiency, however, is a material one: Once the model is deprecated and no longer useful, the custom ASICs need to be discarded – and perhaps new ones etched for the new model version. Given the short lifespans of AI models (currently a few months), this price would be quite substantial, were this paradigm to be deployed at the scale necessary to serve current levels of AI queries.

## 10.3 Data centre growth and the grid

The sustained current growth of data centres is mainly due to accelerated computing and AI. These loads bring three main characteristics, as Section 8.1 discussed: i) a high degree of power density (previously almost unseen, apart from few industrial sites), ii) the unusual hourly flexibility of some of these large loads (thus named “large flexible loads”), and iii) the steep, sub-second power load ramps due to the synchronicity of many accelerated compute loads, in particular AI training.

These characteristics bring lead to novel challenges for the power grid. The main challenges are also threefold: more difficult power provisioning (due to both the new large loads and also the extreme power density, which makes not only the generation of the power challenging, but also its transport), the exacerbation of peaks loads, and grid stability dangers due to the novelty of the steep ramps. The power provisioning challenges lead to further challenging consequences such as increased grid congestion, system-wide marginal-price increases, and increased marginal transmission losses, all of which can in turn increase local energy prices, affecting the local communities in the vicinity of DCs.

These challenges are being addressed via several measures. The most important are: new paradigms of DC workload shifting (for shaving the peak loads), the mitigation of grid stability issues via both algorithms designed to break the synchronicity of AI workloads and battery systems that can absorb the ramps at their source in the servers and not let them propagate to the power grid, and on-site power generation, which alleviates both peaks loads and power provisioning issues.

Besides their primary reasons, these mitigation measures can also yield further benefits, both direct and indirect. Mature workload shifting paradigms can not only address the DCs peak loads, but also offer demand-side management to the grid itself. Similarly, large DC batteries can not only absorb potential grid stability problems of the DC itself, but when not entirely needed by the DC, they can provide additional power to grid to shave off DC-unrelated peak loads. Taken together, these two thus offer flexibility to the grid. More indirectly but not less importantly, both BESS and on-site generation can bring about new innovation for the power grid through energy generation and storage advancements. Finally, similar direct and indirect benefits can come directly from the LFLs themselves, as they can absorb large surpluses of renewable generation and thus lower the amount of renewable curtailments, but also support the development of dearly needed new high-voltage power transmission lines.

Unfortunately, however, the mitigation measures also bring about further negative consequences of their own. Both battery systems and on-site generation are large consumers of materials and because they are also often overprovisioned, this leads to even more (and unnecessary) resource depletion. At the same time, they are also competing with other industries and sectors requiring the same technologies. The energy sector is particularly affected: Data centre BESSs are in direct competition with grid-level batteries required for the storage of renewable energy. And on-site generation in DC campuses is often performed via large gas turbines. As described in Section 8.4.1, these are precisely the turbines typically employed for the power grid baseload. And given the hundreds of billions that large hyperscale operators are currently pouring every year in their DCs, this competition is one that the energy sector cannot win. More direct drawbacks of on-site gas generation are the produced GHGs as well as noise and pollution, the latter of which directly affect local communities.

# 11 Open issues and future research

## 11.1 New metrics for computational efficiency and individual overheads

### 11.1.1. A metric for computational efficiency

Although there is no generally accepted metric for the efficiency of computing, several past efforts have come up with intriguing proposals, and future work should start from these:

- GreenGrid's "data centre energy productivity" (DCeP) (The Green Grid, 2014),
- JouleX's performance-per-Watt (Davidson, 2011),
- EDNA's "data centre functional efficiency" (DCFE) (Maagøe, 2022),
- GreenGrid's more recent "IT work capacity (The Green Grid, 2025).

And nowadays might be the right moment for such an effort, given several converging factors: the required rehaul of existing metrics due to the fast-paced widespread of accelerated computing, the need to better understand and describe the energy consumption of AI loads, and reporting and rating schemes for DCs which require precisely such metrics.

### 11.1.2. Separate metrics for the two major overheads

Section 3.3.2 introduced the proposal of a *comprehensive* and *dedicated* metric for the entire DC power supply and transformation chain, "total power provisioning efficiency" (TPPE):

$$TPPE = \prod_{i \in \{T\&D\}} \eta_i$$

The TPPE could be instantiated for any power provisioning architecture and thus used both within the same architecture and between different ones to assess and compare efficiencies. It would have the advantage of focusing on energy provisioning, without mixing it with the effects of other infrastructural overheads, as the PUE does. As discussed in Section 3.3.3, one open question that requires additional research is whether to include the final, on-motherboard transformation $\eta_{VRM}$ (from 12 V to 1-2 V today), or whether these losses semantically belong indeed to the chips.

In a similar manner, a metric would be required that only assesses the efficiency of the cooling (or perhaps a little more generally, the HVAC) overhead. A central question here would be how to assess the consumption of server fans for air-cooled servers.

Currently, new architectures emerge for both the cooling and the power provisioning infrastructures: liquid cooling for the former, direct current provisioning, higher voltages, and a migration of battery storage from the DC entrance towards server rooms for the latter. It might thus be a good moment for the emergence of new metrics, and the moment could be seized to discriminate between the two most important types of DC infrastructure.

## 11.2 Addressing complex water-energy trade-offs

Section 10.1.2 summarised the two related trade-offs, between local and upstream water consumption, and between total water and energy consumption. As Equation 26 had already shown, the water-water trade-off depends decisively on the water intensity of electricity production and thus of the power sources. Depending on it, the water-energy trade-off is more or less pronounced.

At the same time, Section 4.1 had presented at length the main cooling technologies deployed in data centres. One aspect only superficially discussed there is the adequacy of different technologies for different climates. One likely rewarding avenue for further research would be an in-depth analysis of these five intertwined dimensions: climate, DC cooling technologies, electricity sources, energy consumption, and water consumption.

An investigation of the best metrics for the assessment of water use would be a worthy complement. While Chapter 4 quickly settled on water consumption as the more meaningful indicator than water withdrawal (an assessment that still stands), water is additionally a highly localised resource with different local value depending on its scarcity. Environmental sciences have developed various water scarcity assessment methods, and they could be examined for suitability in the context of DC water consumption.

## 11.3 Advancing circularity

### 11.3.1. The meaningfulness of heat reuse

Section 5.4.1 has shown that both metrics for heart reuse, $ERF$ and $ERE$, only take into consideration the quantity of heat reused, not its quality. In particular, no metric today considers a *counterfactual*. The implicit assumption behind metrics such as $ERF$ and $ERE$ is that by being reused they have substituted energy of the same quality that would otherwise have been spent to create the heating effect. This assumption is generally not true, as the example of air-source heat pumps shows: Thes would generate the same heat at just a fraction of the electricity input.

By contrast, a counterfactual describes "what would have happened otherwise". The concept is well-established in fields such as economics to study, for example, the effects of policy interventions (Caliendo and Hujer, 2006). It has been adopted in the context of net impact assessment of digitalisation, where other terms are being used synonymously, such as "baseline" (Coroamă *et al.*, 2020), "baseline scenario" (ITU-T, 2022), or "reference scenario" (Thieme and Prettejohn, 2024).

A metric more reflective of reality for the circularity of energy reuse would be based on counterfactuals. It would not account for the reused heat but the *replaced* (and thus saved) energy. Sometimes the two might coincide, often probably not. The balance will depend on the source of energy in the counterfactual, possibly its quality (in terms of genericity and low entropy), certainly of its efficiency.

Creating such a metric while keeping it manageable is both a challenging and an intriguing prospect. It would also be conform to the LCA principle of "avoided burden" (ISO, 2006). To account for circularity, the avoided burden principle requires system expansion. The environmental impacts of a co-product or secondary material are then subtracted from the main product system because they avoid the need to produce virgin material elsewhere.

For DCs, that means to expand the system boundaries to include the heat reuse – which already takes place for the $ERF$ and $ERE$ metrics. Instead of computing the avoided burden, however, the current paradigm takes the simplistic approach of postulating the quantity of reused heat as equal to the avoided burden.

### 11.3.2. Establishing circularity for AI and addressing possible trade-offs with efficiency

As addressed at several points throughout this report, the current AI growth leads to substantial changes in DC design at all levels: Server design changes from general-purpose processing units towards accelerated computing processors with high-bandwidth memory. Server racks change to accommodate the new servers and their new power and cooling needs. Power distribution changes to allow for much higher power densities. And to reject the additional amount of energy, cooling systems change as well.

In this context, the reusability of legacy data centre designs becomes challenging and, in many cases, probably impossible. The more important, however, becomes the development of circularity models for the future data centres. These should focus on different topics, depending on the lifespan of components: Server components have a lifespan of a few years. In their case, modularity (allowing for functional upgrades of individual components) and an EoL-planning that prioritises both ease of reuse and the dismantling and material recovery is likely the most important.

As opposed to servers, the DC infrastructure (power provisioning and cooling) has much longer lifetimes, of 10-15 years and possibly more. Modularity is less important than a design taking foreseeable future developments and possible upgrades into account. Power conversion and liquid cooling standards across developers would also increase the likelihood that the infrastructure can be used for its entire technical lifespan.

Finally, it is worth investigating the possible trade-off between resource consumption and energy efficiency highlighted in Section 10.2.3. Purpose-built ASICs could make AI inference much faster and energy efficient; on the other hand, they would need to be discarded after a few months. This trade-off, however, might not be simply between energy and material consumption. As discussed in Section 10.2.3, such ASICs could also reduce the need for more generic components such as high-bandwidth memory or GPUs.

Whether building such ASICs is economically viable, likely depends on various factors such as material versus energy costs, lifespan of models and amount of inferences along this lifespan, or possible hybrid paradigms (in which some parts of a model would be etched while others could still be changed, thus enabling a limited reuse of these integrated circuits at the price of additional energy consumption). If not economically viable, then this phenomenon will likely remain a negligible niche. If viable, however, it might become an increasingly important trend.

An economic assessment would thus first be required to examine the viability of this idea. If viable, to properly understand its sustainability consequences, a comparative LCA of general-compute inference versus inference via dedicated ASICs is likely the most promising avenue. It would need to be a thorough LCA under various usage and hardware lifespan scenarios. And one that would bundle understanding of several domains such as energy consumption of AI compute, lifecycle assessment of microelectronics, future AI developments, and AI demand scenarios.

#### 11.3.3. Better understanding the production impact of DC components

As argued throughout Chapter 6, however, the production impact of microelectronics is poorly understood. Producers publish very little data and even the methodologies (such as allocation between different product systems, as many materials used in microelectronics are by-products of other processes) are not sufficient. This is valid for all three main production phases, but perhaps mostly for the refining and purification stage of materials. A large and coordinated effort is required to increase both data quantity and quality to enable, for example, comparative LCAs such as the one required above.

## 11.4 Shed light on the complex relationship between DC growth and the grid

Chapter 8 discussed the complex relationship between data centre growth and the power grid, a discussion graphically represented in Figure 10 and summarised in Section 10.3 of the Conclusions. Several of those topics deserve further attention.

Section 3.2 presented a new, direct current power distribution paradigm for DCs, which is required for the denser power of modern DCs. Together with the emergence of solid state transformers, it was argued that traditional UPS systems might be replaced by novel BESSs that migrate deeper into the DC and thus closer to the servers where they can both more quickly react and absorb the steep power ramps directly at their source, not letting them propagate to the grid. While this is certainly a desirable property, the batteries are also foreseen as possibly advantageous by providing short-term peak-shaving services to the grid (see Section 10.3). Whether and to which extent this new architecture can still fulfil this other desirable goal, or whether there is a trade-off between the two, is worth investigating.

It has further been argued that DC battery systems can be both a competition to other battery systems that are critical for the green transition (i.e., to batteries of electric vehicles and grid-level batteries alike), but that they could also spur new innovation in the field. These influences are worthy of a further-reaching investigation, including possibly policy measures required.

Finally, a deep dive into on-site generation is highly needed. This is a very new phenomenon, more than 90% of the projects being announced since the beginning of 2025, as described in Section 8.4.1. A of the relevant questions are how low-carbon sources (such as renewables or nuclear) can be incentivised and gas discouraged, but also which other challenges these alternatives would bring (e.g., scalability and intermittence for renewables, security and radioactive waste for nuclear). Irrespective of the technology, most of these behind-the-grid solutions are developed only as temporary crutches as long as grid connectivity is slow are outright unavailable. A pertinent question, however, is how they could be designed for future benefits and reusage: Once grid connectivity would be available, how could these on-site generation solutions be best put to use in the future, e.g. as backup generators for the DCs themselves or as sources of flexibility for the grid?

## 11.5 Cross-cutting topics: DC siting and local communities

Finally, there are two cross-cutting topics that might be of interest for further investigation. One of them are the various drivers for DC development site choice as well as the resulting consequences of this choice. Important criteria for this choice include available grid capacity and the stability of the grid, geopolitical and economic stability, and the availability of qualified workforce. But they also include sustainability-relevant attributes such as the required cooling overhead, carbon intensity and water intensity of the electricity, the potential demand for grid flexibility services that DCs can bring, and the potential for circularity (including the potential for heat reuse and the possible need for reducing renewable curtailment).

Data centre impact on local communities is another such cross-cutting topic. Detrimental consequence can occur due to the DCs water consumption (particularly in water-scarce regions), increased local power prices due to nearby DC development and in the worst case even grid availability or stability issues, and air, noise and soil pollution, even if – or perhaps primarily if – the DC is not connected to the power grid, but powered by on-site gas turbines. On the other hand, DCs can also bring several economic benefits, both directly through e.g. attractive employment opportunities and indirectly, for example via a substantial amount of locally paid taxes.